%% file: main_arXiv.tex
\documentclass[twocolumn]{aastex63}

\usepackage{amsmath,amssymb,natbib}
\usepackage{microtype}
\usepackage{hyperref}
\usepackage{graphicx}

\usepackage{comment}
\usepackage{color}
\usepackage{url}

\usepackage{dcolumn}

\input{commands}


\input{citation_fix}

\newcommand{\beq}{\begin{equation}}
\newcommand{\beqa}{\begin{eqnarray}}
\newcommand{\eeq}{\end{equation}}
\newcommand{\eeqa}{\end{eqnarray}}

\newcommand{\AU}{\mathrm{AU}}
\newcommand{\mas}{\mathrm{mas}}
\newcommand{\rmday}{\mathrm{day}}
\newcommand{\Msun}{M_{\odot}}
\newcommand{\s}{\mathrm{s}}
\newcommand{\km}{\mathrm{km}}

\newcommand{\brak}[1]{\left( #1 \right)}

\newcommand{\sqbrak}[1]{\left[ #1 \right]}
\newcommand{\angbrak}[1]{\left\langle #1 \right\rangle}

\newcommand{\E}{\mathrm{E}}
\newcommand{\rel}{\mathrm{rel}}

\newcommand{\X}{\mathrm{X}}

\newcommand{\N}{\mathrm{N}}

\newcommand{\ds}{d_{\rm s}}
\newcommand{\dl}{d_{\rm l}}

\newcommand{\dsmin}{d_{\rms,\min}}
\newcommand{\dsmax}{d_{\rms,\max}}

\newcommand{\vtheta}{\boldsymbol{\theta}} 
\newcommand{\vv}{\boldsymbol{v}} 

\newcommand{\vu}{\boldsymbol{u}} 
\newcommand{\vs}{\boldsymbol{s}} 

\newcommand{\vd}{\boldsymbol{d}}

\newcommand{\tilvv}{\tilde{\vv}}
\newcommand{\vSigma}{\boldsymbol{\Sigma}}

\newcommand{\rE}{r_{\rm E}} 
\newcommand{\thetaE}{\theta_{\rm E}} 
\newcommand{\tE}{t_{\rm E}} 

\newcommand{\piE}{\pi_{\rm E}} 
\newcommand{\vpiE}{\boldsymbol{\pi}_{\rm E}} 
\newcommand{\piEN}{\pi_{\rm E,N}} 
\newcommand{\piEE}{\pi_{\rm E,E}} 

\shorttitle{Finding GW black holes with parallax microlensing}
\shortauthors{Toki \& Takada}

\begin{document}

\rightline{IPMU21-0019}
\title{Finding gravitational-wave black holes with parallax microlensing}

\email{satoshi.toki@ipmu.jp, masahiro.takada@ipmu.jp}
\author{Satoshi~Toki}
\affiliation{%
Department of Physics,
The University of Tokyo, Bunkyo, Tokyo 113-0031, Japan
}%
\affiliation{%
Kavli Institute for the Physics and Mathematics of the Universe (WPI), The University of Tokyo Institutes for Advanced Study (UTIAS),
The University of Tokyo, 5-1-5 Kashiwanoha, Kashiwa, Chiba, 277-8583, Japan
}
\author[0000-0002-5578-6472]{Masahiro Takada}
\affiliation{%
Kavli Institute for the Physics and Mathematics of the Universe (WPI), The University of Tokyo Institutes for Advanced Study (UTIAS),
The University of Tokyo, 5-1-5 Kashiwanoha, Kashiwa, Chiba, 277-8583, Japan
}%

\date{\today}

\begin{abstract}
The LIGO-Virgo gravitational-wave (GW) observation unveiled the new population of black holes (BHs) that appears to have an extended mass spectrum up to around $70M_\odot$, much heavier than the previously-believed mass range ($\sim 8M_\odot$).
In this paper, we study the capability of a microlensing observation of stars in the Milky Way (MW) bulge region 
to identify BHs of GW mass scales, taking into account the microlensing parallax characterized by the parameter $\piE\propto M^{-1/2}$ ($M$ is the mass of a lens), which is a dimension-less quantity defined by the ratio of the astronomical unit 
to the projected Einstein radius.
First, assuming that BHs follow the same spatial and velocity distributions of stars as predicted by the standard MW model,
we show that microlensing events with long light curve timescales, $\tE\gtrsim 100~{\rm days}$, and small parallax effects, $\piE\sim 10^{-2}$, are dominated by BH lenses compared to stellar-mass lenses. 
Second, using a Markov chain Monte Carlo analysis of the simulated light curve, we show that BH lens candidates are securely identified on individual basis, if the parallax effect is detected or well constrained to the precision of a percent level in $\piE$. We also discuss that a microlensing event of an intermediate-mass BH of $\sim 1000M_\odot$, if it occurs, can be identified in a distinguishable way from stellar-mass BHs. 

\end{abstract}





\section{Introduction}
\label{sec:introduction}

The LIGO-Virgo Gravitational-Wave Transient Catalog 2 \citep[GWTC-2][]{2020arXiv201014527A} revealed the population properties of black holes (BHs) and neutron stars (NSs) in compact 
binary systems from the gravitational wave (GW) observation. The BH population displays the mass spectrum that is fairly well-fitted by a Salpeter-like power-law form, but extends to $\sim 65\,M_\odot$ or beyond \citep{2020arXiv201014533T}, which is different from the previously known BH population in X-ray binary systems that is centered around $8\,M_\odot$.
Further intriguingly,
the GW signal GW190521 \citep{2020PhRvL.125j1102A} indicates the binary BH merger with a total mass of $150\,M_\odot$, whose primary BH mass lies within the mass gap due to pair-instability supernova processes \citep{1967PhRvL..18..379B}.
Thus the LIGO-Virgo GW observation has triggered an intense discussion on the origin and nature of such massive BHs.

There are mainly three channels of GW binary BH formation that have been studied; from binary systems of massive stars, e.g. through common envelope evolution in low-metallicity environments
\citep[e.g.][]{1998ApJ...506..780B,2002ApJ...572..407B,2016Natur.534..512B}; through mergers of lighter mass BHs in star clusters \citep[e.g.][]{2000ApJ...528L..17P,2020arXiv200609744S}
and galactic nuclei \citep[e.g.][]{2012ApJ...757...27A}; 
or through gas drag and stellar scattering in an accretion disk surrounding a super-massive BH at the center of a galaxy \citep[e.g.][]{2012MNRAS.425..460M,2019PhRvL.123r1101Y}.
Alternatively, GW binary BHs could originate from primordial BHs that might have formed in the early universe \citep{Sasakietal:16,Birdetal:16,2020arXiv200109160K}.
Thus the origin of GW BHs involves rich physical processes and an observational exploration of BHs is of critical importance in the next decade. 

Gravitational microlensing \citep{Paczynski:86,Griestetal:91,1991ApJ...374L..37M} provides us with a powerful tool to search for BHs or more generally any compact, invisible objects in the Milky Way (MW) Galaxy, plus the Andromeda galaxy if it is taken as a target of a microlensing observation. 
When a compact object lens approaches and then moves away from a background source star in almost perfect alignment
with the line-of-sight direction of an observer, the flux of the source star is magnified, yielding a characteristic light curve as a function of time. The light curve timescale depends on a combination of the lens mass and the relative velocity between the lens, source and observer \citep{1995ApJ...447...53H,1996ApJ...467..540H}. Various observations/experiments have proven that microlensing can be indeed used to constrain the population of BHs and 
other invisible objects such as exoplanets, brown dwarfs, free-floating planets and primordial BHs \citep{1995ApJ...454L.125A,Alcocketal:00,2002MNRAS.329..349M,2002ApJ...579..639B,Sumietal:03,OGLE:09,OGLE:10,Sumietal:11,2016MNRAS.458.3012W,2018arXiv181100441M,2017arXiv170102151N,2019PhRvD..99h3503N,2019ApJ...885....1W,2020A&A...636A..20W}.

The purpose of this paper is to study the capability of a microlensing
observation of stars in the MW bulge region to find and constrain the BH population of $\sim40\,M_\odot$ mass scales in the MW.
A GW mass-scale BH generally causes a long-timescale microlensing event with $\tE\gtrsim 100~{\rm days}$, compared to a stellar mass ($\sim 1\,M_\odot)$ lens that is a majority of the population of lenses in the MW \citep{1996ApJ...473...57M,2005MNRAS.362..945W,2016ApJ...830...41L,2019PhRvD..99h3503N,2020ApJ...889...31L,2020ApJ...905..121A}.
However, measuring only the timescale of the microlensing light curve does not allow us to obtain a definite mass estimation of the BH lens because of a severe degeneracy between the lens mass and relative velocity.
This difficulty can be resolved by the microlensing parallax \citep{1966MNRAS.134..315R,1986Natur.324..126G,2000ApJ...535..928G,2004ApJ...606..319G} \citep[][for the first observation]{1995ApJ...454L.125A}; 
the microlensing parallax is caused by the orbital motion of the Earth around the Sun and imprints characteristic modifications in the microlensing
light curve. 
The microlensing parallax effect depends on the Einstein radius of a lens relative to the separation between the Earth and the Sun, i.e. the astronomical unit (AU). Hence, GW BHs generally predict small microlensing parallax effects compared to stellar-mass lenses. 
If a sufficiently accurate photometry is available to measure or precisely constrain the parallax effect in
an observed light curve, we can robustly discriminate a GW BH candidate from a stellar mass lens, as we will show below in detail. 
First, we study statistical properties of the microlensing parallax for GW BH lenses for an observation of the Galactic bulge region, assuming that BHs follow the same spatial and velocity distributions as those of main-sequence stars (MSs) in the standard model of the MW bulge and disk. Second, we study how a BH candidate can be robustly identified from an observation of an individual microlensing light curve. To do this, we use a Markov chain Monte Carlo analysis for various simulated light curves to estimate the posterior distributions of lens masses. We will show that the microlensing parallax is indeed very powerful to identify secure candidates of GW BHs. 
We also study how the microlensing observation 
can be used to identify an event by an intermediate-mass BH of $\sim 1000M_\odot$ that might exist in the Galactic bulge region if such an event occurs, and could give a pathway to an understanding of the formation of a super-massive BH at the Galactic center. Our study is relevant to the 10-year monitoring observation with the Vera Rubin Observatory's Legacy Survey of Space and Time (LSST)\footnote{\url{https://www.lsst.org}} \citep{2018arXiv181204445S,2020arXiv200414347G}. 

The structure of this paper is as follows. In Section~\ref{sec:preliminaries} we 
briefly review the theory of the microlensing light curve including the parallax effect. 
In Section~\ref{sec:stat_distribution} we study the statistical properties of BH microlensing events, 
focusing on the parallax effect, and compare to those of stellar lenses, 
using the standard model of the MW to describe the spatial and velocity distributions of lens populations. 
In Section~\ref{sec:mass_estimation} we study
the capability of an observation of an individual microlensing
light curve to identify the candidates 
of BH lenses. Section~\ref{sec:conclusion} is devoted to discussion and conclusion.

\section{Preliminaries: Basics of Microlensing Parallax}
\label{sec:preliminaries}

\subsection{Microlensing light curve}
\label{sec:tE_piE}

When a lens approaches a source star by separation closer than the Einstein radius on the sky, the source star is doubly imaged and its flux is magnified from its original flux
 \citep{Paczynski:86,1992grle.book.....S}. 
While the multiple images cannot be spatially resolved, we can observe a time-varying brightness of the source star -- 
the microlensing light curve 
-- when the lens is moving against the line of sight to the source.
 The Einstein radius is given by 
\begin{align}
r_{\rm E}&\equiv \frac{\sqrt{4 G M d_{\rm l}d_{\rm ls}/d_{\rm s}}}{c},
\label{eq:RE_def}
\end{align}
where $M$ is the lens mass, and $d_{\rm l}$, $d_{\rm s}$ and $d_{\rm ls}$ are the distances to the lens, to the source, and between the lens and the source, respectively.
Throughout this paper we consider an observation of microlensing 
light curve for a source star at distance $8\,\kpc$
in the Galactic bulge region.
The angular extent of the Einstein radius on the sky gives a scale of the angular separation between the double images: 
\begin{align}
\theta_{\rm E}\equiv \frac{r_{\rm E}}{d_{\rm l}}
\simeq 6.4~{\rm mas}\left(\frac{M}{40M_\odot}\right)^{1/2}\left(\frac{D}{8~{\rm kpc}}\right)^{-1/2},
\label{eq:thetaE}
\end{align}
where $D\equiv d_{\rm s}d_{\rm l}/d_{\rm ls}$. Here we assume $\dl=4\,\kpc$ for a lens distance and $M=40M_\odot$ for a typical mass of lens BHs that represent the LIGO-Virgo GW BHs,
and in this case $\theta_{\rm E}\sim 6.4~{\rm mas}$, meaning that the double images cannot be resolved by an optical telescope, but the change in the centroid position of a source star, i.e. astrometric lensing, could be detected if a sufficiently significant detection of the source's flux is obtained \citep[e.g., see][for the astrometry accuracy of a ground-based data]{2020arXiv200412899Q}.

The timescale of a microlensing 
light curve is characterized by the Einstein timescale, which is the time for a source star to cross the Einstein radius of a lens: 
\begin{align}
t_{\rm E}\equiv \frac{r_{\rm E}}{v_{\rm rel}}, \label{eq:tE}
\end{align}
where $v_{\rm rel}$ is the relative velocity between the lens and the source, projected onto the lens plane perpendicular to the line-of-sight 
direction:
\begin{align}
\vv_{\rm rel}=\vv_{\rm l}-\left[\frac{d_{\rm l}}{d_{\rm s}} \vv_{\rm s}+\frac{d_{\rm ls}}{d_{\rm s}}(\tilvv_{\odot}+\tilvv_\oplus)\right].
\label{eq:vrel_total_def}
\end{align}
Here $\vv_{\rm s}$ and $\vv_{\rm l}$ are the velocities of the source and the lens with respect to the Galactic center
(the rest frame of the Galaxy), $\tilvv_\odot$ is the velocity of the Sun, and $\tilvv_\oplus$ is the velocity component of the Earth's orbital motion around the Sun, in the two-dimensional plane perpendicular to the line-of-sight direction.
For convenience of our discussion, we define the timescale of a 
microlensing 
light curve observed from the Sun's position:
\begin{align}
t_{\rm E,\odot}&\equiv \frac{r_{\rm E}}{v_{\rm rel,\odot}}\nonumber\\
&\hspace{-1em}\simeq 221~{\rm days}\left(\frac{M}{40M_\odot}\right)^{1/2}\!\!\left(\frac{d_{\rm l}d_{\rm ls}/d_{\rm s}}{2~{\rm kpc}}\right)^{1/2}
\!\!\left(\frac{v_{\rm rel\odot}}{200~{\rm km~s}^{-1}}\right)^{-1}, 
\label{eq:tE_sun}
\end{align}
where $v_{\rm rel,\odot}$ is the amplitude of the relative velocity without the contribution of the Earth's motion (i.e. the relative velocity when $\tilde{\vv}_\oplus=0$ in Eq.~\ref{eq:vrel_total_def}). 
For the typical relative velocity of the Sun with respect to the lens or the source,
we assume $v_{\rm rel,\odot}=200~{\rm km~s}^{-1}$, which is
taken from the Galactic rotation velocity (see below). 
A BH lens with $\sim 40~M_\odot$ causes  
a microlensing 
event that typically has a timescale of $\sim 221~$days, which is much longer than that of an event caused by 
an MS lens ($\sim M_\odot$) by a factor of $40^{1/2}\simeq 6$. 
However, the above equation shows that the same timescale can be realized from different combinations of lens mass and $v_{\rm rel,\odot}$ (even if the distance is fixed); for example, the lens-source system with $(4M,2v_{\rm rel,\odot})$ or $(M/4,v_{\rm rel,\odot}/2)$
gives the same 
timescale as that with $(M,v_{\rm rel,\odot})$. 
Thus, one cannot uniquely estimate the lens mass solely from the light curve timescale, unless a typical relative velocity is a priori assumed.

Since the proper motions of the Sun, the lens and the source star can be considered to be straight motions in most cases for a period of a microlensing observation, 
we can regard the relative velocity, $\vv_{\rm rel,\odot}$, as a constant (time-independent) vector.
On the other hand, the Earth's Kepler motion around the Sun varies within a year timescale as implied by Eq.~\eqref{eq:vrel_total_def}, and it causes 
a characteristic modification in the 
microlensing 
light curve that is actually seen by an observer on the Earth. 
This is the 
microlensing 
parallax \citep{2004ApJ...606..319G}. In the next section we discuss the parallax effect.

\subsection{Light curve formula with 
microlensing 
parallax}
\label{sec:lightcurve}

When a lens passes through our line of sight direction to a source star, 
the separation between the lens and the source in the lens plane, in units of the Einstein 
radius, is given by
\beq
\vu(t)=\brak{u_{\parallel}(t),u_{\perp}(t)}=\brak{\frac{t-t_0}{t_{{\rm E}}},u_0}, \label{eq:u_wo_parallax}
\eeq
where the quantities with subscripts,
``$\parallel$'' or ``$\perp$'', denote the components parallel or perpendicular to the relative velocity direction, respectively,  
$u_0$ is the impact parameter (the shortest separation between the lens and the source), and 
$t_{{\rm E}}$ is the Einstein timescale (see Eq.~\ref{eq:tE}).
The time-varying magnification, i.e. the 
microlensing light curve, is given by
\beq
\mu(t)\equiv \frac{f(t)}{f_{\rm int}}=\frac{u^2+2}{u\sqrt{u^2+4}}, \label{eq:mu}
\eeq
where $u(t)=\sqrt{u_{\parallel}(t)^2+u_\perp(t)^2}$, $f(t)$ is the observed flux of the source at an epoch $t$, and $f_{\rm int}$ is the intrinsic flux.
When a source and a lens get closer than the Einstein radius, i.e. $u<1$, the flux of the source is magnified by more than 
a factor of 1.34. 

Following \citet{2004ApJ...606..319G}, we now study the 
microlensing light curve seen by an observer on the Earth, taking into account 
the parallax effect due to the Earth's orbital motion around the Sun. 
Let $\vs(t)$ be the three-dimensional Earth-to-Sun position vector in the heliocentric frame. 
Then consider the two trajectories starting from a pivot epoch $t_p$; one is the trajectory taking into account the Earth's rotation around the Sun, and the other is a ``hypothetical'' trajectory where the Earth is considered to have a straight-motion with a constant velocity. The difference position vector between these two trajectories is
\beq
\Delta \vs(t) = \vs(t)-[\vs(t_p)+\vv_{\oplus}(t_p)(t-t_p)],
\label{eq:Ds_3D}
\eeq
where $\vv_\oplus$ is used to denote the three-dimensional velocity of the Earth's Kepler motion; $|\vv_\oplus|\simeq 
30~{\rm km~s}^{-1}$, $t_p$ denotes the reference (pivot) time, and $\vv_\oplus(t_p)$ is the Earth's velocity vector at $t_p$. 
Here $\vs(t)-\vs(t_p)$ is the trajectory including the Kepler motion (the first trajectory in the above two), while $\vv_\oplus(t_p)(t-t_p)$ is
the hypothetical straight-line trajectory (the latter trajectory). The path lengths of the two trajectories are the same, by definition: 
$\left|\int^t_{t_p}\!\mathrm{d}\vs(t) \right|=|\vv_\oplus(t_p)(t-t_p)|$.
In the following we assume the Earth's Kepler motion as a circular orbit for simplicity to model the trajectory $\vs(t)$.
For a given observation epoch, we can precisely know the Earth's motion relative to the Sun, and $\Delta \vs(t)$
is a known quantity, not a parameter. In addition, we can know the components of $\Delta\vs$ projected onto the lens plane (the two-dimensional plane 
perpendicular to the line-of-sight direction to a source star).
The direction of the Galactic bulge is almost in the ecliptic plane. We also note that the direction of $\vv_\oplus$ becomes almost perpendicular to the line-of-sight direction (the Galactic bulge direction) at the solstices, while it is almost parallel to the line-of-sight direction at the equinoxes. 
We use the publicly-available software \textsf{Astropy} to compute the parallax displacement, $\Delta \vs(t)$, for input values of parameters 
(the observation epoch and the RA and dec direction of a target source star).

Projecting Eq.~\eqref{eq:Ds_3D} onto the two-dimensional plane perpendicular to the line-of-sight direction at the observer's position (the observer plane),
we obtain the displacement of the Sun due to the Earth's motion: 
\beq
\Delta \tilde{\vs}(t) = \tilde{\vs}(t)-[\tilde{\vs}(t_p)+\tilde{\vv}_{\oplus}(t_p)(t-t_p)].
\label{eq:Ds_projected}
\eeq
Here we assume that all these quantities are normalized by $\AU$. 
The displacement on the lens plane $\Delta\vu(t)$ at time $t$ is
\beq
\Delta\vu(t)=\piE\Delta\tilde{\vs}(t).
\eeq
Then the orbit of the lens, seen by an observer on the Earth, is modified from Eq.~(\ref{eq:mu}), as
\begin{align}
\vu_{\rm parallax}(t)&=\vu(t)+\Delta\vu(t)\nonumber\\
&=\brak{\frac{t-t_0}{t_{\rm E}}+\Delta
u_\parallel(t),u_0+\Delta u_\perp(t)},
\label{eq:mu_earth}
\end{align}
where
\begin{align}
\Delta u_\parallel(t)= \vpiE\cdot\Delta\tilde{\vs}(t),
\quad \Delta u_\perp(t)=\vpiE\times\Delta\tilde{\vs}(t),
\label{eq:Du}
\end{align}
and the Einstein time scale is given by
\beq
\tE=\frac{\rE}{v_{\rel}(t_p)}. \label{eq:tE_Earth}
\eeq
Here $\vpiE$ is the vector whose amplitude is given by the 
microlensing 
parallax
\begin{align}
\pi_{\rm E}&\equiv \frac{{\rm AU}}{\tilde{r}_{\rm E}}\nonumber \\
	&\simeq 0.020 \left(\frac{M}{40M_\odot}\right)^{-1/2}\left(\frac{D}{8~{\rm kpc}}\right)^{-1/2}, 
	\label{eq:piE_def}
\end{align}
where $\tilde{r}_{\rm E}\equiv D\theta_{\rm E}$. 
The direction of $\vpiE$ is given by 
\beq
\frac{\vpiE}{\piE}\equiv \frac{\vv_{\rel}(t_p)}{v_{\rel}(t_p)}.
\label{eq:piE_vector}
\eeq
The parameter $\piE$ is given by the ratio of the Einstein radius, projected onto the observer plane, to AU (the length of the Earth's orbital displacement), and does not depend on the relative velocity.
If $\pi_{\rm E}$ is larger, the 
microlensing 
parallax effect is greater, as we will show below.
Plugging Eq.~(\ref{eq:mu_earth}) in Eq.~(\ref{eq:mu}) (replacing $u(t)$ with $u_{\rm parallax}(t)$), we can obtain the 
microlensing 
light curve including the parallax effect.

As we will show later quantitatively (e.g. Fig.~\ref{fig:lightcurve_M40D}), the 
microlensing 
parallax effect imprints characteristic signatures in the 
microlensing 
light curve. 
Expanding 
the magnification $\mu[u(t)]$ (Eq.~\ref{eq:mu}) 
to the linear order in $\Delta\vu=\piE\Delta\tilde{\vs}$, we obtain
\beq
\Delta\mu=\mu[u_{\rm parallax}]-\mu[u]=\frac{4}{u^3(u^2+4)^{3/2}}\piE \vu\cdot\Delta\tilde{\vs}.
\label{eq:Delta_mu}
\eeq
Recall that the displacement vector, projected onto the lens plane, $\Delta\tilde{\vs}$ is in units of AU. Here
the displacement of the Earth motion around the Sun is given as
$\vu\cdot\Delta\tilde{\vs}\sim u [4\sin^2\{\Omega_\oplus (t-t_p)/2\}
+\{v_{\oplus} (t-t_p)/\mathrm{AU}\}^{2}
-2 \{v_{\oplus} (t-t_p)/\mathrm{AU}\} \sin \Omega_{\oplus} t]$, where $\Omega_\oplus$ is the angular velocity of the Earth's circular motion around the Sun. 
Eq.~(\ref{eq:Delta_mu}) shows that the parallax effect on the lensing magnification is proportional to the microlensing parallax: $\Delta\mu\propto \piE\propto 1/\tilde{r}_{\rm E}\propto M^{-1/2}$.
Although the parallax effect on the lensing magnification at a certain epoch
cannot be discriminated from the intrinsic lensing magnification, we can extract the parallax information from the {\it shape} of the light curve, i.e. the time-dependent variation in the light curve compared to the light curve without the parallax effect.

\begin{figure}
\centering
\includegraphics[width = \columnwidth]{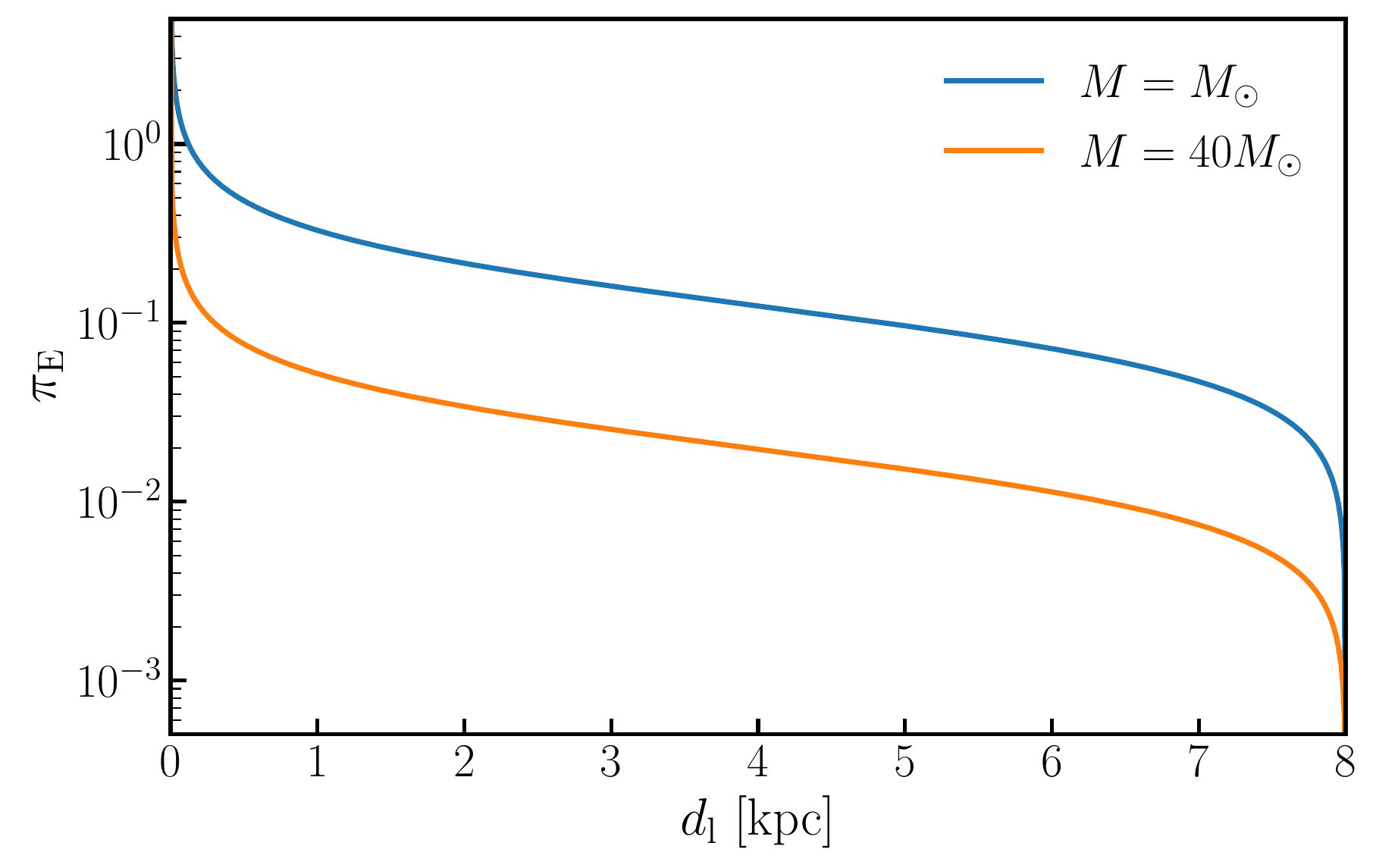}
\caption{The dependence of the 
microlensing 
parallax $\piE$
on the distance ($\dl$) and mass ($M$) of a lens, for a 
microlensing 
event for a source star at $d_{\rm s}=8\,{\rm kpc}$ in the Galactic bulge region. 
Due to the dependence $\piE\propto M^{-1/2}$ (see Eq.~\ref{eq:piE_def}), a BH of $40\,M_\odot$ predicts a much smaller value of $\piE$ than a 
star lens of $1\,M_\odot$. 
}
\label{fig:piE_vs_dl}
\end{figure}
Fig.~\ref{fig:piE_vs_dl} shows the dependence of the microlensing parallax $\piE$ on the mass and distance of a lens (we throughout this paper assume a source star at $d_{\rm s}=8\,{\rm kpc}$ in the bulge region).
A BH of $40\,M_\odot$ or heavier masses leads to the small microlensing parallax effect; such a BH lens gives $\piE\sim 10^{-2}$ at almost all the distances, and even $\piE\sim 10^{-3}$ if the lens is closer to the bulge. On the other hand, an MS lens of $\sim 1\,M_\odot$ gives the larger parallax effect, $\piE\sim 10^{-1}$, in most cases, unless the lens is very close to the source star
(we will come back to this question later).

From Eq.~(\ref{eq:tE_sun}) and Fig.~\ref{fig:piE_vs_dl}, we find that 
an event having a long timescale ($\tE\gtrsim 100~{\rm day}$) and a small parallax effect is a good BH candidate.

\section{Statistical properties of microlensing parallax  for BH lenses}
\label{sec:stat_distribution}

In this section, assuming the standard model of the MW to model the spatial and velocity distributions of BH lenses, 
we study statistical characteristics of BH microlensing events, compared to events due to MSs. 

\subsection{Coordinate system}
\label{sec:coordinates}

For convenience of our discussion, we use the same coordinate system as that used in \citet{2019PhRvD..99h3503N}. 
We choose the Galactic center to be the origin of the coordinates, the direction from the origin towards the Sun to be the $x$-axis, the direction of the Galactic rotation to be the $y$-axis, and the direction perpendicular to the Galactic disk which makes the coordinates right-handed to be the $z$-axis.
Here we consider a 
microlensing 
observation towards the Galactic bulge, and assume that the observation regions are approximately along the $x$-axis. 

\subsection{Microlensing event rate}
\label{sec:eventrate}

Following \citet{Griestetal:91}, 
we consider
the differential rate of 
microlensing 
 events for a single source star, due to 
lens objects with mass $M$ at distance $\dl$, which have relative velocity $\vv_{\rm rel}$ and
enter an infinitesimally-thin circular-ring
of radius $\rE$ 
centered at a source star \citep[see Fig.~7 in the supplementary material of][]{2017arXiv170102151N}:
\beq
d\Gamma_i = \sum_{\rm X} n_{i,\rm X}(\dl;M)\rE v_{\rel}^2\cos\theta f_i(\vv_{\rel})dv_{\rel}d\theta d\alpha 
d[\dl]dM. \label{eq:dGamma}
\eeq
Here $n_{X, i}(d_{\rm l}; M)dM$ is the number density of lens objects with masses in the range $[M,M+dM]$, 
and 
its subscript ``X'' denotes different lens populations; either of main-sequence stars (MSs), white dwarfs (WDs), neutron stars (NSs) or black holes (BHs).
The subscript ``$i$'' stands for lens objects in either of the ``disk'' or ``bulge'' region because we need to consider the different velocity and spatial distributions
in the respective region even for the same population of lens objects, as we will show below.
The function
$f_i(\vv_{\rel})$ is the probability density function for lenses that have 
the relative velocity vector $\vv_{\rm rel}$
in the region $i$, where $f_i$ satisfies the normalization condition $\int\!d^2\vv_{\rel}~f_i(\vv_{\rel})=1$;
$\alpha$ denotes the angle between the encounter point of the lens trajectory with the thin ring of radius $\rE$ and the Galactic-plane direction on the sky (i.e. the $y$-direction in our coordinate system);
$\theta$ is the angle between $\vv_{\rel}$ and the line connecting the source and the encounter point.
Hence the two components of relative velocity is given as
\begin{align}
\vv_{\rel}=(v_{{\rm rel},y}, v_{{\rm rel},z})=-v_{\rel}\brak{\cos(\theta+\alpha),\sin(\theta+\alpha)}.
\label{eq:vrel}
\end{align}

We now study the differential 
microlensing 
event rate as a function of the 
microlensing 
parallax $\pi_{\rm E}$.
Using the relations, 
\begin{align}
v_{\rel}=\frac{\rE}{\tE},\quad M=\frac{c^2\AU^2}{4G\piE^2D},\quad u_0=\sin\theta,
\end{align}
we can change the variables ($v_{\rm rel},M,\theta$) in Eq.~(\ref{eq:dGamma}) to $(\tE,\piE,u_0)$ \citep[also see][]{2019PhRvD..99h3503N}
to rewrite $d\Gamma_i$ as
\beq
d\Gamma_i =\sum_{\rm X}n_{i,\rm X}(\dl;M)\frac{c^2\AU^2}{2G\piE^3 D}v_{\rel}^4f_i(\vv_{\rel})
d\tE du_0 d\alpha d[\dl]d\piE. \label{eq:dGamma_piEtEu0}
\eeq
Hence we obtain the differential event rate via 
\begin{align}
\frac{d^2\Gamma_i}{d\tE d\piE}
&=\int_{\dsmin}^{\dsmax}\!d[\ds]\frac{n_{\rms}(\ds)}{N_{\rms}}\sum_{\rm X}\int_{0}^{\ds}d[\dl]\,n_{i,\rm X}(\dl;M) \notag \\
&
\times\frac{c^2\AU^2}{2G\piE^3 D}v_{\rel}^4 \int_{0}^{2\pi}d\alpha\int_{-1}^{1}\!du_0~ f_i(\vv_{\rel}), \label{eq:event_rate_tE_piE_u0}
\end{align}
where $n_{\rms}(\ds)$ is the number density distribution 
of source stars at distance $\ds$
and $N_{\rms}$ is the normalization factor so as to satisfy 
$\int_{d_{\rm s,min}}^{d_{\rm s,max}}\!d[d_{\rm s}]~n_{\rm s}(\ds)=N_{\rm s}$; 
we set $d_{\rm s,min}=4\,{\rm kpc}$ and $d_{\rm s,max}=12\,{\rm kpc}$, meaning that we assume the size of the Galactic bulge to be $8\,\kpc$ in diameter.
Note that $\vv_{\rm rel}$ is given as a function of $\dl, \alpha$ and $u_0$ in the above integration, because 
$t_{\rm E}=r_{\rm E}/v_{\rm rel}$.
For a case that a lens is 
in the disk region, we simply assume that $n_{\rms}(\ds)=N_{\rms}\delta_D(\ds-\bar{d}_{\rms})$, where $\bar{d}_{\rms}=8\,\kpc$.
The total event rate including the contributions  of lenses in the Galactic bulge and disk regions
can be obtained by their sum 
$\Gamma=\Gamma_{\rm disk}+\Gamma_{\rm bulge}$. 
The event rate gives the expected number of 
microlensing 
events for a single star, per unit observation time, per unit interval of the 
microlensing 
parallax $\pi_{\rm E}$ and per the unit interval of the 
microlensing 
light curve timescale $t_{\rm E}$; 
the dimension $[d^2\Gamma/d\tE d\piE]=[{\rm event}~{\rm star}^{-1}~{\rm day}^{-2}]$ (because $\piE$ is a dimension-less quantity).

\subsection{The Milky Way model}
\label{sec:MW_model}

To compute the microlensing event rate 
for a source star in the Galactic bulge, we need to model the spatial
and velocity distributions of lens populations in the MW bulge and disk regions. Here we employ the standard models of the MW in 
\citet{1995ApJ...447...53H} \citep[also see][]{1996ApJ...467..540H,2019PhRvD..99h3503N,2020ApJ...905..121A}, and in this subsection we briefly 
review the model.

First, we introduce the stellar mass density distribution in the MW.
For the bulge region, we employ the ellipsoidal mass profile given in 
\cite{1992ApJ...387..181K}:
\begin{align}
&\rho_{\rmb}(x,y,z) \notag \\
&=\left\{\begin{array}{ll}
1.04\times10^6\brak{\frac{s}{0.482\,\pc}}^{-1.85}\Msun\,\pc^{-3}, & (s<938\,\pc) \\
3.53 K_0\brak{\frac{s}{667\,\pc}}\Msun\,\pc^{-3}, & (s\geq 938\,\pc).
\end{array}\right. \label{eq:mass_dist_Kent}
\end{align}
where $K_0(x)$ is the modified Bessel function, $s^4\equiv R^4+\brak{z/0.61\,\pc}^4$ and $R^2=x^2+y^2$. For the disk region, we employ the 
exponential mass profile
that is given by 
\cite{1986ARA&A..24..577B}:
\begin{align}
&\rho_{\rmd}(x,y,z)\notag \\
&=0.06\exp\sqbrak{-\brak{\frac{R-8000\,\pc}{3500\,\pc}+\frac{|z|}{325\,\pc}}}\Msun\,\pc^{-3}. \label{eq:mass_dist_Bahcall}
\end{align}
These models give $M_{\rm disk}\simeq 2.0\times 10^{10}M_\odot $ or $M_{\rm bulge}\simeq 1.8\times 10^{10}M_\odot$ for the total stellar mass of disk or bulge region, respectively.

For the velocity distribution, we assume the Gaussian distribution around the mean bulk motion of stars:
\beq
f_i(\vv_{\rel})=
\frac{\exp\sqbrak{-\frac{1}{2}
(\vv_{\rel}-\bar{\vv}_i)^{\rm T}\vSigma_{v,i}^{-1}(\vv_{\rel}-\bar{\vv}_i)}}{\sqrt{(2\pi)^2|\vSigma_{v,i}|}},
\eeq
where the covariance matrix is $\vSigma_{v,i}=\mathrm{diag}(\sigma^2_{i,y},\sigma^2_{i,z})$.
The mean velocity 
is given by 
the ensemble average $\bar{\vv}_{i}=\angbrak{\vv_{\rel}}$, where $\vv_{\rel}$ is given by Eq.\eqref{eq:vrel_total_def}. Assuming that 
stars in the bulge region are at rest with respect to the Galactic center on average and 
stars in the disk region have the flat rotational velocity, 
we employ the model given by
\begin{align}
&\bar{\vv}_\rmb=(\angbrak{v_{\rel,y}},\angbrak{v_{\rel,z}})=(-220(1-r)\,\km~\s^{-1},0), \nonumber \\
&\bar{\vv}_\rmd=(\angbrak{v_{\rel,y}},\angbrak{v_{\rel,z}})=(220r\,\km~\s^{-1},0)),
\label{eq:mean_relative_motion}
\end{align}
where $r\equiv \dl/\ds$ and $220~{\rm km~s}^{-1}$ is the Galactic rotational velocity. 
The above equation shows that
lenses in the disk region  have the mean relative motion along the Galactic rotation ($\angbrak{v_{\rel,y}}>0$), 
while lenses in the bulge region  have the mean motion in the opposite direction to the Galactic rotation ($\angbrak{v_{\rel,y}}<0$).

We further assume that the velocity dispersion of stars 
in the bulge region is $100\,\km\,\s^{-1}$. Then we find
\beq
\sigma^2_{\rmb,y}=\sigma^2_{\rmb,z}=(1+r^2)(100\,\km~\s^{-1})^2.
\eeq
For the velocity dispersion of stars in the disk region, we employ the model in \cite{1995ApJ...447...53H}:
\begin{align}
&\sigma^2_{\rmd,y}=(\kappa\dl+30)^2+(100r)^2(\km~\s^{-1})^2, \\
&\sigma^2_{\rmd,z}=(\lambda\dl+20)^2+(100r)^2(\km~\s^{-1})^2,
\end{align}
where $\kappa=5.625\times10^{-3}\,\km~\s^{-1}\,\pc^{-1}$ and $\lambda=3.75\times10^{-3}\,\km~\s^{-1}\,\pc^{-1}$.

\subsection{Mass spectrum of BHs}
\label{sec:mass_spectrum_bh}

In the preceding section we introduced the models to describe the spatial and velocity distributions of stars in the bulge and disk regions. 
Since the Einstein timescale also depends on lens mass, we further need to model the mass distribution of lens objects.
Following the method in \citet{2019PhRvD..99h3503N}, we consider MSs, WDs, NSs and BHs for lens populations. 
For MSs, which we can directly see with a telescope, 
we employ the Kroupa-like mass function \citep{2001MNRAS.322..231K}, which includes the mass spectrum over masses ranging from $0.08\,M_\odot$ to $1\,M_\odot$. 
Note that we assume a Salpeter-like mass function with $\alpha_{\rm MS}=2$
for MSs with $M\ge 0.5\,M_\odot$.
We assume that 
MSs have the universal Kroupa-like mass function in both the bulge and disk regions.

We also assume that heavier stars at birth, i.e. zero-age MSs (ZAMS) with $\ge 1\,M_\odot$, have already evolved into the stellar remnants by today: ZAMSs with $1\le M_{\rm ZAMS}/M_\odot\le 8$ evolved into WDs, ZAMSs with $8\le M_{\rm ZAMS}\le 20$ into NSs, and ZAMSs with $M_{\rm ZAMS}\ge 20\,M_\odot$ into BHs. 
Assuming the number conservation between ZAMSs and the respective stellar remnants, we can obtain the total number of each stellar remnant, relative to the number of MSs. This gives the number ratios for each of the stellar remnants, relative to the number of MSs, today: 
\begin{align}
{\rm MS:WD:NS:BH}=1:0.15:0.013:0.0068.
\label{eq:number_ratio}
\end{align}
That is, we assume 0.0068 BHs per MS. 

For the calculation of 
microlensing 
event rates, we further need to assume the mass distribution of each stellar remnant. For WDs, we adopt a simple mass conversion between each ZAMS progenitor and WD, given by 
$M_{\rm WD}=0.339+0.129M_{\rm ZAMS}$ \citep{2009ApJ...693..355W}. For NSs, we assume a Gaussian mass function with mean $\bar{M}_{\rm NS}=1.33M_\odot$ 
and width $\sigma_{\rm NS}=0.12M_\odot$.

The mass function of BHs, which is the main interest of this paper, is poorly known. Motivated by the recent LIGO/Virgo GW observation, we assume 
that BHs follow a Salpeter-like mass function over the range of $8\le M_{\rm BH}/M_\odot \le 70$, given by 
\begin{align}
\frac{d\phi_{\rm BH}}{d\ln M_{\rm BH}}\propto M_{\rm BH}^{1-\alpha_{\rm BH}}
\end{align}
with $\alpha_{\rm BH}=2$. We normalize the mass function so as to reproduce the number ratio (Eq.~\ref{eq:number_ratio}) 
relative to the abundance of MSs. 

\subsection{Statistical properties of BH microlensing 
events}
\label{sec:statproperties_BHevents}
\begin{figure}
\centering
\includegraphics[width = \columnwidth]{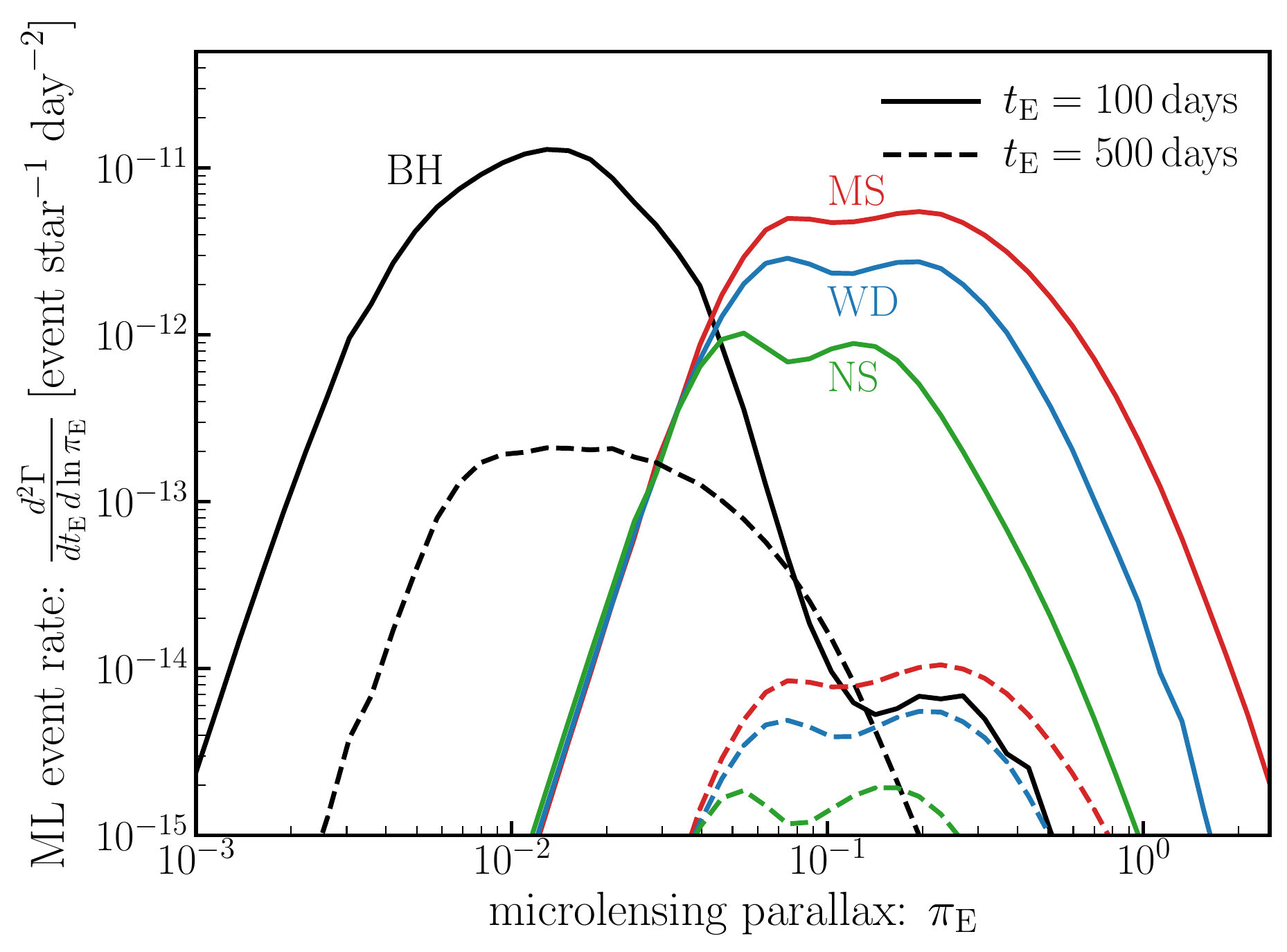}
\caption{The 
microlensing event rate, $d^2\Gamma/dt_{\rm E}\,d\ln\pi_E$, which gives the expected number 
of 
microlensing 
events for a single star in the Galactic bulge, per unit observation time [day], per unit interval of the microlensing timescale $t_{\rm E}$, and per logarithmic interval of the 
microlensing 
parallax $\pi_{\rm E}$. The different color curves show the event rates for 
different populations of lens objects: BH, MS, WD and NS, respectively. We employ the standard bulge and disk models for the spatial and 
velocity distributions of lens populations. The solid and dashed curves show the results for long timescale lensing events of $t_{\rm E}=100~{\rm days}$ and $t_{\rm E}=500~{\rm days}$, respectively. 
}
\label{fig:eventrate}
\end{figure}%

\begin{figure}
\centering
\includegraphics[width = \columnwidth]{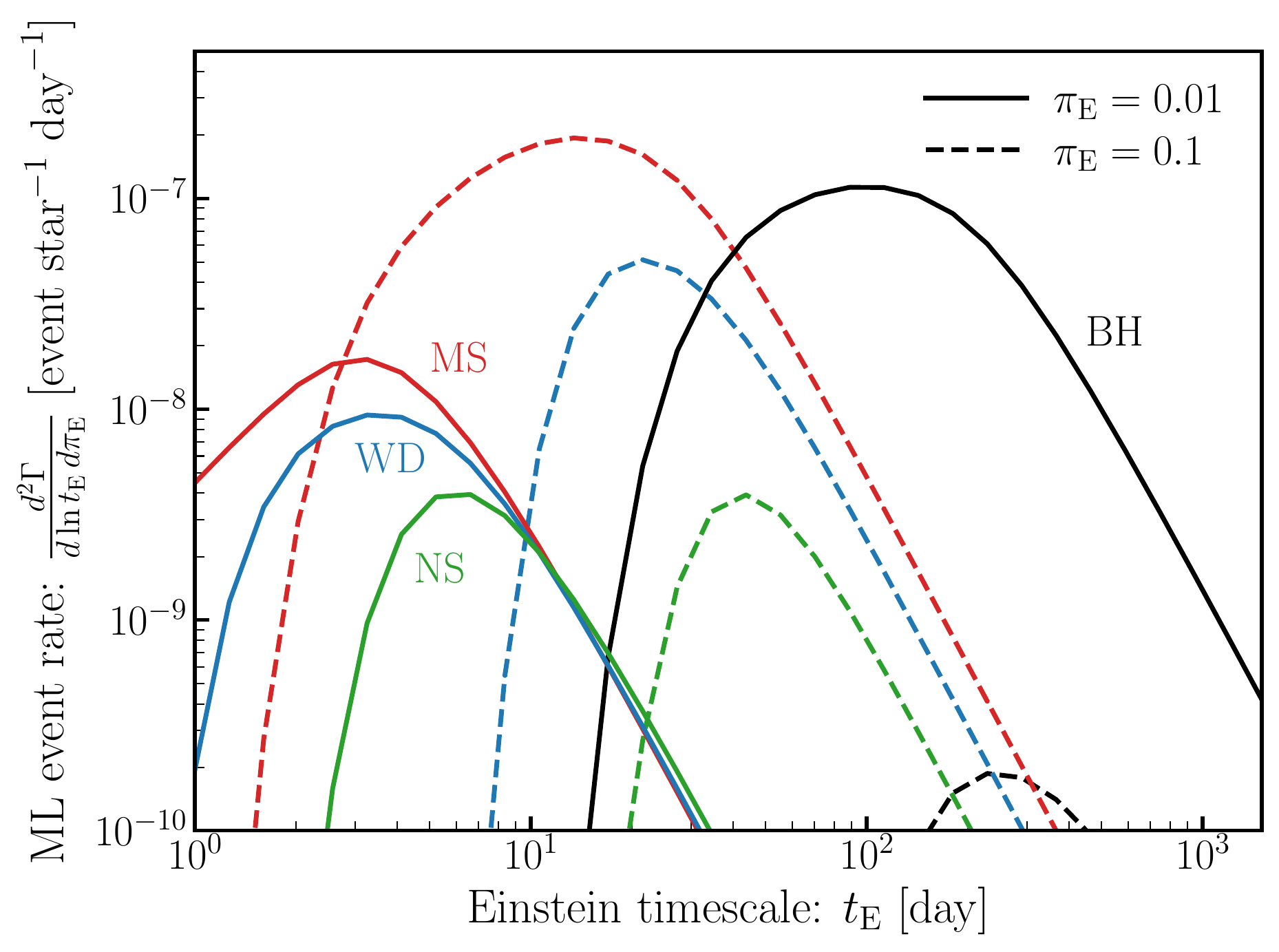}
\caption{Similar to the previous figure, but the plot shows the 
microlensing 
event rate, $d^2\Gamma/d\ln t_{\rm E}\,d\pi_E$ 
per unit observation time, per logarithmic interval of the 
microlensing light curve
timescale $\tE$.
The solid curves show the rate for events of the fixed 
microlensing 
parallax $\piE=0.01$, while the dashed curves are the results for $\piE=0.1$. }
\label{fig:eventrate_fixed_piE}
\end{figure}%

As we described in the preceding section, we model the spatial, velocity and mass distributions of BHs, relative to those of MSs, in the MW bulge and 
disk regions. In our model we assume that massive
lens objects with $M\ge 8M_\odot$ arise only from BHs.
As shown in \citet{2020ApJ...905..121A}, long timescale 
microlensing 
events with $t_{\rm E}\gtrsim 100~{\rm days}$ are dominated by BH lenses due to the boosted dependence 
of the lensing cross section on lens masses, even if BHs are much less abundant than MSs (only 0.0068~BHs per MS). 
This clearly shows the power of 
a microlensing observation to search for BHs in the MW. 

Fig.~\ref{fig:eventrate} shows the event rate plotted against $\piE$ for events that have a fixed Einstein timescale of 
$\tE=100$ or 500~days, respectively.
The differential event rate for BHs is of the order of $10^{-11}\,[{\rm event\,star}^{-1}\,{\rm day}^{-2}]$. As discussed in \citet{2020ApJ...905..121A}, 
if we can conduct a monitoring observation of about $10^{10}$ stars in the bulge region over a 10-year time scale, which can be performed
with the LSST bulge observation \citep{2018arXiv181204445S}, we expect about $3.7\times 10^4$ events for BH lenses with $\tE\sim 100~{\rm days}$ and $\piE\sim 10^{-2}$, which is obtained by 
$N_{\rm s}\times d^2\Gamma/d\tE d\ln\piE \times t_{\rm obs} \times \tE\sim 10^{10}\times 10^{-11}\times 10\,{\rm years}\times 365\,{\rm days/yr}
\times 100\,{\rm days}\simeq 3.7\times 10^4$.
Thus an observation by LSST can find many BH events if BHs follow the spatial and velocity distributions of stars, which would be the case if GW BHs originate from massive stars.
Fig.~\ref{fig:eventrate} also shows that, 
if the microlensing events of $\tE=100$ or 500 days have small microlensing parallaxes of $\piE\sim 10^{-2}$ or smaller, those are very likely to be BH events
because of the dependence
$\piE\propto M^{-1/2}$ (see Eq.~\ref{eq:piE_def} and Fig.~\ref{fig:piE_vs_dl}).

Fig.~\ref{fig:eventrate_fixed_piE} gives a slightly different view of the characteristic properties of BH events,
which is
the $\tE$ distribution of the 
microlensing 
events that have a fixed value of the 
microlensing 
parallax, $\piE=0.01$ or $0.1$. 
It is clear that long timescale events with $\tE\gtrsim 100$ days and a small parallax of $\piE=0.01$ are dominated by BH events.
On the other hand, stellar-mass lenses give a dominant source of events with $\piE=0.1$ for almost all the timescales.
Thus, if we can find long timescale events that have $\tE\gtrsim 100~$days as well as the small parallax effect or even a null detection of the parallax effect, those events are good candidates of BHs.

\section{BH mass estimation from an observation of individual microlensing light curves}
\label{sec:mass_estimation}

In this section we study the capability of an observation of individual 
microlensing 
events to identify the candidates of BHs with $\gtrsim 40\,M_\odot$. To do this, we simulate a light curve of 
microlensing event including the parallax effect for an observation of a source star in the Galactic bulge region, and then 
perform the model fitting to assess the performance of lens mass estimation.
In particular we will below focus on the usefulness of the parallax effect for the lens mass estimation. 

\subsection{Method}

\begin{figure}
\centering
\includegraphics[width = 0.98\columnwidth]{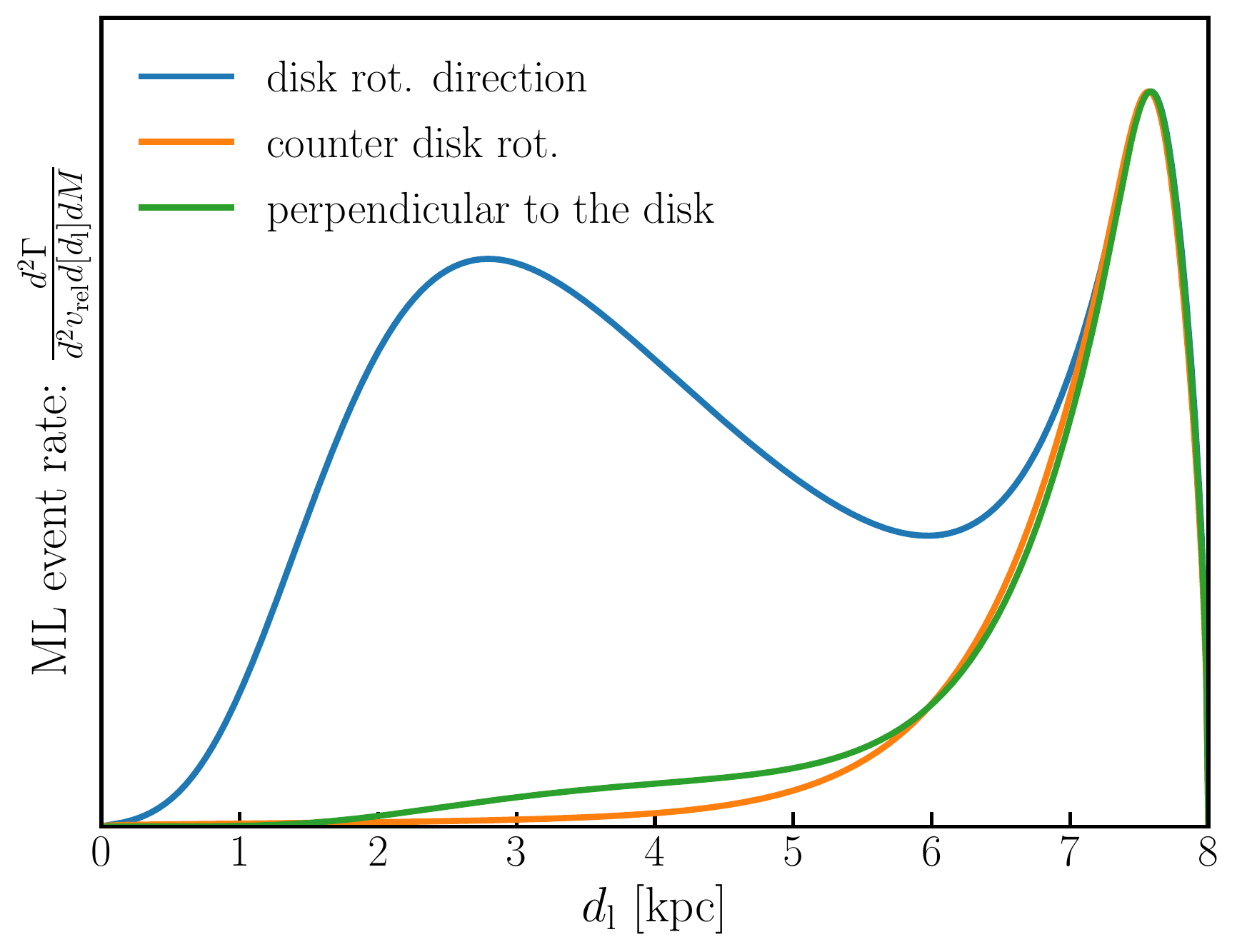}
\includegraphics[width = \columnwidth]{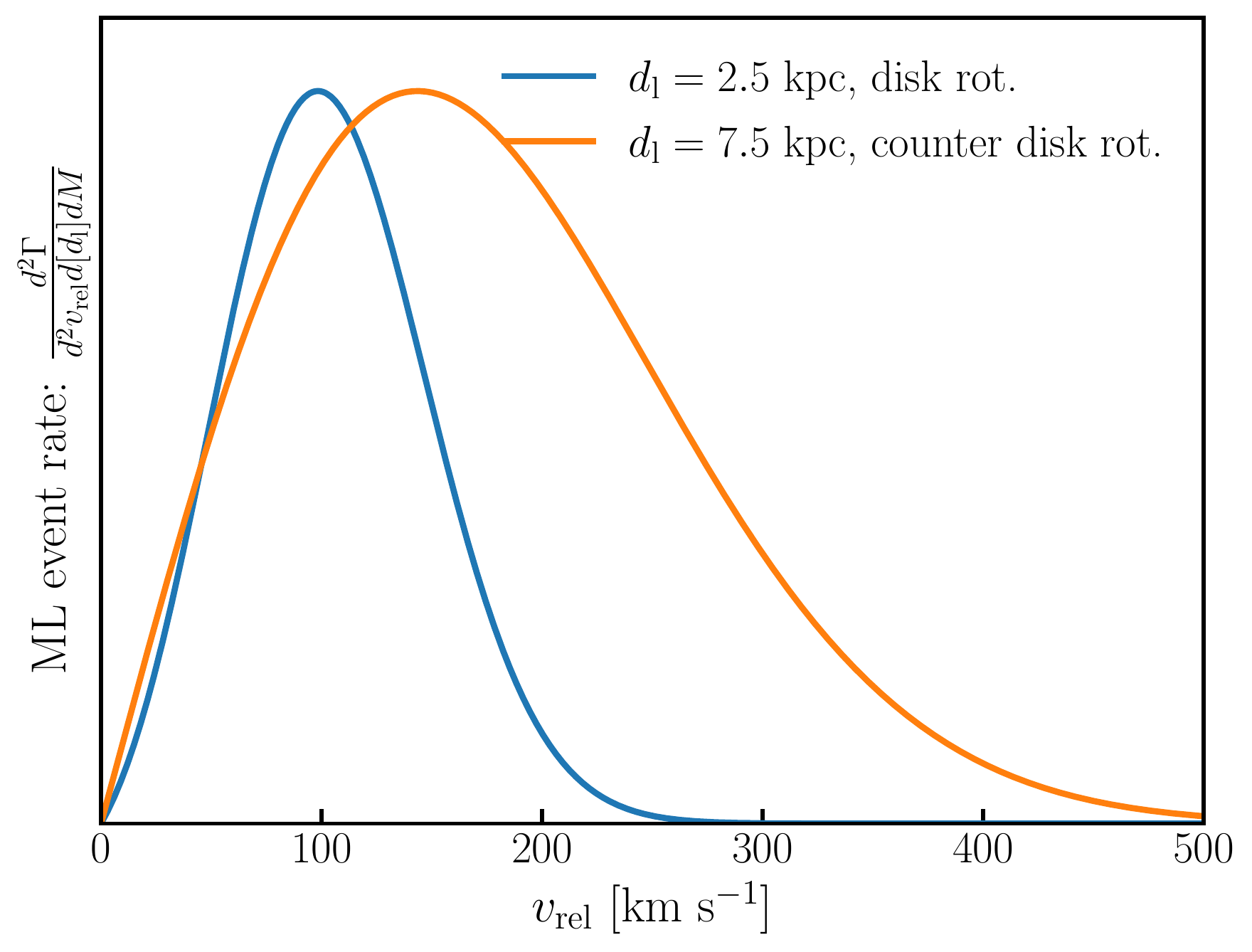}
\caption{The differential event rate $d^4\Gamma/d^2v_{\rel}d[\dl]dM$ in the prior (Eq.~\ref{eq:prior_def}) is given 
as a function of the distance to a lens ($\dl$) and the relative velocity ($\vv_{\rm rel}$), assuming 
the standard MW model for the spatial and velocity distributions of lens populations. 
The upper plot shows the event rate as a function of $\dl$, for events that have a fixed relative velocity amplitude of
$v_{\rel}=100~{\rm km~s}^{-1}$.
For the velocity direction, we consider 
the direction along the disk rotation, the counter disk-rotation direction, and the vertical direction to the disk plane, respectively. 
The lower plot shows the similar result, but the event rate as a function of the relative velocity amplitude, for events for lenses at the fixed distance $\dl=2.5$ (in the disk region) or 7.5~kpc (in the bulge), respectively. 
}
\label{fig:prior}
\end{figure}%

A microlensing light curve 
is specified by 5 parameters, ($t_0,\vpiE,\tE,u_0$), where $t_0$ is the fiducial epoch of the light curve for which we take the epoch of the light curve peak throughout this paper. For a hypothetical observation of a 
microlensing 
event, we assume that 
the ratio of an observed flux of a source star to the original flux, i.e. the lensing magnification, at each epoch is measured, and then the time-varying
lensing magnification sampled at regularly-revisited epochs, i.e. the 
microlensing 
light curve, is obtained from a monitoring observation of the same source star.  
In this setting, we assume that the likelihood for comparing the observed light curve to its theoretical template, $\mathcal{L}$,  
is given by
\begin{align}
-2\ln{\cal L}({\bf d}|\vtheta)&=\sum_{i}\frac{\left[\mu^{\rm sim}(t_i)-\mu^{\rm model}(t_i|\vtheta)\right]^2}{
\sigma_\mu^2},
\label{eq:likelihood}
\end{align}
where ${\bf d}$ is the data vector that consists of the observed magnification at the observation epoch $t_i$, i.e. $\mu(t_i)$, for which 
we use the data of a simulated light curve, 
$\mu^{\rm model}(t_i|\vtheta)$ is the model magnification at $t_i$, and $\sigma_\mu$ is the error at $t_i$. 
In the following we assume that the peak epoch of the light curve, $t_0$, is well determined from the observation, and we do not include 
it as a free parameter for simplicity. Hence, the theoretical template is given by 4 parameters, $\vtheta=(\vpiE,\tE,u_0)$. 
Throughout this paper we assume $\sigma_\mu=0.01$, meaning a
1\% accuracy in the flux measurement of the source star at each epoch. 
In addition we consider a noise-less light curve for the input data vector and  assume that a hypothetical observation of the 
microlensing light curve is done under 
a cadence of 10 days
(the light curve is sampled 
every 10~days) 
and that the fitting is done over the  period of $t\in[t_0-\tE,t_0+\tE]$. 
We checked that the following results are almost unchanged even if we consider a longer period than our fiducial range.

As we discuss below
we can estimate lens mass only from a combination of $\theta_{\rm E}$ and $\piE$. However, $\theta_{\rm E}$ cannot be directly estimated from a 
microlensing light curve. To resolve this difficulty, we employ the MW standard model for the spatial and velocity distributions of lens populations as discussed in Section~\ref{sec:stat_distribution}, and then adopt the Bayesian inference to estimate physical parameters of a lens:
\beq
P(\vpiE,\tE,\thetaE,u_0 |\bd)\propto 
\mathcal{L}(\bd| \vpiE,\tE,u_0)\Pi(\vpiE,\tE,u_0,\thetaE),
\label{eq:bayes}
\eeq
where $P(\vpiE,\tE,\thetaE,u_0|\vd)$ is the posterior distribution of the parameters and 
$\Pi(\vpiE,\tE,u_0,\thetaE)$ is the prior of the parameters. 
With this Bayesian method, we treat the 6 parameters $(\vpiE,\tE,\thetaE,u_0)$ as ``observable'' parameters estimated from the fitting of a 
microlensing light curve. 
We can then find that the physical parameters of a lens are given in terms of these observable parameters as
\begin{align}
M&=\frac{\thetaE}{\kappa\piE}\nonumber\\
\dl&=\frac{\rm AU}{\piE\thetaE+{\rm AU}/d_{\rm s}}\nonumber\\
\frac{v_{\rm rel}}{\dl}&=\frac{\thetaE}{t_E}\nonumber\\
\hat{\vv}_{\rm rel}&=\hat{\boldsymbol{\pi}}_{\rm E},
\label{eq:physical_parameters_vs_ML_parameters}
\end{align}
where $\kappa\equiv 4G/[c^2{\rm AU}]$, $\hat{\vv}_{\rm rel}\equiv \vv_{\rm rel}/v_{\rm rel}$, and $\hat{\boldsymbol{\pi}}_{\rm E}\equiv \vpiE/\piE$. 
In the above equations, the left-hand side of each equation gives the physical parameters (or derived parameters), while 
the right-hand side gives the parameters inferred from the Bayesian inference. 
These equations show that, unless the prior on $\thetaE$ is assumed, 
we cannot fully constrain the 5 physical parameters $(M,\dl,v_{\rm rel},\hat{\vv}_{\rm rel})$ from the microlensing observables
$(\piE,\tE,\hat{\boldsymbol{\pi}}_{\rm E})$. However, we note that, if the astrometric lensing is measured, it gives a constraint on 
$\thetaE$ and then enables constraining the lens mass, when combined with the microlensing light curve constraints, as we will discuss later. 

Hence, once the posterior distribution of the microlensing parameters is obtained, we can compute the posterior distribution of lens mass
in the mass bin $M_i$
from the projection: 
\begin{align}
P(M_i)&\propto \int\!d\vpiE d\tE d\thetaE du_0~ P(\vpiE,\tE,\thetaE,u_0 |\vd)\nonumber\\
&\hspace{1em}\times\delta_D[M(\piE,\thetaE)-M_i]. 
\end{align}
This is the method to obtain the posterior distribution of lens mass from the observed microlensing light curve.

To model the prior in Eq.~(\ref{eq:bayes}), we employ the same MW model in Section~\ref{sec:stat_distribution}
and assume that the prior is proportional to the 
microlensing 
event rate: 
\beq
\Pi(\vpiE,\tE,u_0,\thetaE)\propto\frac{d^4\Gamma}{d^2\piE d\tE d\thetaE}.
\eeq
This prior can be interpreted as the probability of 
microlensing 
events that have given values of the parameters
($\piE,\tE,\thetaE$). We can obtain the prior from a change of variables: 
\beq
\frac{d^4\Gamma}{d^2\piE d\tE d\thetaE} = \frac{d^4\Gamma}{d^2v_{\rel} d[\dl] dM}\left|\frac{\partial(\vv_{\rel},\dl,M)}{\partial(\vpiE,\tE,\thetaE)}\right|,
\label{eq:prior_def}
\eeq
where $|\partial(\vv_{\rel},\dl,M)/\partial(\vpiE,\tE,\thetaE)|$ is the Jacobian of the change of variables:
\begin{align}
\left|\frac{\partial(\vv_{\rel},\dl,M)}{\partial(\vpiE,\tE,\thetaE)}\right|=
\frac{2{\rm AU}^3\, \ds^4\thetaE^3}{\kappa\piE^2\tE^3({\rm AU}+\ds\piE\thetaE)^4}.
\label{eq:Jacobian}
\end{align}
Note that we do not assume any mass spectra of lenses.
Recalling that the differential event rate is given as
\beq
d\Gamma_i\propto n_i(\dl)\rE v_{\rel}\cos\theta f_i(\vv_{\rel})
d^2\vv_{\rel} d\alpha d[\dl] dM.
\eeq
and 
integrating this by $\alpha$, we can obtain the first factor on the right-hand side of Eq.~(\ref{eq:prior_def}):
\begin{align}
\frac{d^4\Gamma_i}{d^2\vv_{\rel}d[\dl]dM}&\propto n_{i}(\dl)\rE v_{\rel} f_i\brak{\vv_{\rel}} \notag \\ &\quad\times\int_{\arg(-\vv_\rel)-\frac{\pi}{2}}^{\arg(-\vv_{\rel})+\frac{\pi}{2}}d\alpha\,\cos\brak{\arg(-\vv_{\rel})-\alpha}. \label{eq:dGamma_vrel_dl_M}
\end{align}
Here $\arg(-\vv_{\rel})$ is the angle between the vector $-\vv_{\rel}$ and the $y$-axis on the lens plane. Since $\vv_{\rel}=-v_{\rel}(\cos(\theta+\alpha),\sin(\theta+\alpha))$, we have $\theta+\alpha=\arg(-\vv_{\rel})$. The range of the angle $\theta$ is $-\pi/2\leq\theta\leq\pi/2$ (that is, we only count ``in-going'' lenses), and hence the integration range of $\alpha$ is as in Eq.~\eqref{eq:dGamma_vrel_dl_M}.
We assume the ellipsoidal-bulge and exponential-disk models for the number density distribution of lenses, $n_i$ in the above equation, as discussed for MSs in Section~\ref{sec:MW_model}. 
In other words, in the following parameter estimation, we implicitly assume that BHs in the MW follow the same spatial and velocity distributions as those of MSs. 

Fig.~\ref{fig:prior} shows the dependences of the prior, 
$d^4\Gamma/d^2v_{\rm rel}d[\dl]dM$, on the distance ($\dl$) or on the 
relative velocity amplitude ($v_{\rm rel}$), for events where we fix either $\vv_{\rm rel}$ (its amplitude and direction) or $\dl$. 
As can be found from the upper panel, the MW standard model predicts a higher probability for lenses at $\dl \sim 2.5~{\rm kpc}$ in the disk region or
at $\dl\sim 7.5~{\rm kpc}$ in the bulge region, respectively. 
A closer look reveals that lenses in the disk region tend to have the relative-velocity direction along the Galactic rotation, because lenses in the disk region co-rotate with the Sun with respect to the Galactic center as can be found from Eq.~(\ref{eq:mean_relative_motion}). 
On the other hand, all the three curves in the bulge region ($6\lesssim \dl/[{\rm kpc}]\lesssim 8$) have the similar amplitudes, meaning that lenses in all the relative-velocity directions have similar probabilities, because sources and lenses in the bulge have random motions on average.

The lower panel shows that, for lenses in the disk or bulge region, the lenses tend to have the relative velocity of $v_{\rm rel}\simeq 100\,{\rm km\,s}^{-1}$ in the Galactic rotation direction or 200~${\rm km~s}^{-1}$ in the counter-rotation direction, respectively. 
Thus the prior or more exactly the MW model gives the information on $\dl$, $v_{\rm rel}$ and the direction of $\vv_{\rm rel}$,  and in turn $\thetaE$ is inferred from Eq.~\eqref{eq:physical_parameters_vs_ML_parameters} as $\thetaE = v_{\rm rel}\tE/\dl$, where $\tE$ can be estimated from the microlensing light curve.

\subsection{BH mass estimation from an observation of microlensing light curves}
\label{sec:lightcurve_simulations_results}

\begin{table*}
\caption{Model parameters used in the simulated microlensing light curves \label{tab:MLsettings}}
\begin{tabular}{l|cccccccc} \hline\hline
name & $M\,[M_{\odot}]$ & $\dl\,[\kpc]$ & $\vv_{\rm rel}\,[\km~\s^{-1}]$ & 
$t_{{\rm E},\odot}\,[\rmday]$ & $\theta_{\rm E}$
& $\piE$ & $t_0$-epoch & $u_0$ \\ 
\hline
diskBH40 &  40 & 2.5 & $(100,0)$ & $409.5$ 
& 9.46
& 0.0291 & March~1 & 0.1\\
${\rm diskMS1_{BH40}}$  &   1.0 & 2.5 & $(100/\sqrt{40},0)$ & $409.5$ 
& 1.50
& 0.1844 & March~1 & 0.1\\
diskBH70     & 70 & 2.5 & $(100,0)$ & $541.7$ 
& 12.5
& 0.0220 & March~1 & 0.1\\
${\rm diskMS1_{BH70}}$  &  1.0 & 2.5 & $(100/\sqrt{70},0)$ &  $541.7$ 
& 1.50
& 0.1844 & March~1 & 0.1\\
bulgeBH70  & 70 & 7.5 & $(-150,0)$ & $188.6$  
& 2.18
& 0.0038 & March~1 & 0.1\\
${\rm bulgeMS1_{BH70}}$ & 1.0 & 7.5 & $(-150/\sqrt{40},0)$ & $188.6$ 
& 0.26
& 0.0321 & March~1 & 0.1\\ \hline\hline
\end{tabular}
    \tablecomments{The lens mass ($M$), the distance to a lens ($d_{\rm l}$), the relative velocity 
    ($\vv_{\rm rel}$), the Einstein timescale ($t_{{\rm E},\odot}$), 
    the Einstein radius ($\theta_{\rm E}$), the microlensing parallax ($\piE$), the epoch of the light curve peak, and the impact parameter ($u_0$), 
    used in each simulation of the microlensing light curve for a source star 
    at $d_{\rm s}=8~$kpc in the Galactic bulge. The relative velocity vector is given in two dimension as $\vv_{\rm ref}=(v_y,v_z)$, where $v_y$ is along the Galactic rotation direction and $v_z$ is along the direction perpendicular to the Galactic disk. Here we consider 
    microlensing events due to BHs with masses of $40$ or $70\,M_\odot$ in the Galactic disk or bulge region. Then we consider microlensing events due to an MS of $1\,M_\odot$, which have the same light curve timescale as that of the BH event. For example, we name each simulation as ``diskBH40'' for a microlensing event due to the $40\,M_\odot$ BH or ``diskMS1$_{\rm BH40}$'' for the $1\,M_\odot$ MS lens that has the same timescale as that of the BH lens.
    }
\end{table*}

\begin{figure}
\centering
\includegraphics[width = 0.45\textwidth]{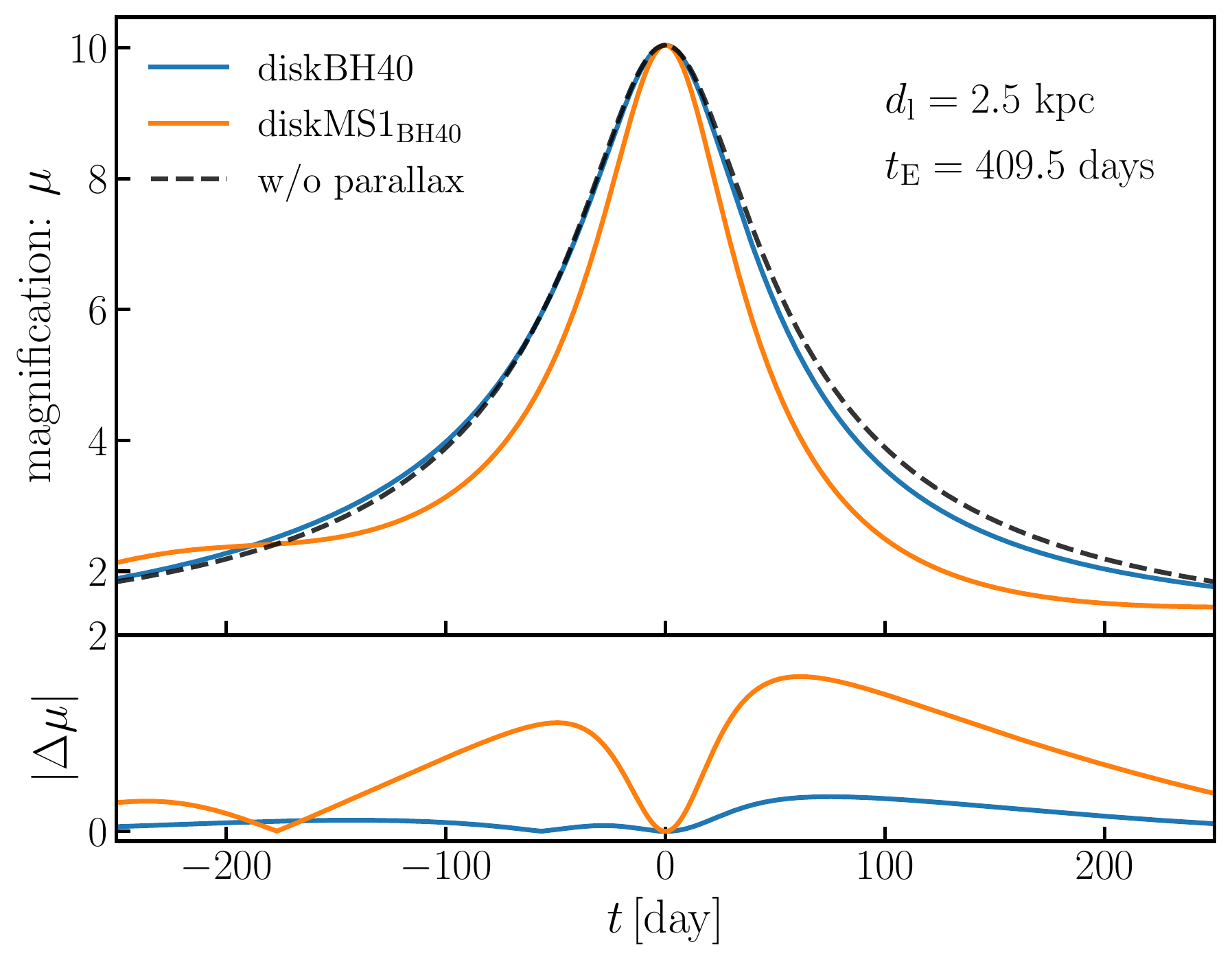}
\includegraphics[width = 0.45\textwidth]{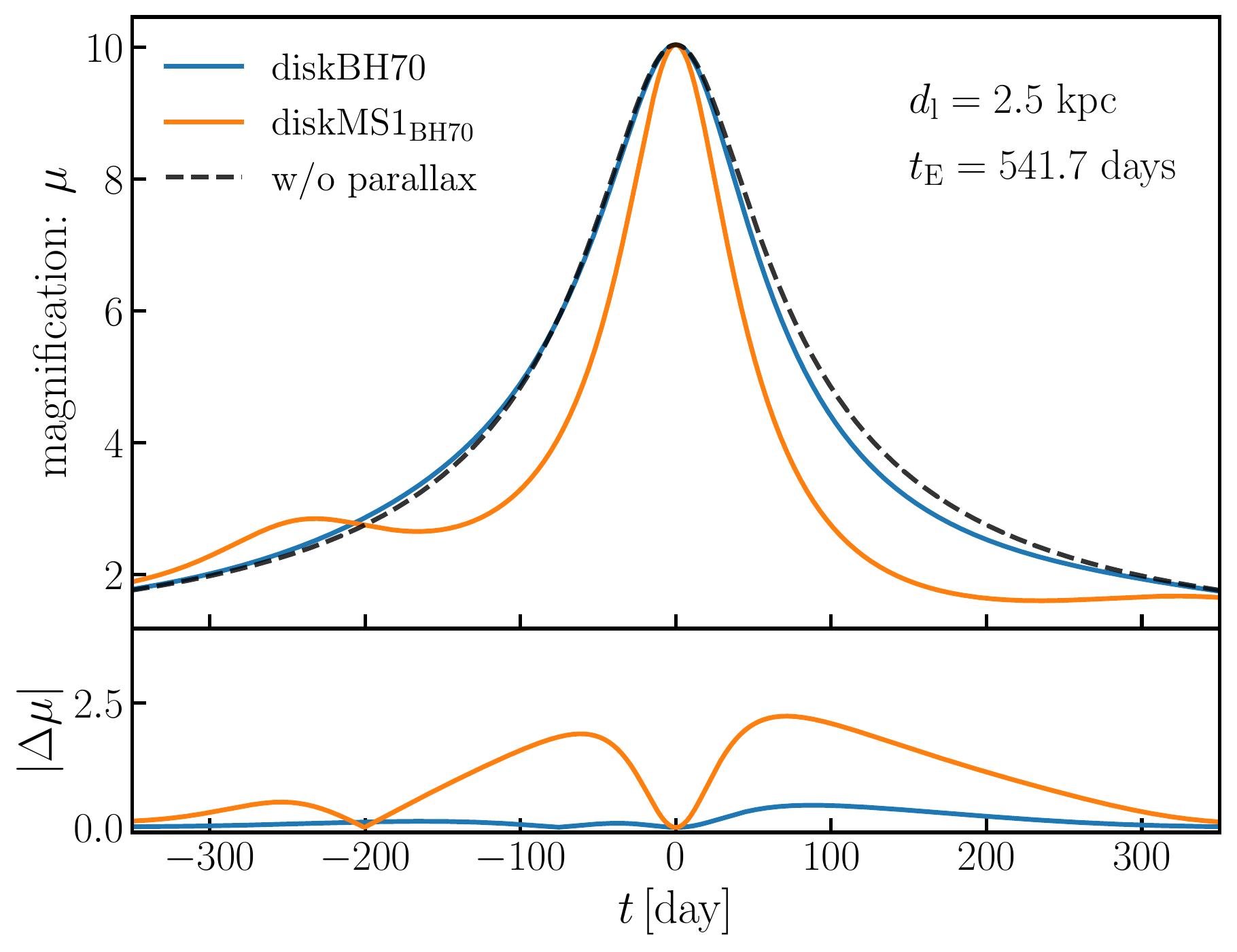}
\includegraphics[width = 0.45\textwidth]{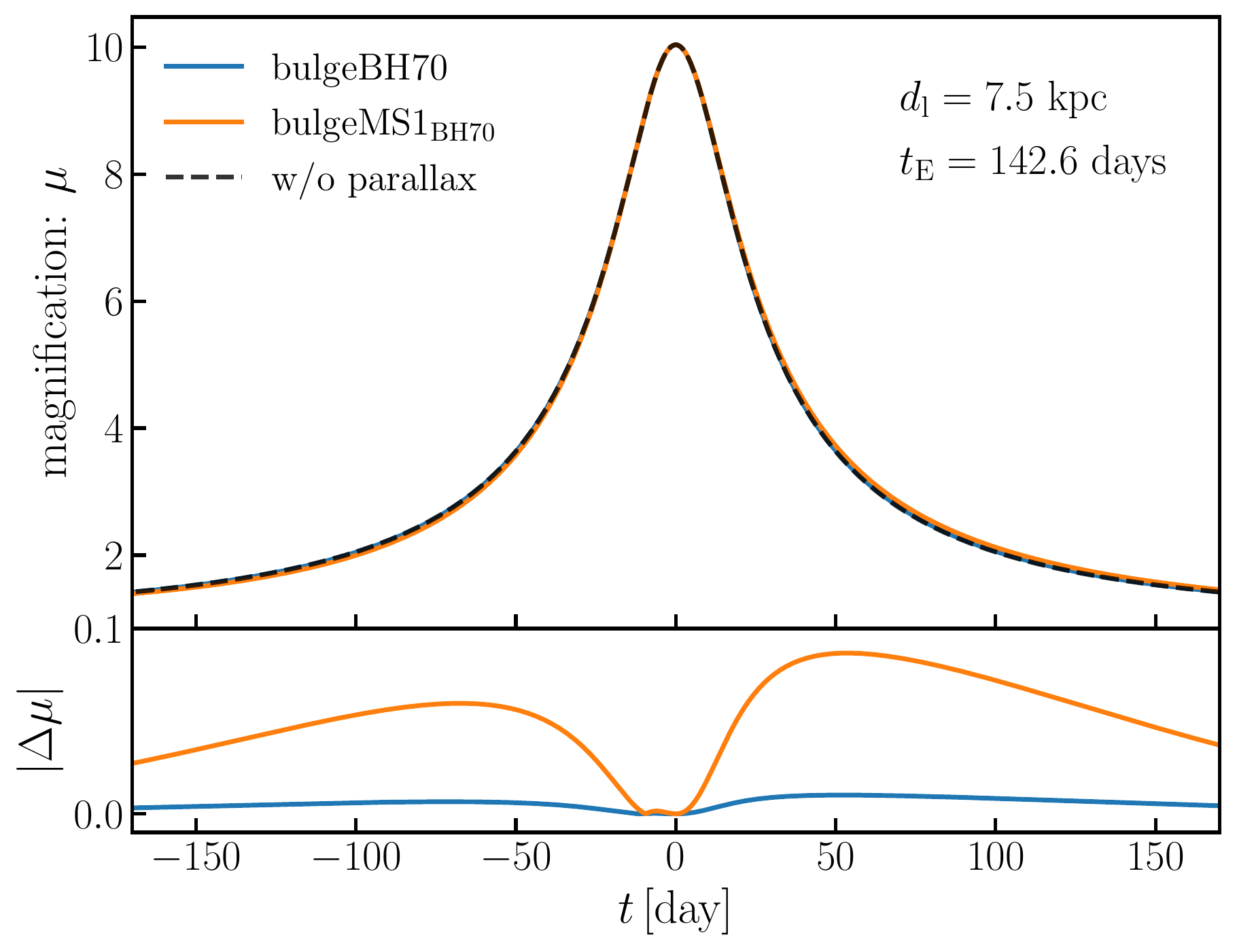}
\caption{The microlensing light curves for a source star at $d_{\rm s}=8~{\rm kpc}$, due to a BH or an MS lens. For the microlensing light curve parameters,
we adopt the setting in Table~\ref{tab:MLsettings}. In each panel, we consider the microlensing events due to a BH and an MS lens, which have the same light curve when the parallax effects are ignored. The label for each curve is as in Table~\ref{tab:MLsettings}: e.g. ``diskBH40'' stands for a BH lens with mass $40M_\odot$ and at distance (in the disk region). 
The lower plot in each panel shows the absolute value of the magnification difference between the light curves with and without microlensing parallax.
}
\label{fig:lightcurve_M40D}
\end{figure}%

\begin{figure}
\centering
\includegraphics[width=\columnwidth]{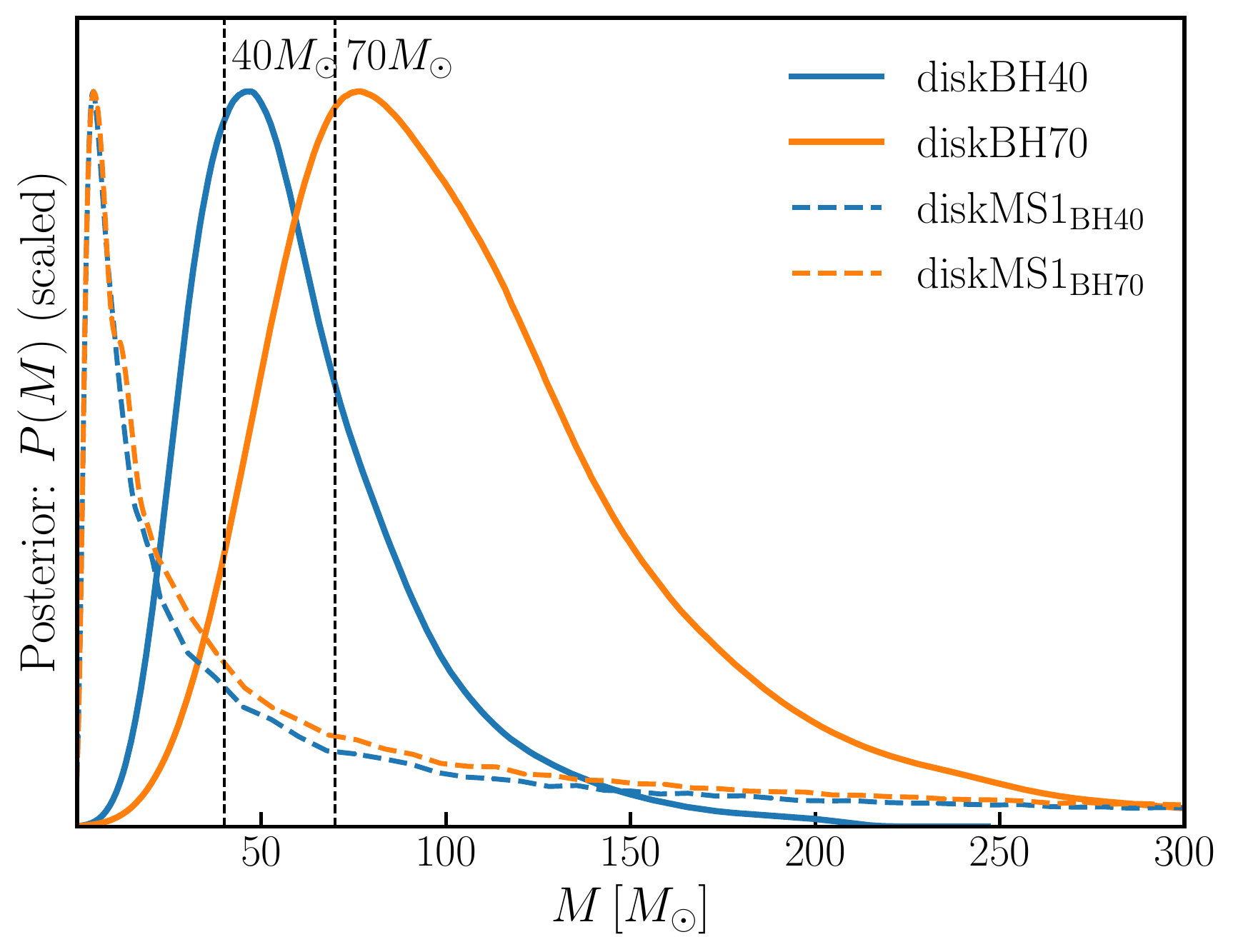}
\caption{The posterior distributions of lens mass ($M$) obtained from the MCMC analysis of the simulated light curves, for the four cases of the BH or MS lenses at $\dl=2.5~{\rm kpc}$, i.e. the disk lenses shown in Table~\ref{tab:MLsettings} and Fig.~\ref{fig:lightcurve_M40D}.
For illustrative purpose we scale all the posterior distributions vertically so as to have the same 
amplitudes at their peaks.
The solid curves show the results for the BH lenses of 
$40$ and $70\,M_\odot$, 
while the dashed curves are for the MS lenses of $1\,M_\odot$ that have the same Einstein timescales as those for the corresponding BH lens cases, as
shown in Table~\ref{tab:MLsettings}. The two vertical lines denote the input values of BH mass, i.e. $40M_{\odot}$ and $70\,M_\odot$.
}
\label{fig:posterior_disk}
\end{figure}%
\begin{figure}
\centering
\includegraphics[width = \columnwidth]{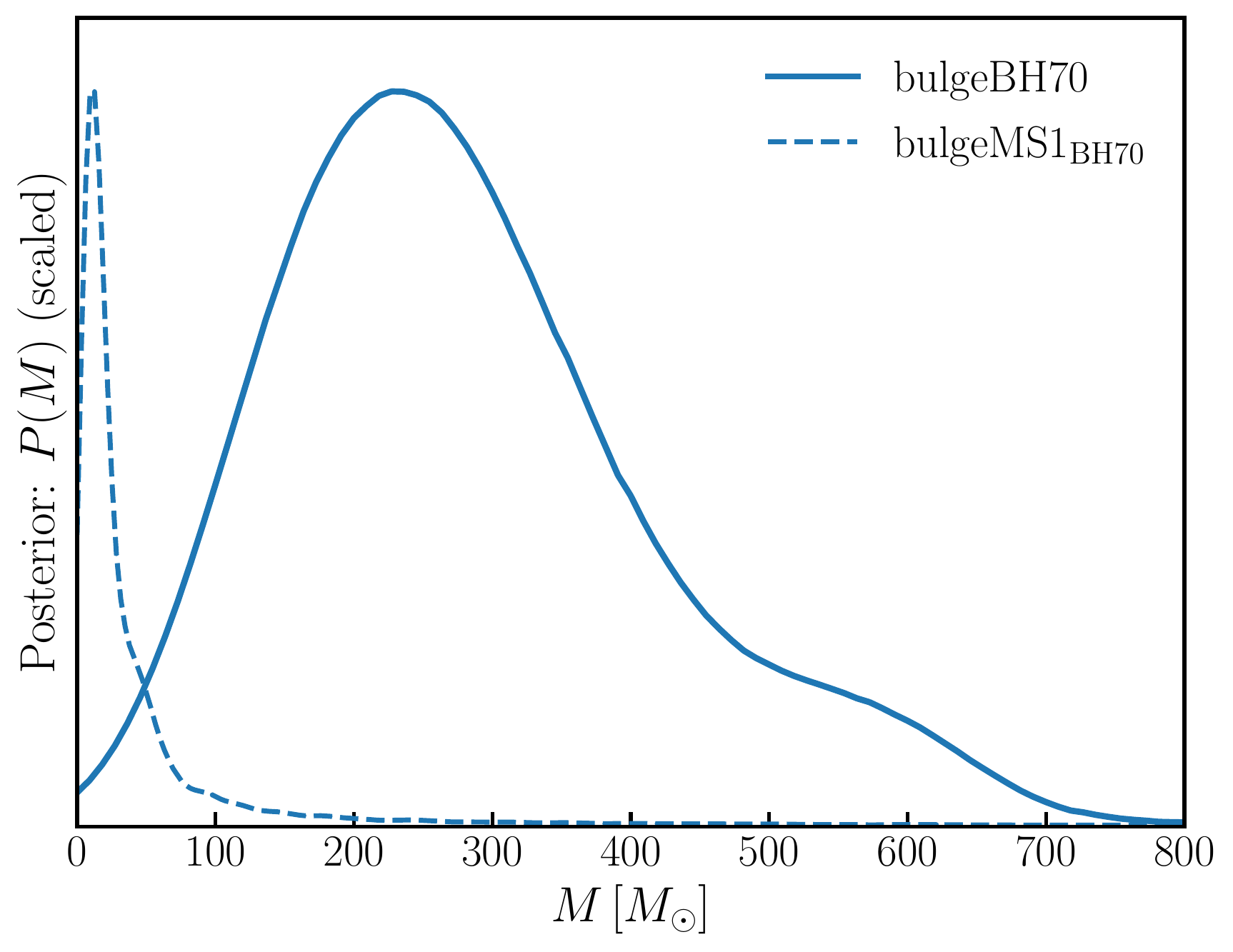}
\caption{Similar to the previous figure, but here we show the results for a BH and an MS lens at $7.5~{\rm kpc}$, i.e. the bulge lenses shown in Table~\ref{tab:MLsettings} and Fig.~\ref{fig:lightcurve_M40D}.
}
\label{fig:posterior_bulge}
\end{figure}%

We now study the expected precision of estimating a lens mass, especially the mass of a BH lens, from an observation of the microlensing light curve including the parallax microlensing. As described in the preceding section, we employ the Bayesian inference to estimate the parameters of microlensing events. We use the sampler module \textsf{emcee} 
\citep{2013PASP..125..306F}
 to perform the Markov chain Monte Carlo (MCMC) analysis of the parameters.

Here we consider hypothetical microlensing events for {\it typical} lenses in the Galactic disk or bulge region as 
listed in Table~\ref{tab:MLsettings}. 
For a disk lens, we consider a case that a BH lens is at the distance $\dl=2.5~{\rm kpc}$ and has $v_{\rm rel}=100~{\rm km~s}^{-1}$ for the 
relative velocity amplitude, which corresponds to a peak of the prior function (Fig.~\ref{fig:prior}), i.e. one of the most frequent events according to the standard MW model. We study two cases of BH lenses with $M_{\rm BH}=40$ and $70~M_\odot$. For each case, we consider an MS lens of $1~M_\odot$ that has exactly the same light curve as that of BH lens if the parallax effect is ignored; this can be done by reducing the relative velocity amplitude by a factor of $\sqrt{M_{\rm BH}/M_{\rm MS}}$ ($\sqrt{40}$ or $\sqrt{70}$ in our case). 
For convenience, we call the two BH events ``diskBH40'' and ``diskBH70'', and the two corresponding MS events ``diskMS1$_{\rm BH40}$'' and ``diskMS1$_{\rm BH70}$''.
For a bulge BH lens, we consider a case that a lens is at $\dl=7.5\,{\rm kpc}$ and has $v_{\rel}=150~{\rm km~s}^{-1}$, which corresponds to the peak for the bulge lens as shown in Fig.~\ref{fig:prior}. Similarly to the disk lens cases, we consider a BH lens of $70\,M_\odot$, and then an MS lens that has the same light curve without the parallax effect.

Fig.~\ref{fig:lightcurve_M40D} displays the simulated light curves for the lenses in Table~\ref{tab:MLsettings}.
As can be found, all cases show that the BH lenses predict the smaller parallax effects, i.e. the smaller deviation from the lensing magnifications without the parallax effect, than the respective MS case does. 

Figs.~\ref{fig:posterior_disk} and \ref{fig:posterior_bulge} show the main results of this paper, which display 
the posterior distributions of lens mass, obtained from the MCMC analysis for each of the 6 simulated light curves in Fig.~\ref{fig:lightcurve_M40D} (see Eqs.~\ref{eq:likelihood} and \ref{eq:prior_def} for the likelihood and the prior, respectively). 
The posterior distributions $P(M)$
for all the BH lens cases are clearly different from those for the MS lenses, meaning that the microlensing observation, with the parallax information, can discriminate BH events of GW mass scales from MS-like events. 
Fig.~\ref{fig:posterior_disk} shows that,
for the two disk BH cases, the posterior nicely recovers the input mass, although the posterior has a wide distribution extending to $M>100\,M_\odot$. 
The relatively broad width of the posterior $P(M)$ is due to parameter degeneracies. 
Nevertheless it is important to notice that masses around $M\sim 0$, i.e. small masses like MS lenses, are strongly disfavored by the posteriors.
The posterior distributions in a full space of the microlensing parameters and the derived physical parameters are given in Appendix~\ref{sec:fullcorner}.
On the other hand, the posterior distributions for the MS lenses display the peaks around $\sim M_\odot$, implying that the microlensing can properly identify these MS-like events.

The bulge BH result in Fig.~\ref{fig:posterior_bulge} shows that a BH lens can be recognized,
but the posterior 
tends to predict a larger mass than the input value, because the bulge BH lens predicts too small microlensing parallax to be detected for the assumed photometry accuracy ($1\%$ accuracy) and thus heavier BHs, which predict such a small parallax, are
all allowed, as implied from Fig.~\ref{fig:piE_vs_dl}. 
We should note that there is a non-zero probability around $M\sim 0$ in this case, meaning that an MS-like lens is not completely ruled out compared to Fig.~\ref{fig:posterior_disk}.
On the other hand, the posterior for the MS lens has a much narrower distribution including a mass of $\sim 1\,M_\odot$, but extends to GW-BH mass scales of $40\,M_\odot$ or greater.

If a lens is not very close to a source, or more specifically $\dl\lesssim7~\kpc$ as we assume $d_{\rm s}=8\,{\rm kpc}$ for the distance to the source, the posteriors of MS-like events have the peaks around $\sim M_{\odot}$ not only in the particular example in Fig.~\ref{fig:posterior_disk} but in general (if the photometry of the light curve observation is sufficiently accurate).
In our parameter inference, the angular Einstein radius $\thetaE$ is mainly constrained by the prior, 
and hence  the relative angular velocity $\mu_{\rel}\equiv \thetaE/\tE$ is heavily prior-dependent. As we can see from Fig.~\ref{fig:prior}, the typical relative velocity of the event rate is $v_{\rel}\sim100~\km~\s^{-1}$, yielding
$\mu_{\rel}\sim 10^{-7}~\mas~\s^{-1}$ as the typical relative angular velocity for events with $\tE\sim10^2~{\rm days}$.
The estimator of the relative angular velocity $\hat{\mu}_{\rel}$ appears around this typical value because of the heavy prior dependence. 
For MS-like lenses with $M\sim M_{\odot}$, $\piE$ is typically $\piE\sim0.1$ as can be found from Fig.~\ref{fig:piE_vs_dl}. Hence
the posterior of lens mass for MS lenses tends to peak around
\beq
M=\frac{\mu_{\rel}\tE}{\kappa\piE}\sim
M_{\odot}\left(\frac{\hat{\mu}_{\rel}}{10^{-7}~\mas~\s^{-1}}\right)\left(\frac{\tE}{100~\rm days}\right)\left(\frac{\piE}{0.1}\right)^{-1}.
\eeq
That is, if we can ``detect'' the microlensing parallax to be $\piE\sim 0.1$ even from a long timescale event, we can safely recognize 
 such an event as an MS lens.

On the other hand, a BH event has $\piE\sim 10^{-2}$ or smaller, which is one order of magnitude smaller than the parallax of an MS lens. 
Hence we can identify a strong candidate of BH events if the microlensing parallax of $\piE\sim 10^{-2}$ is {\it detected}, and the mass estimate would be fairly accurate. In addition, events with a  null detection of $\piE$, to the precision $\sigma(\piE)\sim 10^{-2}$, are also 
strong candidates of BHs, but the mass estimation basically gives only a lower limit, allowing a larger mass than the true mass. 
To be more precise, this argument only applies to events at
$\dl\lesssim7\,\kpc$, where $\piE\sim0.1$ for MS lenses as can be found from Fig.~\ref{fig:piE_vs_dl}. 
If an MS lens is located at
$\dl\gtrsim 7\,\kpc$, its microlensing parallax becomes $\piE\sim10^{-2}$, which is as small as that for disk BHs (BHs in $\dl\lesssim4\,\kpc$).
Thus such MS lenses could mimic a light curve of a disk BH, 
leading to a misidentification of BH candidate.
However, such a misidentification occurs
only if $\piE\sim10^{-2}$, and, as we showed in Section~\ref{sec:statproperties_BHevents},
the microlensing events with $\piE\sim10^{-2}$ are dominated by BHs.
From a statistical point of view, a misidentification of MS-like lenses at $\dl\simeq \ds$ as BH candidates is negligible.

\subsection{BH identification from atypical BH lens events}
\label{sec:discussion}
\begin{figure}
\centering
\includegraphics[width = \columnwidth]{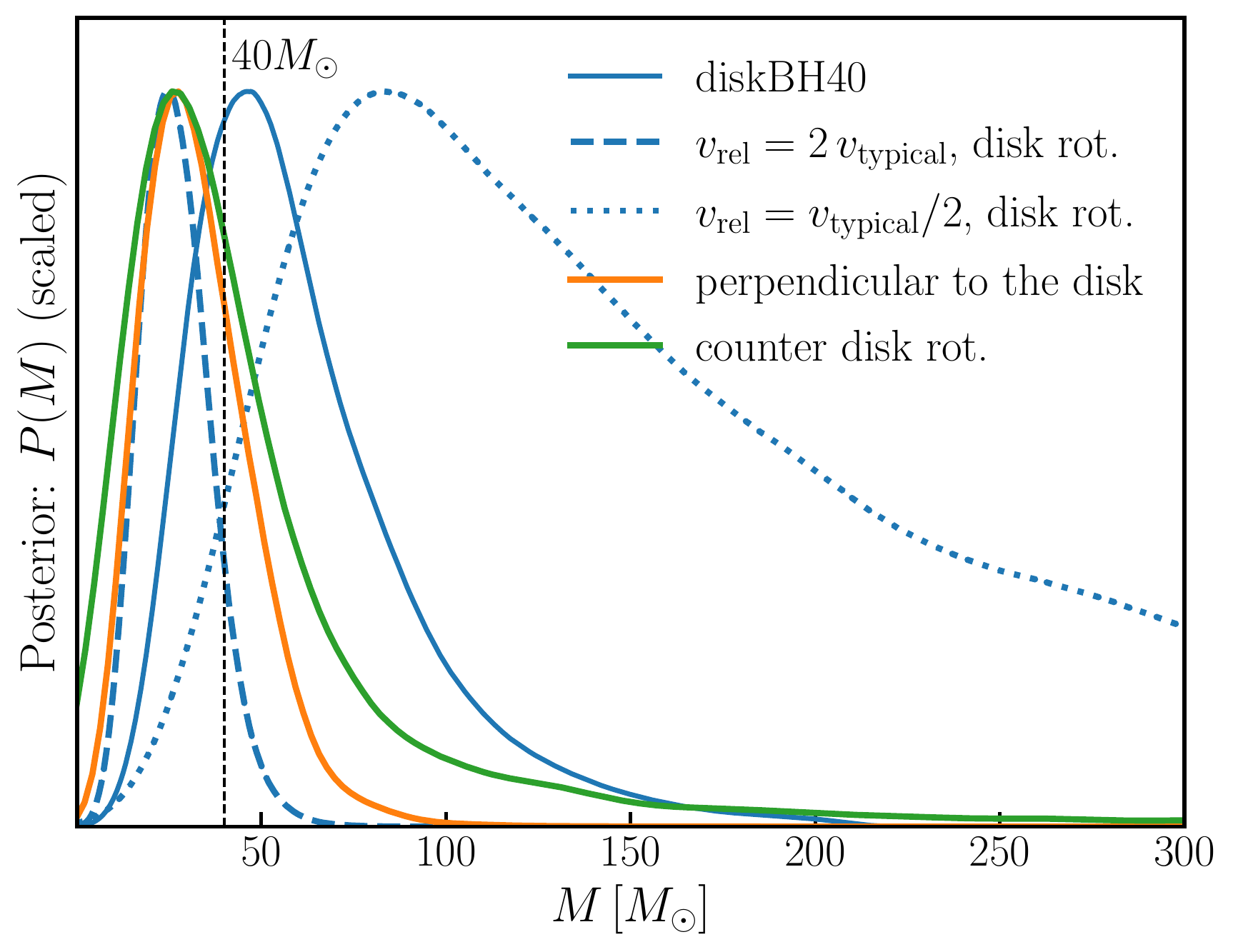}
\caption{Similar to Fig.~\ref{fig:posterior_disk}, but we show the results for the disk BH lens of $40\,M_\odot$,
 where we consider atypical values for the relative velocity amplitude or the relative velocity direction that are not favored by the input prior as shown in 
Fig.~\ref{fig:prior}. More specifically, we here consider four cases compared to the case ``diskBH40'' in Table~\ref{tab:MLsettings} and Fig.~\ref{fig:lightcurve_M40D}.
For the two cases we assume 
a doubled or halved relative velocity amplitude compared to the typical velocity ($v_{\rm typical}=100~{\rm km~s}^{-1}$) assumed for ``diskBH40'' with the same direction along the Galactic rotation; 
the light curve timescales are $t_{\rm E,\odot}\simeq 205$ or 819~days, respectively, while $t_{\rm E,\odot}=409.5\,{\rm days}$ for diskBH40. 
For the other two cases, we assume $v_{\rm rel}=100~{\rm km~s}^{-1}$, i.e. the same velocity amplitude as 
that of ``diskBH40'', but in the different directions, along the counter-disk rotation or along the direction perpendicular to the Galactic disk.
In these cases the light curve timescales $t_{\rm E,\odot}$ are the same as that of diskBH40.
}
\label{fig:posterior_atypical}
\end{figure}%
\begin{figure}
\centering
\includegraphics[width = \columnwidth]{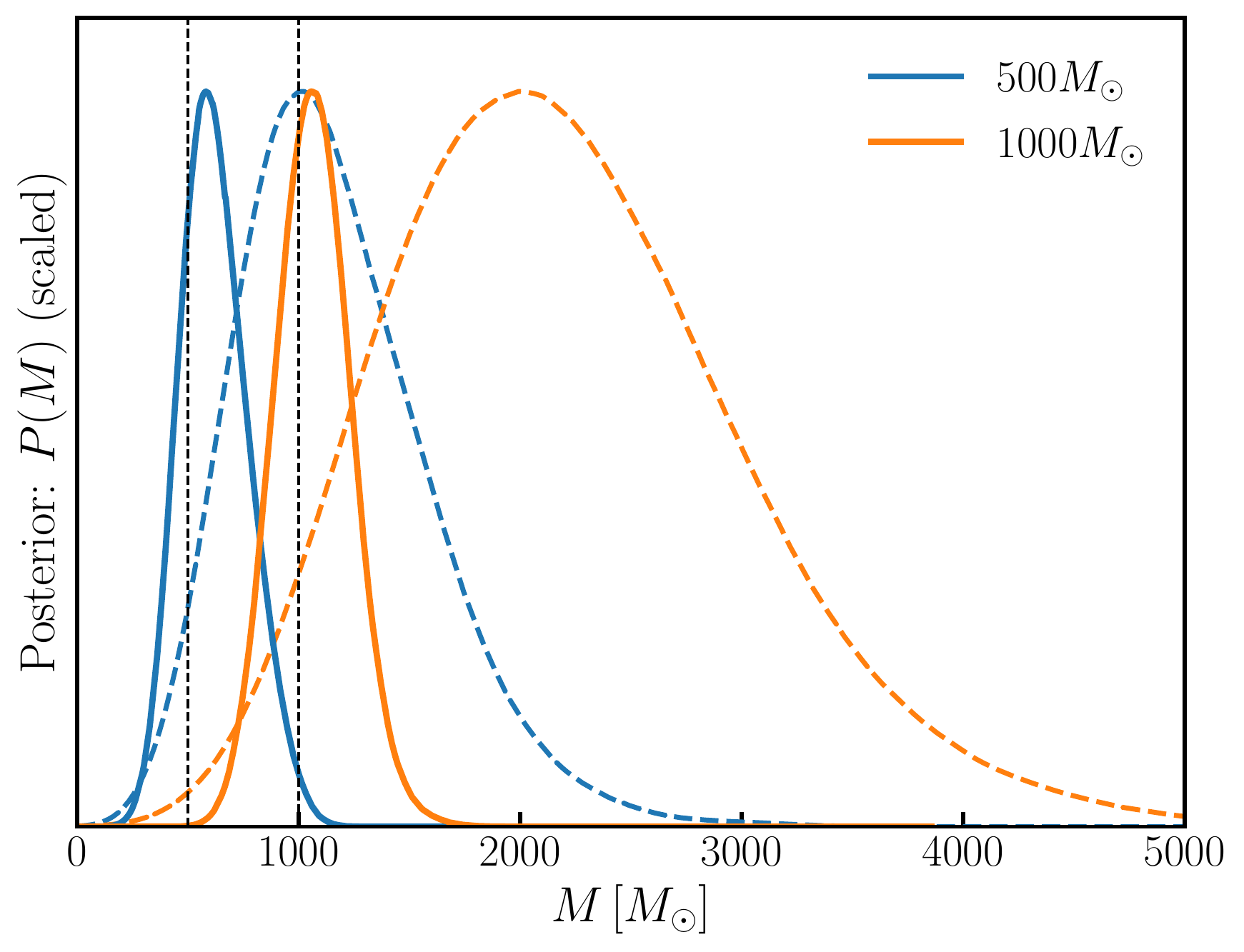}
\caption{Similar to Fig.~\ref{fig:posterior_disk}, but the dashed lines show the results assuming a microlensing event due to an intermediate mass BH of $500$ or $1000\,M_\odot$ at $\dl=7.5~{\rm kpc}$, i.e. in the bulge region (we always assume a source star at $d_{\rm s}=8~{\rm kpc}$). 
We employed the same parameters to simulate the microlensing light curve as those used for ``bulgeBH70'' in Table~\ref{tab:MLsettings} and \ref{fig:posterior_bulge}. 
The light curve timescales are $t_{\rm E,\odot}=504$ days and 713 days for the BH lenses of 500 and 1000$~M_\odot$, respectively. 
Note that we employ the same prior, i.e. the prediction of microlensing events for the standard MW model, as in the other figures. 
The solid lines show the results when further including the astrometric lensing information, assuming the accuracy of 
$\sigma(\thetaE)=1\,{\rm mas}$.
}
\label{fig:posterior_IMBH}
\end{figure}%

We have so far considered typical lenses in terms of their distance and relative velocity (its amplitude and direction) to study the capability of 
a microlensing light curve observation to identify a BH candidate on individual basis. 
In this section we study whether a BH event can be identified from the microlensing light curve with atypical lens parameters.

Fig.~\ref{fig:posterior_atypical} displays the posterior distributions of lens mass for a hypothetical observation of the microlensing light curves for BH lenses with $40M_\odot$ and at $\dl=2.5~{\rm kpc}$, where we employed different lens parameters from those of ``diskBH40'' in Fig.~\ref{fig:posterior_disk} (or Table~\ref{tab:MLsettings}). 
For two cases we consider a BH lens with the doubled or halved velocity amplitude, $v_{\rm rel}=200$ or $50~{\rm km~s}^{-1}$, compared to the typical velocity of $v_{\rm rel}=100~{\rm km~s}^{-1}$ of diskBH40, 
but assume that the velocity direction is along the Galactic rotation direction, i.e. the same direction as in diskBH40. These cases lead to a halved or doubled light curve timescale $t_{\E,\odot}$ from that of diskBH40. 
For other two cases, we fix the velocity amplitude to $v_{\rm rel}=100~{\rm km~s}^{-1}$, but assume that its 
direction is along the counter Galactic rotation or along the direction perpendicular to the Galactic disk plane. In these cases the light curve timescale $t_{\rm E,\odot}$ is kept fixed. 
For other parameters such as $t_0$ and $u_0$, we assume the same values as those of diskBH40.
As can be found from Fig.~\ref{fig:prior}, these BH events have much lower probabilities than the typical BH events in Fig.~\ref{fig:posterior_disk}. 
The figure clearly shows that a microlensing observation can distinguish these events from MS-like events and identify these to be likely BH lenses. 
In particular, BH lenses, which have the relative velocity direction in the Galactic rotation direction, have almost a null probability around $M\sim 0$, because such events prefer disk lenses as implied from Fig.~\ref{fig:prior} to bulge MS-like lenses that tend to predict small microlensing parallaxes as in BH lenses.
For other events that have the relative velocity in the counter-disk direction or the direction perpendicular to the Galactic disk, the posterior has a peak at a lower mass than the true mass and includes a non-negligible probability around an MS-like lens. From Figs.~\ref{fig:posterior_disk}--\ref{fig:posterior_atypical}, we can conclude that lens events, which have long light curve timescales and relative velocities in the Galactic rotation direction, are secure candidates of BH lenses.

Another interesting possibility is to find a candidate of intermediate-mass BHs (IMBHs) of $\sim 1000\,M_\odot$ mass scales.
Fig.~\ref{fig:posterior_IMBH} shows the posteriors for two IMBH microlensing events, which have long tails towards very high masses, meaning that we can identify an IMBH microlensing event if it occurs.
IMBHs might exist in the Galactic bulge region, due to dynamical friction, or more generally in all galactic centers as a pathway to a formation of the supermassive BH at the nucleus \citep[e.g.][]{2021arXiv210102727M}. However, IMBHs have not yet been found with certainty.
Hence, if the microlensing due to an IMBH were to occur, its observation could be a discovery of the existence of IMBHs in the Galactic bulge region.
An IMBH, thanks to its large mass, has a relatively large astrometric lensing: $\thetaE\simeq 8.2\,{\rm mas}$ for a $1000\,M_\odot$ IMBH at $\dl=7.5\,\kpc$ (see Eq.~\ref{eq:thetaE}).
Such an astrometric lensing can be detected with a ground-based telescope with the aid of adaptive optics or with a spaced-based telescope, and therefore a detection of the astrometric lensing for a good IMBH candidate can improve the precision of mass estimation. 

\section{Discussion and Conclusion}
\label{sec:conclusion}

In this paper we have studied the capability of a microlensing observation of stars in the Galactic bulge region to identify GW BH lenses of $40M_\odot$ or greater masses.
Our results are summarized as follows. 
First, assuming that BH lenses follow the same spatial and velocity distributions of stars in the MW bulge and disk regions, we studied statistical properties of microlensing events due to BH lenses, compared to those of stellar lenses. We found that microlensing events with long Einstein timescales
$\tE\gtrsim 100~{\rm days}$ and small parallax $\piE\lesssim 10^{-2}$ are dominated by BH lenses (Figs.~\ref{fig:eventrate} and \ref{fig:eventrate_fixed_piE}).
Second, we have shown that, even for an observation of individual microlensing light curves with long timescales, each BH event can be identified, or at least be discriminated from a main-sequence star lens, if the microlensing parallax is constrained to the precision of $\sigma(\piE)\simeq 10^{-2}$ (Figs.~\ref{fig:posterior_disk}--\ref{fig:posterior_atypical}). 
More precisely, BH lenses in the disk region are secure candidates because the microlensing parallax is relatively large to be detected and such events lead to a relatively accurate estimation of lens mass. 
Even if a BH lens has an atypical value or direction of the relative velocity compared to those of stars, a BH lens in the disk region can be securely identified, given that the parallax effect is constrained (Fig.~\ref{fig:posterior_atypical}).
On the other hand, for 
BH lenses in the bulge region, the microlensing parallax is small and not easy to detect, so only an identification of BH events is possible; the mass estimation is not accurate, and gives only a lower limit on the lens mass that is nevertheless clearly distinguishable from a stellar lens. 
We also studied how a microlensing event due to an intermediate-mass BH of $1000\,M_\odot$, which might exist around the Galactic center, can be securely identified, if it occurs (Fig.~\ref{fig:posterior_IMBH}). Finally we comment that microlensing can probe both isolated 
or binary BHs including wide-orbit binary systems, 
and therefore is complementary to other methods such as the GW observation that is sensitive to BHs in close binary systems.

However, to fully exploit the power of microlensing to find BH lenses, we need a sufficiently accurate photometry such as a few percent accuracy in the flux measurement as well as a long-term monitoring observation that at least spans longer than a year timescale.
A large-aperture, wide field-of-view telescope such as the VRO LSST, the Subaru Telescope or the Roman Space Telescope could achieve the required precision. For example, 
LSST can easily reach the depth of $g,r\sim 26$ ($5\sigma$) for a point source, with a reasonable exposure time ($\sim 10~{\rm min}$), at each visit, as implied by the Subaru Hyper Suprime-Cam observation \citep{HSCOverView:17} that has the similar capability to LSST. Hence the LSST observation can reach a $\sim 100\sigma$ detection for 
stars with $g,r\sim 22.5$, leading to $1\%$ flux measurements that we assumed throughout this paper (since the distance modulus to the MW bulge $\mu\simeq 14.5$, the depth of $22.5$ allows one to use many stars down to G or K-type stars for the microlensing observation).
However, since the Galactic bulge region suffers from high dust extinction, especially in optical wavelengths, a careful study of the field selection in the Galactic bulge region is needed. For this, an observation in infrared wavelengths would allow for a more efficient microlensing observation. For example, 
the Roman Space Telescope can be powerful, although the long-term monitoring observation longer than a year timescale can be challenging. 
The ULTIMATE-Subaru, on the other hand, which is a next-generation, wide-field near-infrared imager and multi-object spectrograph with the aid of a ground-layer adaptive optics system at the Subaru Telescope \citep{2020SPIE11450E..0OM,2020SPIE11447E..0NM} would allow for the required microlensing observation, although the Galactic center has a limited accessibility (only 3-4 months per year) from the Subaru Telescope. 

Another challenge for a microlensing observation of the Galactic bulge is the systematic effect caused by a possible confusion of many stars in each CCD pixel, i.e. pixel lensing. For this, the image difference technique \citep{2017arXiv170102151N} can be used to properly measure the time-varying flux of an individual source star. Or a space-based telescope such as the Roman Space Telescope has an advantage 
due to its exquisite angular resolution. 
In addition, the space-based telescope could give us a useful information on  astrometric lensing (and/or proper motions that are useful to constrain distances to lens or source). A further careful study to attain the full potential of microlensing observation is needed and will be our future work.

\section*{Acknowledgments}
We would like to thank the committee members of ST's master thesis defense, Profs.~Naoteru~Gouda and Hideyuki~Tagoshi, for useful comments on the
thesis on which this study is based. We also thank Sunao~Sugiyama, Toshiki~Kurita and Masamune~Oguri for useful discussion. 
This work was supported in part by World Premier International Research Center Initiative (WPI Initiative), MEXT, Japan, and JSPS KAKENHI Grant Numbers JP18H04350, JP18H04358, JP19H00677, JP20H05850, and JP20H05855.

\noindent{\it Software}: \textit{Astropy} \citep{2013A&A...558A..33A,2018AJ....156..123A}, \textit{IPython} \citep{4160251}, 
\textit{matplotlib} \citep{4160265}, 
\textit{numpy} \citep{Harris:2020xlr}, \textit{scipy} \citep{Virtanen:2019joe},
\textit{emcee} \citep{2013PASP..125..306F}, \textit{GetDist} \citep{2019arXiv191013970L}.


\appendix

\section{The posterior distributions in a full parameter space for the MCMC analysis of the microlensing light curve}
\label{sec:fullcorner}
\begin{figure*}
\centering
\includegraphics[width = \textwidth]{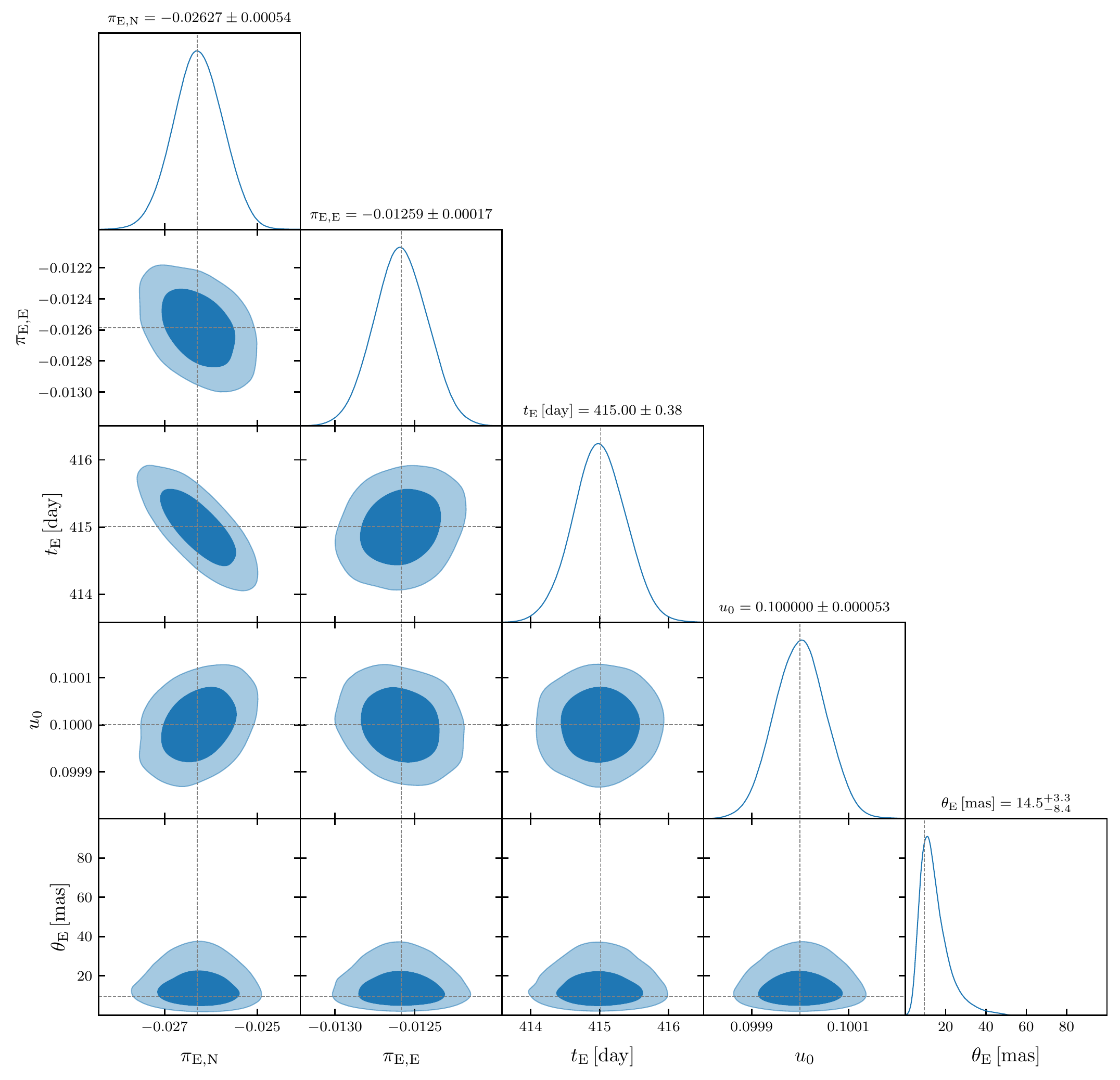}
\caption{The posterior distributions in each two-parameter space of the microlensing parameters ($\vpiE,\tE,u_0, \thetaE$) used in our MCMC analysis of the simulated light curve for 
the $40M_\odot $ BH lens, ``diskBH40'', in Table~\ref{tab:MLsettings} and Fig.~\ref{fig:lightcurve_M40D}.
Here $\piEN$ and $\piEE$ are the ``N'' (north) and ``E'' (east) components of the vector microlensing parallax $\vpiE$, respectively, where we defined the N-axis on the two-dimensional lens plane as the direction of \textit{ecliptic} north, and E-axis as the direction perpendicular to the N-direction that makes the $\rm (N,E)$ coordinates right-handed on the sky (when seen from an observer).
The lighter and darker shaded regions in each space correspond to the 
68\% and 95\% credible regions, respectively. The horizontal or vertical dashed line in each panel denotes the input value of each parameter used in the simulated microlensing light curve. 
The value for each parameter 
shown above in the 1D posterior plot 
denotes 
the median of the posterior distribution, 
and the positive and negative errors
denote the 16 and 84 percentiles. 
\label{fig:full_corner_microlensing_parameters}
}
\end{figure*}

\begin{figure*}
\centering
\includegraphics[width = 0.8\textwidth]{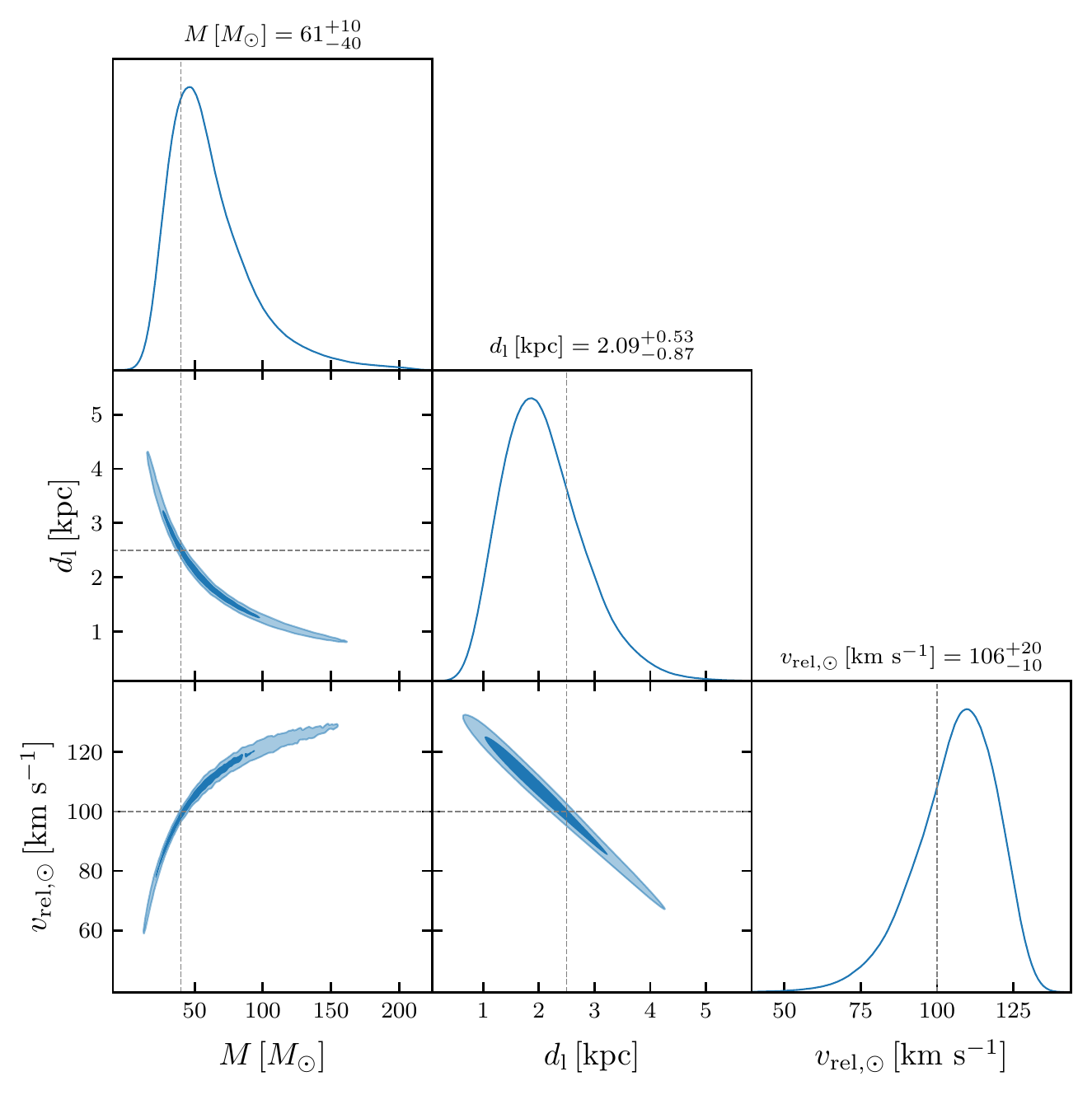}
\caption{The posterior distributions for the derived, physical parameters of the lens, obtained by projecting the posterior distributions in 
Fig.~\ref{fig:full_corner_microlensing_parameters} to the physical parameters 
via Eq.~(\ref{eq:physical_parameters_vs_ML_parameters}).
\label{fig:full_corner_for_physical_parameters}
}
\end{figure*}

In this appendix we show the posterior distributions for the microlensing parameters in a full parameter space in Fig.~\ref{fig:full_corner_microlensing_parameters}, and show 
the posterior distributions for the derived parameters in Fig.~\ref{fig:full_corner_for_physical_parameters}.



\bibliographystyle{aasjournal}
\bibliography{refs}

\end{document}

%% file: commands.tex
\newcommand{\pc}{\>{\rm pc}}
\newcommand{\kpc}{\>{\rm kpc}}

\def\gcm3{\mathrm{g} / \mathrm{cm}^3}
\def\m200m{M_{\rm 200m}}

\def\gtsima{$\; \buildrel > \over \sim \;$}
\def\ltsima{$\; \buildrel < \over \sim \;$}
\def\prosima{$\; \buildrel \propto \over \sim \;$}
\def\gsim{\lower.7ex\hbox{\gtsima}}
\def\lsim{\lower.7ex\hbox{\ltsima}}
\def\simgt{\lower.7ex\hbox{\gtsima}}
\def\simlt{\lower.7ex\hbox{\ltsima}}
\def\simpr{\lower.7ex\hbox{\prosima}}

\def\rmb{{\rm b}}

\def\rmd{{\rm d}}

\def\rms{{\rm s}}

\def\bd{{\bf d}}

%% file: citation_fix.tex
\usepackage{etoolbox}
\makeatletter
\patchcmd{\NAT@citex}
  {\@citea\NAT@hyper@{\NAT@nmfmt{\NAT@nm}\NAT@date}}
  {\@citea\NAT@nmfmt{\NAT@nm}\NAT@hyper@{\NAT@date}}
  {}
  {}
\patchcmd{\NAT@citex}
  {\@citea\NAT@hyper@{%
     \NAT@nmfmt{\NAT@nm}%
     \hyper@natlinkbreak{\NAT@aysep\NAT@spacechar}{\@citeb\@extra@b@citeb}%
     \NAT@date}}
  {\@citea\NAT@nmfmt{\NAT@nm}%
   \NAT@aysep\NAT@spacechar%
   \NAT@hyper@{\NAT@date}}
  {}
  {}
\patchcmd{\NAT@citex}
  {\@citea\NAT@hyper@{%
     \NAT@nmfmt{\NAT@nm}%
     \hyper@natlinkbreak{\NAT@spacechar\NAT@@open\if*#1*\else#1\NAT@spacechar\fi}%
       {\@citeb\@extra@b@citeb}%
     \NAT@date}}
  {\@citea\NAT@nmfmt{\NAT@nm}%
   \NAT@spacechar\NAT@@open\if*#1*\else#1\NAT@spacechar\fi%
   \NAT@hyper@{\NAT@date}}
  {}
  {}
\makeatother

%% file: main_arXiv.bbl
\begin{thebibliography}{}
\expandafter\ifx\csname natexlab\endcsname\relax\def\natexlab#1{#1}\fi
\providecommand{\url}[1]{\href{#1}{#1}}
\providecommand{\dodoi}[1]{doi:~\href{http://doi.org/#1}{\nolinkurl{#1}}}
\providecommand{\doeprint}[1]{\href{http://ascl.net/#1}{\nolinkurl{http://ascl.net/#1}}}
\providecommand{\doarXiv}[1]{\href{https://arxiv.org/abs/#1}{\nolinkurl{https://arxiv.org/abs/#1}}}

\bibitem[{{Abbott} {et~al.}(2020{\natexlab{a}}){Abbott}, {Abbott}, {Abraham},
  {Acernese}, {Ackley}, {Adams}, {Adams}, {Adhikari}, {Adya}, {Affeldt},
  {Agathos}, {Agatsuma}, {Aggarwal}, {Aguiar}, {Aiello}, {Ain}, {Ajith},
  {Akcay}, {Allen}, {Allocca}, {Altin}, {Amato}, {Anand}, {Ananyeva},
  {Anderson}, {Anderson}, {Angelova}, {Ansoldi}, {Antelis}, {Antier}, {Appert},
  {Arai}, {Araya}, {Areeda}, {Ar{\`e}ne}, {Arnaud}, {Aronson}, {Arun}, {Asali},
  {Ascenzi}, {Ashton}, {Aston}, {Astone}, {Aubin}, {Aufmuth}, {AultONeal},
  {Austin}, {Avendano}, {Babak}, {Badaracco}, {Bader}, {Bae}, {Baer},
  {Bagnasco}, {Baird}, {Ball}, {Ballardin}, {Ballmer}, {Bals}, {Balsamo},
  {Baltus}, {Banagiri}, {Bankar}, {Bankar}, {Barayoga}, {Barbieri}, {Barish},
  {Barker}, {Barneo}, {Barnum}, {Barone}, {Barr}, {Barsotti}, {Barsuglia},
  {Barta}, {Bartlett}, {Bartos}, {Bassiri}, {Basti}, {Bawaj}, {Bayley},
  {Bazzan}, {Becher}, {B{\'e}csy}, {Bedakihale}, {Bejger}, {Belahcene},
  {Beniwal}, {Benjamin}, {Bennett}, {Bentley}, {Bergamin}, {Berger},
  {Bergmann}, {Bernuzzi}, {Berry}, {Bersanetti}, {Bertolini}, {Betzwieser},
  {Bhandare}, {Bhandari}, {Bhattacharjee}, {Bidler}, {Bilenko}, {Billingsley},
  {Birney}, {Birnholtz}, {Biscans}, {Bischi}, {Biscoveanu}, {Bisht}, {Bitossi},
  {Bizouard}, {Blackburn}, {Blackman}, {Blair}, {Blair}, {Blair}, {Blanch},
  {Bobba}, {Bode}, {Boer}, {Boetzel}, {Bogaert}, {Boldrini}, {Bondu},
  {Bonnand}, {Bonilla}, {Booker}, {Boom}, {Bork}, {Boschi}, {Bose},
  {Bossilkov}, {Boudart}, {Bouffanais}, {Bozzi}, {Bradaschia}, {Brady},
  {Bramley}, {Branchesi}, {Brau}, {Breschi}, {Briant}, {Briggs}, {Brighenti},
  {Brillet}, {Brinkmann}, {Brockill}, {Brooks}, {Brooks}, {Brown}, {Brunett},
  {Bruno}, {Bruntz}, {Buikema}, {Bulik}, {Bulten}, {Buonanno}, {Buscicchio},
  {Buskulic}, {Byer}, {Cabero}, {Cadonati}, {Caesar}, {Cagnoli}, {Cahillane},
  {Calder{\'o}n Bustillo}, {Callaghan}, {Callister}, {Calloni}, {Camp},
  {Canepa}, {Cannon}, {Cao}, {Cao}, {Carapella}, {Carbognani}, {Carney},
  {Carpinelli}, {Carullo}, {Carver}, {Casanueva Diaz}, {Casentini}, {Caudill},
  {Cavagli{\`a}}, {Cavalier}, {Cavalieri}, {Cella}, {Cerd{\'a}-Dur{\'a}n},
  {Cesarini}, {Chaibi}, {Chakravarti}, {Chan}, {Chan}, {Chandra}, {Chanial},
  {Chao}, {Charlton}, {Chase}, {Chassande-Mottin}, {Chatterjee},
  {Chattopadhyay}, {Chaturvedi}, {Chatziioannou}, {Chen}, {Chen}, {Chen},
  {Chen}, {Cheng}, {Cheong}, {Chia}, {Chiadini}, {Chierici}, {Chincarini},
  {Chiummo}, {Cho}, {Cho}, {Cho}, {Choate}, {Christensen}, {Chu}, {Chua},
  {Chung}, {Chung}, {Ciani}, {Ciecielag}, {Cie{\'s}lar}, {Cifaldi}, {Ciobanu},
  {Ciolfi}, {Cipriano}, {Cirone}, {Clara}, {Clark}, {Clark}, {Clarke},
  {Clearwater}, {Clesse}, {Cleva}, {Coccia}, {Cohadon}, {Cohen}, {Colleoni},
  {Collette}, {Collins}, {Colpi}, {Constancio}, {Conti}, {Cooper}, {Corban},
  {Corbitt}, {Cordero-Carri{\'o}n}, {Corezzi}, {Corley}, {Cornish}, {Corre},
  {Corsi}, {Cortese}, {Costa}, {Cotesta}, {Coughlin}, {Coughlin}, {Coulon},
  {Countryman}, {Cousins}, {Couvares}, {Covas}, {Coward}, {Cowart}, {Coyne},
  {Coyne}, {Creighton}, {Creighton}, {Croquette}, {Crowder}, {Cudell},
  {Cullen}, {Cumming}, {Cummings}, {Cunningham}, {Cuoco}, {Curylo}, {Dal
  Canton}, {D{\'a}lya}, {Dana}, {DaneshgaranBajastani}, {D'Angelo}, {Danila},
  {Danilishin}, {D'Antonio}, {Danzmann}, {Darsow-Fromm}, {Dasgupta}, {Datrier},
  {Dattilo}, {Dave}, {Davier}, {Davies}, {Davis}, {Daw}, {Dean}, {DeBra},
  {Deenadayalan}, {Degallaix}, {De Laurentis}, {Del{\'e}glise}, {Del Favero},
  {De Lillo}, {De Lillo}, {Del Pozzo}, {DeMarchi}, {De Matteis}, {D'Emilio},
  {Demos}, {Denker}, {Dent}, {Depasse}, {De Pietri}, {De Rosa}, {De Rossi},
  {DeSalvo}, {de Varona}, {Dhurandhar}, {D{\'\i}az}, {Diaz-Ortiz}, {Didio},
  {Dietrich}, {Di Fiore}, {DiFronzo}, {Di Giorgio}, {Di Giovanni}, {Di
  Giovanni}, {Di Girolamo}, {Di Lieto}, {Ding}, {Di Pace}, {Di Palma}, {Di
  Renzo}, {Divakarla}, {Dmitriev}, {Doctor}, {D'Onofrio}, {Donovan}, {Dooley},
  {Doravari}, {Dorrington}, {Downes}, {Drago}, {Driggers}, {Du}, {Ducoin},
  {Dupej}, {Durante}, {D'Urso}, {Duverne}, {Dwyer}, {Easter}, {Eddolls},
  {Edelman}, {Edo}, {Edy}, {Effler}, {Eichholz}, {Eikenberry}, {Eisenmann},
  {Eisenstein}, {Ejlli}, {Errico}, {Essick}, {Estell{\'e}s}, {Estevez},
  {Etienne}, {Etzel}, {Evans}, {Evans}, {Ewing}, {Fafone}, {Fair}, {Fairhurst},
  {Fan}, {Farah}, {Farinon}, {Farr}, {Farr}, {Fauchon-Jones}, {Favata}, {Fays},
  {Fazio}, {Feicht}, {Fejer}, {Feng}, {Fenyvesi}, {Ferguson},
  {Fernandez-Galiana}, {Ferrante}, {Ferreira}, {Fidecaro}, {Figura}, {Fiori},
  {Fiorucci}, {Fishbach}, {Fisher}, {Fishner}, {Fittipaldi}, {Fitz-Axen},
  {Fiumara}, {Flaminio}, {Floden}, {Flynn}, {Fong}, {Font}, {Forsyth},
  {Fournier}, {Frasca}, {Frasconi}, {Frei}, {Freise}, {Frey}, {Frey},
  {Fritschel}, {Frolov}, {Fronz{\'e}}, {Fulda}, {Fyffe}, {Gabbard}, {Gadre},
  {Gaebel}, {Gair}, {Gais}, {Galaudage}, {Gamba}, {Ganapathy}, {Ganguly},
  {Gaonkar}, {Garaventa}, {Garc{\'\i}a-Quir{\'o}s}, {Garufi}, {Gateley},
  {Gaudio}, {Gayathri}, {Gemme}, {Gennai}, {George}, {George}, {George},
  {Gergely}, {Ghonge}, {Ghosh}, {Ghosh}, {Ghosh}, {Giacomazzo}, {Giacoppo},
  {Giaime}, {Giardina}, {Gibson}, {Gier}, {Gill}, {Giri}, {Glanzer}, {Gleckl},
  {Godwin}, {Goetz}, {Goetz}, {Gohlke}, {Goncharov}, {Gonz{\'a}lez},
  {Gopakumar}, {Gossan}, {Gosselin}, {Gouaty}, {Grace}, {Grado}, {Granata},
  {Granata}, {Grant}, {Gras}, {Grassia}, {Gray}, {Gray}, {Greco}, {Green},
  {Green}, {Gretarsson}, {Griggs}, {Grignani}, {Grimaldi}, {Grimes}, {Grimm},
  {Grote}, {Grunewald}, {Gruning}, {Guerrero}, {Guidi}, {Guimaraes},
  {Guix{\'e}}, {Gulati}, {Guo}, {Gupta}, {Gupta}, {Gupta}, {Gustafson},
  {Gustafson}, {Guzman}, {Haegel}, {Halim}, {Hall}, {Hamilton}, {Hammond},
  {Haney}, {Hanke}, {Hanks}, {Hanna}, {Hannam}, {Hannuksela}, {Hannuksela},
  {Hansen}, {Hansen}, {Hanson}, {Harder}, {Hardwick}, {Haris}, {Harms},
  {Harry}, {Harry}, {Hartwig}, {Hasskew}, {Haster}, {Haughian}, {Hayes},
  {Healy}, {Heidmann}, {Heintze}, {Heinze}, {Heinzel}, {Heitmann}, {Hellman},
  {Hello}, {Helmling-Cornell}, {Hemming}, {Hendry}, {Heng}, {Hennes}, {Hennig},
  {Hennig}, {Hernandez Vivanco}, {Heurs}, {Hild}, {Hill}, {Hines}, {Hochheim},
  {Hofgard}, {Hofman}, {Hohmann}, {Holgado}, {Holland}, {Hollows}, {Holmes},
  {Holt}, {Holz}, {Hopkins}, {Horst}, {Hough}, {Howell}, {Hoy}, {Hoyland},
  {Huang}, {H{\"u}bner}, {Huddart}, {Huerta}, {Hughey}, {Hui}, {Husa},
  {Huttner}, {Hutzler}, {Huxford}, {Huynh-Dinh}, {Idzkowski}, {Iess},
  {Imperato}, {Inchauspe}, {Ingram}, {Intini}, {Isi}, {Iyer},
  {JaberianHamedan}, {Jacqmin}, {Jadhav}, {Jadhav}, {James}, {Jani},
  {Janssens}, {Janthalur}, {Jaranowski}, {Jariwala}, {Jaume}, {Jenkins},
  {Jeunon}, {Jiang}, {Johns}, {Johnson-McDaniel}, {Jones}, {Jones}, {Jones},
  {Jones}, {Jones}, {Jonker}, {Ju}, {Junker}, {Kalaghatgi}, {Kalogera},
  {Kamai}, {Kandhasamy}, {Kang}, {Kanner}, {Kapadia}, {Kapasi}, {Karathanasis},
  {Karki}, {Kashyap}, {Kasprzack}, {Kastaun}, {Katsanevas}, {Katsavounidis},
  {Katzman}, {Kawabe}, {K{\'e}f{\'e}lian}, {Keitel}, {Key}, {Khadka},
  {Khalili}, {Khan}, {Khan}, {Khazanov}, {Khetan}, {Khursheed}, {Kijbunchoo},
  {Kim}, {Kim}, {Kim}, {Kim}, {Kim}, {Kim}, {Kimball}, {King}, {Kinley-Hanlon},
  {Kirchhoff}, {Kissel}, {Kleybolte}, {Klimenko}, {Knowles}, {Knyazev}, {Koch},
  {Koehlenbeck}, {Koekoek}, {Koley}, {Kolstein}, {Komori}, {Kondrashov},
  {Kontos}, {Koper}, {Korobko}, {Korth}, {Kovalam}, {Kozak}, {Kr{\"a}mer},
  {Kringel}, {Krishnendu}, {Kr{\'o}lak}, {Kuehn}, {Kumar}, {Kumar}, {Kumar},
  {Kumar}, {Kuns}, {Kwang}, {Lackey}, {Laghi}, {Lalande}, {Lam}, {Lamberts},
  {Landry}, {Lane}, {Lang}, {Lange}, {Lantz}, {Lanza}, {La Rosa},
  {Lartaux-Vollard}, {Lasky}, {Laxen}, {Lazzarini}, {Lazzaro}, {Leaci},
  {Leavey}, {Lecoeuche}, {Lee}, {Lee}, {Lee}, {Lee}, {Lehmann}, {Leon},
  {Leroy}, {Letendre}, {Levin}, {Li}, {Li}, {Li}, {Li}, {Li}, {Linde},
  {Linker}, {Linley}, {Littenberg}, {Liu}, {Liu}, {Llorens-Monteagudo}, {Lo},
  {Lockwood}, {London}, {Longo}, {Lorenzini}, {Loriette}, {Lormand}, {Losurdo},
  {Lough}, {Lousto}, {Lovelace}, {L{\"u}ck}, {Lumaca}, {Lundgren}, {Ma},
  {Macas}, {MacInnis}, {Macleod}, {MacMillan}, {Macquet}, {Maga{\~n}a
  Hernandez}, {Maga{\~n}a-Sandoval}, {Magazz{\`u}}, {Magee}, {Majorana},
  {Maksimovic}, {Maliakal}, {Malik}, {Man}, {Mandic}, {Mangano}, {Mansell},
  {Manske}, {Mantovani}, {Mapelli}, {Marchesoni}, {Marion}, {M{\'a}rka},
  {M{\'a}rka}, {Markakis}, {Markosyan}, {Markowitz}, {Maros}, {Marquina},
  {Marsat}, {Martelli}, {Martin}, {Martin}, {Martinez}, {Martinez}, {Martynov},
  {Masalehdan}, {Mason}, {Massera}, {Masserot}, {Massinger}, {Masso-Reid},
  {Mastrogiovanni}, {Matas}, {Mateu-Lucena}, {Matichard}, {Matiushechkina},
  {Mavalvala}, {Maynard}, {McCann}, {McCarthy}, {McClelland}, {McCormick},
  {McCuller}, {McGuire}, {McIsaac}, {McIver}, {McManus}, {McRae}, {McWilliams},
  {Meacher}, {Meadors}, {Mehmet}, {Mehta}, {Melatos}, {Melchor}, {Mendell},
  {Menendez-Vazquez}, {Mercer}, {Mereni}, {Merfeld}, {Merilh}, {Merritt},
  {Merzougui}, {Meshkov}, {Messenger}, {Messick}, {Metzdorff}, {Meyers},
  {Meylahn}, {Mhaske}, {Miani}, {Miao}, {Michaloliakos}, {Michel}, {Middleton},
  {Milano}, {Miller}, {Millhouse}, {Mills}, {Milotti}, {Milovich-Goff},
  {Minazzoli}, {Minenkov}, {Mir}, {Mishkin}, {Mishra}, {Mistry}, {Mitra},
  {Mitrofanov}, {Mitselmakher}, {Mittleman}, {Mo}, {Mogushi}, {Mohapatra},
  {Mohite}, {Molina}, {Molina-Ruiz}, {Mondin}, {Montani}, {Moore}, {Moraru},
  {Morawski}, {Moreno}, {Morisaki}, {Mours}, {Mow-Lowry}, {Mozzon},
  {Muciaccia}, {Mukherjee}, {Mukherjee}, {Mukherjee}, {Mukherjee}, {Mukund},
  {Mullavey}, {Munch}, {Mu{\~n}iz}, {Murray}, {Nadji}, {Nagar}, {Nardecchia},
  {Naticchioni}, {Nayak}, {Neil}, {Neilson}, {Nelemans}, {Nelson}, {Nery},
  {Neunzert}, {Nitz}, {Ng}, {Ng}, {Nguyen}, {Nguyen}, {Nguyen}, {Nichols},
  {Nissanke}, {Nocera}, {Noh}, {North}, {Nothard}, {Nuttall}, {Oberling},
  {O'Brien}, {O'Dell}, {Oganesyan}, {Ogin}, {Oh}, {Oh}, {Ohme}, {Ohta},
  {Okada}, {Olivetto}, {Oppermann}, {Oram}, {O'Reilly}, {Ormiston}, {Ortega},
  {O'Shaughnessy}, {Ossokine}, {Osthelder}, {Ottaway}, {Overmier}, {Owen},
  {Pace}, {Pagano}, {Page}, {Pagliaroli}, {Pai}, {Pai}, {Palamos}, {Palashov},
  {Palomba}, {Pan}, {Panda}, {Pang}, {Pankow}, {Pannarale}, {Pant}, {Paoletti},
  {Paoli}, {Paolone}, {Parker}, {Pascucci}, {Pasqualetti}, {Passaquieti},
  {Passuello}, {Patel}, {Patricelli}, {Payne}, {Pechsiri}, {Pedraza},
  {Pegoraro}, {Pele}, {Penn}, {Perego}, {Perez}, {P{\'e}rigois}, {Perreca},
  {Perri{\`e}s}, {Petermann}, {Petterson}, {Pfeiffer}, {Pham}, {Phukon},
  {Piccinni}, {Pichot}, {Piendibene}, {Piergiovanni}, {Pierini}, {Pierro},
  {Pillant}, {Pilo}, {Pinard}, {Pinto}, {Piotrzkowski}, {Pirello}, {Pitkin},
  {Placidi}, {Plastino}, {Pluchar}, {Poggiani}, {Polini}, {Pong}, {Ponrathnam},
  {Popolizio}, {Porter}, {Poverman}, {Powell}, {Pracchia}, {Prajapati},
  {Prasai}, {Prasanna}, {Pratten}, {Prestegard}, {Principe}, {Prodi},
  {Prokhorov}, {Prosposito}, {Prudenzi}, {Puecher}, {Punturo}, {Puosi},
  {Puppo}, {P{\"u}rrer}, {Qi}, {Quetschke}, {Quinonez}, {Quitzow-James},
  {Raab}, {Raaijmakers}, {Radkins}, {Radulesco}, {Raffai}, {Rafferty}, {Rail},
  {Raja}, {Rajan}, {Rajbhandari}, {Rakhmanov}, {Ramirez}, {Ramirez},
  {Ramos-Buades}, {Rana}, {Rao}, {Rapagnani}, {Rapol}, {Ratto}, {Raymond},
  {Razzano}, {Read}, {Regimbau}, {Rei}, {Reid}, {Reitze}, {Rettegno}, {Ricci},
  {Richardson}, {Richardson}, {Richardson}, {Ricker}, {Riemenschneider},
  {Riles}, {Rizzo}, {Robertson}, {Robinet}, {Rocchi}, {Rocha}, {Rodriguez},
  {Rodriguez-Soto}, {Rolland}, {Rollins}, {Roma}, {Romanelli}, {Romano},
  {Romel}, {Romero}, {Romero-Shaw}, {Romie}, {Ronchini}, {Rose}, {Rose},
  {Rose}, {Rosell}, {Rosi{\'n}ska}, {Rosofsky}, {Ross}, {Rowan}, {Rowlinson},
  {Roy}, {Roy}, {Ruggi}, {Ryan}, {Sachdev}, {Sadecki}, {Sadiq},
  {Sakellariadou}, {Salafia}, {Salconi}, {Saleem}, {Samajdar}, {Sanchez},
  {Sanchez}, {Sanchez}, {Sanchis-Gual}, {Sanders}, {Sandles}, {Santiago},
  {Santos}, {Saravanan}, {Sarin}, {Sassolas}, {Sathyaprakash}, {Sauter},
  {Savage}, {Savant}, {Sawant}, {Sayah}, {Schaetzl}, {Schale}, {Scheel},
  {Scheuer}, {Schindler-Tyka}, {Schmidt}, {Schnabel}, {Schofield},
  {Sch{\"o}nbeck}, {Schreiber}, {Schulte}, {Schutz}, {Schwarm}, {Schwartz},
  {Scott}, {Scott}, {Seglar-Arroyo}, {Seidel}, {Sellers}, {Sengupta},
  {Sennett}, {Sentenac}, {Sequino}, {Sergeev}, {Setyawati}, {Shaffer},
  {Shahriar}, {Sharifi}, {Sharma}, {Sharma}, {Shawhan}, {Shen}, {Shikauchi},
  {Shink}, {Shoemaker}, {Shoemaker}, {Shukla}, {ShyamSundar}, {Sieniawska},
  {Sigg}, {Singer}, {Singh}, {Singh}, {Singha}, {Singhal}, {Sintes}, {Sipala},
  {Skliris}, {Slagmolen}, {Slaven-Blair}, {Smetana}, {Smith}, {Smith},
  {Somala}, {Son}, {Soni}, {Soni}, {Sorazu}, {Sordini}, {Sorrentino},
  {Sorrentino}, {Soulard}, {Souradeep}, {Sowell}, {Spencer}, {Spera},
  {Srivastava}, {Srivastava}, {Staats}, {Stachie}, {Steer}, {Steinhoff},
  {Steinke}, {Steinlechner}, {Steinlechner}, {Steinmeyer}, {Stevenson},
  {Stolle-McAllister}, {Stops}, {Stover}, {Strain}, {Stratta}, {Strunk},
  {Sturani}, {Stuver}, {S{\"u}dbeck}, {Sudhagar}, {Sudhir}, {Suh},
  {Summerscales}, {Sun}, {Sun}, {Sunil}, {Sur}, {Suresh}, {Sutton}, {Swinkels},
  {Szczepa{\'n}czyk}, {Tacca}, {Tait}, {Talbot}, {Tanasijczuk}, {Tanner},
  {Tao}, {Tapia}, {Tapia San Martin}, {Tasson}, {Taylor}, {Tenorio},
  {Terkowski}, {Thirugnanasambandam}, {Thomas}, {Thomas}, {Thomas}, {Thompson},
  {Thondapu}, {Thorne}, {Thrane}, {Tiwari}, {Tiwari}, {Tiwari}, {Toland},
  {Tolley}, {Tonelli}, {Tornasi}, {Torres-Forn{\'e}}, {Torrie}, {Melo},
  {T{\"o}yr{\"a}}, {Tran}, {Trapananti}, {Travasso}, {Traylor}, {Tringali},
  {Tripathee}, {Trovato}, {Trudeau}, {Tsai}, {Tsang}, {Tse}, {Tso}, {Tsukada},
  {Tsuna}, {Tsutsui}, {Turconi}, {Ubhi}, {Udall}, {Ueno}, {Ugolini},
  {Unnikrishnan}, {Urban}, {Usman}, {Utina}, {Vahlbruch}, {Vajente}, {Vajpeyi},
  {Valdes}, {Valentini}, {Valsan}, {van Bakel}, {van Beuzekom}, {van den
  Brand}, {Van Den Broeck}, {Vander-Hyde}, {van der Schaaf}, {van Heijningen},
  {Vardaro}, {Vargas}, {Varma}, {Vass}, {Vas{\'u}th}, {Vecchio}, {Vedovato},
  {Veitch}, {Veitch}, {Venkateswara}, {Venneberg}, {Venugopalan}, {Verkindt},
  {Verma}, {Veske}, {Vetrano}, {Vicer{\'e}}, {Viets}, {Vijaykumar},
  {Villa-Ortega}, {Vinet}, {Vitale}, {Vo}, {Vocca}, {Vorvick}, {Vyatchanin},
  {Wade}, {Wade}, {Wade}, {Walet}, {Walker}, {Wallace}, {Wallace}, {Walsh},
  {Wang}, {Wang}, {Wang}, {Wang}, {Ward}, {Warner}, {Was}, {Washington},
  {Watchi}, {Weaver}, {Wei}, {Weinert}, {Weinstein}, {Weiss}, {Wellmann},
  {Wen}, {We{\ss}els}, {Westhouse}, {Wette}, {Whelan}, {White}, {White},
  {Whiting}, {Whittle}, {Wilken}, {Williams}, {Williams}, {Williamson},
  {Willis}, {Willke}, {Wilson}, {Wimmer}, {Winkler}, {Wipf}, {Woan}, {Woehler},
  {Wofford}, {Wong}, {Wrangel}, {Wright}, {Wu}, {Wysocki}, {Xiao}, {Yamamoto},
  {Yang}, {Yang}, {Yang}, {Yap}, {Yeeles}, {Yoon}, {Yu}, {Yu}, {Yuen},
  {Zadro{\.z}ny}, {Zanolin}, {Zelenova}, {Zendri}, {Zevin}, {Zhang}, {Zhang},
  {Zhang}, {Zhang}, {Zhao}, {Zhao}, {Zheng}, {Zhou}, {Zhou}, {Zhu},
  {Zimmerman}, {Zlochower}, {Zucker}, \& {Zweizig}}]{2020arXiv201014527A}
{Abbott}, R., {Abbott}, T.~D., {Abraham}, S., {et~al.} 2020{\natexlab{a}},
  arXiv e-prints, arXiv:2010.14527.
\newblock \doarXiv{2010.14527}

\bibitem[{{Abbott} {et~al.}(2020{\natexlab{b}}){Abbott}, {Abbott}, {Abraham},
  {Acernese}, {Ackley}, {Adams}, {Adams}, {Adhikari}, {Adya}, {Affeldt},
  {Agathos}, {Agatsuma}, {Aggarwal}, {Aguiar}, {Aiello}, {Ain}, {Ajith},
  {Allen}, {Allocca}, {Altin}, {Amato}, {Anand}, {Ananyeva}, {Anderson},
  {Anderson}, {Angelova}, {Ansoldi}, {Antelis}, {Antier}, {Appert}, {Arai},
  {Araya}, {Areeda}, {Ar{\`e}ne}, {Arnaud}, {Aronson}, {Arun}, {Asali},
  {Ascenzi}, {Ashton}, {Aston}, {Astone}, {Aubin}, {Aufmuth}, {AultONeal},
  {Austin}, {Avendano}, {Babak}, {Badaracco}, {Bader}, {Bae}, {Baer},
  {Bagnasco}, {Baird}, {Ball}, {Ballardin}, {Ballmer}, {Bals}, {Balsamo},
  {Baltus}, {Banagiri}, {Bankar}, {Bankar}, {Barayoga}, {Barbieri}, {Barish},
  {Barker}, {Barneo}, {Barnum}, {Barone}, {Barr}, {Barsotti}, {Barsuglia},
  {Barta}, {Bartlett}, {Bartos}, {Bassiri}, {Basti}, {Bawaj}, {Bayley},
  {Bazzan}, {Becher}, {B{\'e}csy}, {Bedakihale}, {Bejger}, {Belahcene},
  {Beniwal}, {Benjamin}, {Bennett}, {Bentley}, {Bergamin}, {Berger},
  {Bergmann}, {Bernuzzi}, {Berry}, {Bersanetti}, {Bertolini}, {Betzwieser},
  {Bhandare}, {Bhandari}, {Bhattacharjee}, {Bidler}, {Bilenko}, {Billingsley},
  {Birney}, {Birnholtz}, {Biscans}, {Bischi}, {Biscoveanu}, {Bisht}, {Bitossi},
  {Bizouard}, {Blackburn}, {Blackman}, {Blair}, {Blair}, {Blair}, {Blanch},
  {Bobba}, {Bode}, {Boer}, {Boetzel}, {Bogaert}, {Boldrini}, {Bondu},
  {Bonilla}, {Bonnand}, {Booker}, {Boom}, {Bork}, {Boschi}, {Bose},
  {Bossilkov}, {Boudart}, {Bouffanais}, {Bozzi}, {Bradaschia}, {Brady},
  {Bramley}, {Branchesi}, {Brau}, {Breschi}, {Briant}, {Briggs}, {Brighenti},
  {Brillet}, {Brinkmann}, {Brockill}, {Brooks}, {Brooks}, {Brown}, {Brunett},
  {Bruno}, {Bruntz}, {Buikema}, {Bulik}, {Bulten}, {Buonanno}, {Buscicchio},
  {Buskulic}, {Byer}, {Cabero}, {Cadonati}, {Caesar}, {Cagnoli}, {Cahillane},
  {Calder{\'o}n Bustillo}, {Callaghan}, {Callister}, {Calloni}, {Camp},
  {Canepa}, {Cannon}, {Cao}, {Cao}, {Carapella}, {Carbognani}, {Carney},
  {Carpinelli}, {Carullo}, {Carver}, {Casanueva Diaz}, {Casentini}, {Caudill},
  {Cavagli{\`a}}, {Cavalier}, {Cavalieri}, {Cella}, {Cerd{\'a}-Dur{\'a}n},
  {Cesarini}, {Chaibi}, {Chakravarti}, {Chan}, {Chan}, {Chandra}, {Chanial},
  {Chao}, {Charlton}, {Chase}, {Chassande-Mottin}, {Chatterjee},
  {Chattopadhyay}, {Chaturvedi}, {Chatziioannou}, {Chen}, {Chen}, {Chen},
  {Chen}, {Cheng}, {Cheong}, {Chia}, {Chiadini}, {Chierici}, {Chincarini},
  {Chiummo}, {Cho}, {Cho}, {Cho}, {Choate}, {Christensen}, {Chu}, {Chua},
  {Chung}, {Chung}, {Ciani}, {Ciecielag}, {Cie{\'s}lar}, {Cifaldi}, {Ciobanu},
  {Ciolfi}, {Cipriano}, {Cirone}, {Clara}, {Clark}, {Clark}, {Clarke},
  {Clearwater}, {Clesse}, {Cleva}, {Coccia}, {Cohadon}, {Cohen}, {Colleoni},
  {Collette}, {Collins}, {Colpi}, {Constancio}, {Conti}, {Cooper}, {Corban},
  {Corbitt}, {Cordero-Carri{\'o}n}, {Corezzi}, {Corley}, {Cornish}, {Corre},
  {Corsi}, {Cortese}, {Costa}, {Cotesta}, {Coughlin}, {Coughlin}, {Coulon},
  {Countryman}, {Couvares}, {Covas}, {Coward}, {Cowart}, {Coyne}, {Coyne},
  {Creighton}, {Creighton}, {Croquette}, {Crowder}, {Cudell}, {Cullen},
  {Cumming}, {Cummings}, {Cunningham}, {Cuoco}, {Cury\{l\}o}, {Dal Canton},
  {D{\'a}lya}, {Dana}, {DaneshgaranBajastani}, {D'Angelo}, {Danilishin},
  {D'Antonio}, {Danzmann}, {Darsow-Fromm}, {Dasgupta}, {Datrier}, {Dattilo},
  {Dave}, {Davier}, {Davies}, {Davis}, {Daw}, {Dean}, {DeBra}, {Deenadayalan},
  {Degallaix}, {De Laurentis}, {Del{\'e}glise}, {Del Favero}, {De Lillo}, {De
  Lillo}, {Del Pozzo}, {DeMarchi}, {De Matteis}, {D'Emilio}, {Demos}, {Denker},
  {Dent}, {Depasse}, {De Pietri}, {De Rosa}, {De Rossi}, {DeSalvo}, {de
  Varona}, {Dhurandhar}, {D{\'\i}az}, {Diaz-Ortiz}, {Didio}, {Dietrich}, {Di
  Fiore}, {DiFronzo}, {Di Giorgio}, {Di Giovanni}, {Di Giovanni}, {Di
  Girolamo}, {Di Lieto}, {Ding}, {Di Pace}, {Di Palma}, {Di Renzo},
  {Divakarla}, {Dmitriev}, {Doctor}, {D'Onofrio}, {Donovan}, {Dooley},
  {Doravari}, {Dorrington}, {Downes}, {Drago}, {Driggers}, {Du}, {Ducoin},
  {Dupej}, {Durante}, {D'Urso}, {Duverne}, {Dwyer}, {Easter}, {Eddolls},
  {Edelman}, {Edo}, {Edy}, {Effler}, {Eichholz}, {Eikenberry}, {Eisenmann},
  {Eisenstein}, {Ejlli}, {Errico}, {Essick}, {Estell{\'e}s}, {Estevez},
  {Etienne}, {Etzel}, {Evans}, {Evans}, {Ewing}, {Fafone}, {Fair}, {Fairhurst},
  {Fan}, {Farah}, {Farinon}, {Farr}, {Farr}, {Fauchon-Jones}, {Favata}, {Fays},
  {Fazio}, {Feicht}, {Fejer}, {Feng}, {Fenyvesi}, {Ferguson},
  {Fernandez-Galiana}, {Ferrante}, {Ferreira}, {Fidecaro}, {Figura}, {Fiori},
  {Fiorucci}, {Fishbach}, {Fisher}, {Fishner}, {Fittipaldi}, {Fitz-Axen},
  {Fiumara}, {Flaminio}, {Floden}, {Flynn}, {Fong}, {Font}, {Forsyth},
  {Fournier}, {Frasca}, {Frasconi}, {Frei}, {Freise}, {Frey}, {Frey},
  {Fritschel}, {Frolov}, {Fronz{\'e}}, {Fulda}, {Fyffe}, {Gabbard}, {Gadre},
  {Gaebel}, {Gair}, {Gais}, {Galaudage}, {Gamba}, {Ganapathy}, {Ganguly},
  {Gaonkar}, {Garaventa}, {Garc{\'\i}a-Quir{\'o}s}, {Garufi}, {Gateley},
  {Gaudio}, {Gayathri}, {Gemme}, {Gennai}, {George}, {George}, {Gergely},
  {Ghonge}, {Ghosh}, {Ghosh}, {Ghosh}, {Giacomazzo}, {Giacoppo}, {Giaime},
  {Giardina}, {Gibson}, {Gier}, {Gill}, {Giri}, {Glanzer}, {Gleckl}, {Godwin},
  {Goetz}, {Goetz}, {Gohlke}, {Goncharov}, {Gonz{\'a}lez}, {Gopakumar},
  {Gossan}, {Gosselin}, {Gouaty}, {Grace}, {Grado}, {Granata}, {Granata},
  {Grant}, {Gras}, {Grassia}, {Gray}, {Gray}, {Greco}, {Green}, {Green},
  {Gretarsson}, {Griggs}, {Grignani}, {Grimaldi}, {Grimes}, {Grimm}, {Grote},
  {Grunewald}, {Gruning}, {Guerrero}, {Guidi}, {Guimaraes}, {Guix{\'e}},
  {Gulati}, {Guo}, {Gupta}, {Gupta}, {Gupta}, {Gustafson}, {Gustafson},
  {Guzman}, {Haegel}, {Halim}, {Hall}, {Hamilton}, {Hammond}, {Haney}, {Hanke},
  {Hanks}, {Hanna}, {Hannuksela}, {Hansen}, {Hansen}, {Hanson}, {Harder},
  {Hardwick}, {Haris}, {Harms}, {Harry}, {Harry}, {Hartwig}, {Hasskew},
  {Haster}, {Haughian}, {Hayes}, {Healy}, {Heidmann}, {Heintze}, {Heinze},
  {Heinzel}, {Heitmann}, {Hellman}, {Hello}, {Helmling-Cornell}, {Hemming},
  {Hendry}, {Heng}, {Hennes}, {Hennig}, {Hennig}, {Hernandez Vivanco}, {Heurs},
  {Hild}, {Hill}, {Hines}, {Hochheim}, {Hofgard}, {Hofman}, {Hohmann},
  {Holgado}, {Holland}, {Hollows}, {Holmes}, {Holt}, {Holz}, {Hopkins},
  {Horst}, {Hough}, {Howell}, {Hoy}, {Hoyland}, {Huang}, {H{\"u}bner},
  {Huddart}, {Huerta}, {Hughey}, {Hui}, {Husa}, {Huttner}, {Hutzler},
  {Huxford}, {Huynh-Dinh}, {Idzkowski}, {Iess}, {Imperato}, {Inchauspe},
  {Ingram}, {Intini}, {Isi}, {Iyer}, {JaberianHamedan}, {Jacqmin}, {Jadhav},
  {Jadhav}, {James}, {Jani}, {Janssens}, {Janthalur}, {Jaranowski}, {Jariwala},
  {Jaume}, {Jenkins}, {Jeunon}, {Jiang}, {Johns}, {Jones}, {Jones}, {Jones},
  {Jones}, {Jones}, {Jonker}, {Ju}, {Junker}, {Kalaghatgi}, {Kalogera},
  {Kamai}, {Kandhasamy}, {Kang}, {Kanner}, {Kapadia}, {Kapasi}, {Karathanasis},
  {Karki}, {Kashyap}, {Kasprzack}, {Kastaun}, {Katsanevas}, {Katsavounidis},
  {Katzman}, {Kawabe}, {K{\'e}f{\'e}lian}, {Keitel}, {Key}, {Khadka},
  {Khalili}, {Khan}, {Khan}, {Khazanov}, {Khetan}, {Khursheed}, {Kijbunchoo},
  {Kim}, {Kim}, {Kim}, {Kim}, {Kim}, {Kim}, {Kimball}, {King}, {Kinley-Hanlon},
  {Kirchhoff}, {Kissel}, {Kleybolte}, {Klimenko}, {Knowles}, {Knyazev}, {Koch},
  {Koehlenbeck}, {Koekoek}, {Koley}, {Kolstein}, {Komori}, {Kondrashov},
  {Kontos}, {Koper}, {Korobko}, {Korth}, {Kovalam}, {Kozak}, {Kr{\"a}mer},
  {Kringel}, {Krishnendu}, {Kr{\'o}lak}, {Kuehn}, {Kumar}, {Kumar}, {Kumar},
  {Kumar}, {Kuns}, {Kwang}, {Lackey}, {Laghi}, {Lalande}, {Lam}, {Lamberts},
  {Landry}, {Lane}, {Lang}, {Lange}, {Lantz}, {Lanza}, {La Rosa},
  {Lartaux-Vollard}, {Lasky}, {Laxen}, {Lazzarini}, {Lazzaro}, {Leaci},
  {Leavey}, {Lecoeuche}, {Lee}, {Lee}, {Lee}, {Lee}, {Lehmann}, {Leon},
  {Leroy}, {Letendre}, {Levin}, {Li}, {Li}, {Li}, {Li}, {Li}, {Linde},
  {Linker}, {Linley}, {Littenberg}, {Liu}, {Liu}, {Llorens-Monteagudo}, {Lo},
  {Lockwood}, {London}, {Longo}, {Lorenzini}, {Loriette}, {Lormand}, {Losurdo},
  {Lough}, {Lousto}, {Lovelace}, {L{\"u}ck}, {Lumaca}, {Lundgren}, {Ma},
  {Macas}, {MacInnis}, {Macleod}, {MacMillan}, {Macquet}, {Maga{\~n}a
  Hernandez}, {Maga{\~n}a-Sandoval}, {Magazz{\`u}}, {Magee}, {Majorana},
  {Maksimovic}, {Maliakal}, {Malik}, {Man}, {Mandic}, {Mangano}, {Mansell},
  {Manske}, {Mantovani}, {Mapelli}, {Marchesoni}, {Marion}, {M{\'a}rka},
  {M{\'a}rka}, {Markakis}, {Markosyan}, {Markowitz}, {Maros}, {Marquina},
  {Marsat}, {Martelli}, {Martin}, {Martin}, {Martinez}, {Martinez}, {Martynov},
  {Masalehdan}, {Mason}, {Massera}, {Masserot}, {Massinger}, {Masso-Reid},
  {Mastrogiovanni}, {Matas}, {Mateu-Lucena}, {Matichard}, {Matiushechkina},
  {Mavalvala}, {Maynard}, {McCann}, {McCarthy}, {McClelland}, {McCormick},
  {McCuller}, {McGuire}, {McIsaac}, {McIver}, {McManus}, {McRae}, {McWilliams},
  {Meacher}, {Meadors}, {Mehmet}, {Mehta}, {Melatos}, {Melchor}, {Mendell},
  {Menendez-Vazquez}, {Mercer}, {Mereni}, {Merfeld}, {Merilh}, {Merritt},
  {Merzougui}, {Meshkov}, {Messenger}, {Messick}, {Metzdorff}, {Meyers},
  {Meylahn}, {Mhaske}, {Miani}, {Miao}, {Michaloliakos}, {Michel}, {Middleton},
  {Milano}, {Miller}, {Miller}, {Millhouse}, {Mills}, {Milotti},
  {Milovich-Goff}, {Minazzoli}, {Minenkov}, {Mir}, {Mishkin}, {Mishra},
  {Mistry}, {Mitra}, {Mitrofanov}, {Mitselmakher}, {Mittleman}, {Mo},
  {Mogushi}, {Mohapatra}, {Mohite}, {Molina}, {Molina-Ruiz}, {Mondin},
  {Montani}, {Moore}, {Moraru}, {Morawski}, {Moreno}, {Morisaki}, {Mours},
  {Mow-Lowry}, {Mozzon}, {Muciaccia}, {Mukherjee}, {Mukherjee}, {Mukherjee},
  {Mukherjee}, {Mukund}, {Mullavey}, {Munch}, {Mu{\~n}iz}, {Murray}, {Nadji},
  {Nagar}, {Nardecchia}, {Naticchioni}, {Nayak}, {Neil}, {Neilson}, {Nelemans},
  {Nelson}, {Nery}, {Neunzert}, {Ng}, {Ng}, {Nguyen}, {Nguyen}, {Nguyen},
  {Nichols}, {Nissanke}, {Nocera}, {Noh}, {North}, {Nothard}, {Nuttall},
  {Oberling}, {O'Brien}, {O'Dell}, {Oganesyan}, {Ogin}, {Oh}, {Oh}, {Ohme},
  {Ohta}, {Okada}, {Olivetto}, {Oppermann}, {Oram}, {O'Reilly}, {Ormiston},
  {Ormsby}, {Ortega}, {O'Shaughnessy}, {Ossokine}, {Osthelder}, {Ottaway},
  {Overmier}, {Owen}, {Pace}, {Pagano}, {Page}, {Pagliaroli}, {Pai}, {Pai},
  {Palamos}, {Palashov}, {Palomba}, {Pan}, {Panda}, {Pang}, {Pankow},
  {Pannarale}, {Pant}, {Paoletti}, {Paoli}, {Paolone}, {Parker}, {Pascucci},
  {Pasqualetti}, {Passaquieti}, {Passuello}, {Patel}, {Patricelli}, {Payne},
  {Pechsiri}, {Pedraza}, {Pegoraro}, {Pele}, {Penn}, {Perego}, {Perez},
  {P{\'e}rigois}, {Perreca}, {Perri{\`e}s}, {Petermann}, {Petterson},
  {Pfeiffer}, {Pham}, {Phukon}, {Piccinni}, {Pichot}, {Piendibene},
  {Piergiovanni}, {Pierini}, {Pierro}, {Pillant}, {Pilo}, {Pinard}, {Pinto},
  {Piotrzkowski}, {Pirello}, {Pitkin}, {Placidi}, {Plastino}, {Pluchar},
  {Poggiani}, {Polini}, {Pong}, {Ponrathnam}, {Popolizio}, {Porter},
  {Poverman}, {Powell}, {Pracchia}, {Prajapati}, {Prasai}, {Prasanna},
  {Pratten}, {Prestegard}, {Principe}, {Prodi}, {Prokhorov}, {Prosposito},
  {Puecher}, {Punturo}, {Puosi}, {Puppo}, {P{\"u}rrer}, {Qi}, {Quetschke},
  {Quinonez}, {Quitzow-James}, {Raab}, {Raaijmakers}, {Radkins}, {Radulesco},
  {Raffai}, {Rafferty}, {Rail}, {Raja}, {Rajan}, {Rajbhandari}, {Rakhmanov},
  {Ramirez}, {Ramirez}, {Ramos-Buades}, {Rana}, {Rao}, {Rapagnani}, {Rapol},
  {Ratto}, {Raymond}, {Razzano}, {Read}, {Regimbau}, {Rei}, {Reid}, {Reitze},
  {Rettegno}, {Ricci}, {Richardson}, {Richardson}, {Richardson}, {Ricker},
  {Riemenschneider}, {Riles}, {Rizzo}, {Robertson}, {Robinet}, {Rocchi},
  {Rocha}, {Rodriguez}, {Rodriguez-Soto}, {Rolland}, {Rollins}, {Roma},
  {Romanelli}, {Romano}, {Romel}, {Romero}, {Romero-Shaw}, {Romie}, {Ronchini},
  {Rose}, {Rose}, {Rose}, {Rosell}, {Rosi{\'n}ska}, {Rosofsky}, {Ross},
  {Rowan}, {Rowlinson}, {Roy}, {Roy}, {Ruggi}, {Ryan}, {Sachdev}, {Sadecki},
  {Sadiq}, {Sakellariadou}, {Salafia}, {Salconi}, {Saleem}, {Samajdar},
  {Sanchez}, {Sanchez}, {Sanchez}, {Sanchis-Gual}, {Sanders}, {Santiago},
  {Santos}, {Saravanan}, {Sarin}, {Sassolas}, {Sathyaprakash}, {Sauter},
  {Savage}, {Savant}, {Sawant}, {Sayah}, {Schaetzl}, {Schale}, {Scheel},
  {Scheuer}, {Schindler-Tyka}, {Schmidt}, {Schnabel}, {Schofield},
  {Sch{\"o}nbeck}, {Schreiber}, {Schulte}, {Schutz}, {Schwarm}, {Schwartz},
  {Scott}, {Scott}, {Seglar-Arroyo}, {Seidel}, {Sellers}, {Sengupta},
  {Sennett}, {Sentenac}, {Sequino}, {Sergeev}, {Setyawati}, {Shaffer},
  {Shahriar}, {Sharifi}, {Sharma}, {Sharma}, {Shawhan}, {Shen}, {Shikauchi},
  {Shink}, {Shoemaker}, {Shoemaker}, {Shukla}, {ShyamSundar}, {Sieniawska},
  {Sigg}, {Singer}, {Singh}, {Singh}, {Singha}, {Singhal}, {Sintes}, {Sipala},
  {Skliris}, {Slagmolen}, {Slaven-Blair}, {Smetana}, {Smith}, {Smith},
  {Somala}, {Son}, {Soni}, {Sorazu}, {Sordini}, {Sorrentino}, {Sorrentino},
  {Soulard}, {Souradeep}, {Sowell}, {Spencer}, {Spera}, {Srivastava},
  {Srivastava}, {Staats}, {Stachie}, {Steer}, {Steinke}, {Steinlechner},
  {Steinlechner}, {Steinmeyer}, {Stevenson}, {Stolle-McAllister}, {Stops},
  {Stover}, {Strain}, {Stratta}, {Strunk}, {Sturani}, {Stuver}, {S{\"u}dbeck},
  {Sudhagar}, {Sudhir}, {Suh}, {Summerscales}, {Sun}, {Sun}, {Sunil}, {Sur},
  {Suresh}, {Sutton}, {Swinkels}, {Szczepa{\'n}czyk}, {Tacca}, {Tait},
  {Talbot}, {Tanasijczuk}, {Tanner}, {Tao}, {Tapia}, {Tapia San Martin},
  {Tasson}, {Taylor}, {Tenorio}, {Terkowski}, {Thirugnanasambandam}, {Thomas},
  {Thomas}, {Thomas}, {Thompson}, {Thondapu}, {Thorne}, {Thrane}, {Tiwari},
  {Tiwari}, {Tiwari}, {Toland}, {Tolley}, {Tonelli}, {Tornasi},
  {Torres-Forn{\'e}}, {Torrie}, {Melo}, {T{\"o}yr{\"a}}, {Tran}, {Trapananti},
  {Travasso}, {Traylor}, {Tringali}, {Tripathee}, {Trovato}, {Trudeau}, {Tsai},
  {Tsang}, {Tse}, {Tso}, {Tsukada}, {Tsuna}, {Tsutsui}, {Turconi}, {Ubhi},
  {Udall}, {Ueno}, {Ugolini}, {Unnikrishnan}, {Urban}, {Usman}, {Utina},
  {Vahlbruch}, {Vajente}, {Vajpeyi}, {Valdes}, {Valentini}, {Valsan}, {van
  Bakel}, {van Beuzekom}, {van den Brand}, {Van Den Broeck}, {Vander-Hyde},
  {van der Schaaf}, {van Heijningen}, {Vardaro}, {Vargas}, {Varma}, {Vass},
  {Vas{\'u}th}, {Vecchio}, {Vedovato}, {Veitch}, {Veitch}, {Venkateswara},
  {Venneberg}, {Venugopalan}, {Verkindt}, {Verma}, {Veske}, {Vetrano},
  {Vicer{\'e}}, {Viets}, {Villa-Ortega}, {Vinet}, {Vitale}, {Vo}, {Vocca},
  {Vorvick}, {Vyatchanin}, {Wade}, {Wade}, {Wade}, {Walet}, {Walker},
  {Wallace}, {Wallace}, {Walsh}, {Wang}, {Wang}, {Wang}, {Wang}, {Ward},
  {Warner}, {Was}, {Washington}, {Watchi}, {Weaver}, {Wei}, {Weinert},
  {Weinstein}, {Weiss}, {Wellmann}, {Wen}, {We{\ss}els}, {Westhouse}, {Wette},
  {Whelan}, {White}, {White}, {Whiting}, {Whittle}, {Wilken}, {Williams},
  {Williams}, {Williamson}, {Willis}, {Willke}, {Wilson}, {Wimmer}, {Winkler},
  {Wipf}, {Woan}, {Woehler}, {Wofford}, {Wong}, {Wrangel}, {Wright}, {Wu},
  {Wysocki}, {Xiao}, {Yamamoto}, {Yang}, {Yang}, {Yang}, {Yap}, {Yeeles},
  {Yoon}, {Yu}, {Yu}, {Yuen}, {Zadro{\.z}ny}, {Zanolin}, {Zelenova}, {Zendri},
  {Zevin}, {Zhang}, {Zhang}, {Zhang}, {Zhang}, {Zhao}, {Zhao}, {Zhou}, {Zhou},
  {Zhu}, {Zimmerman}, {Zucker}, \& {Zweizig}}]{2020arXiv201014533T}
---. 2020{\natexlab{b}}, arXiv e-prints, arXiv:2010.14533.
\newblock \doarXiv{2010.14533}

\bibitem[{{Abbott} {et~al.}(2020{\natexlab{c}}){Abbott}, {Abbott}, {Abraham},
  {Acernese}, {Ackley}, {Adams}, {Adhikari}, {Adya}, {Affeldt}, {Agathos},
  {Agatsuma}, {Aggarwal}, {Aguiar}, {Aich}, {Aiello}, {Ain}, {Ajith}, {Akcay},
  {Allen}, {Allocca}, {Altin}, {Amato}, {Anand}, {Ananyeva}, {Anderson},
  {Anderson}, {Angelova}, {Ansoldi}, {Antier}, {Appert}, {Arai}, {Araya},
  {Areeda}, {Ar{\`e}ne}, {Arnaud}, {Aronson}, {Arun}, {Asali}, {Ascenzi},
  {Ashton}, {Aston}, {Astone}, {Aubin}, {Aufmuth}, {AultONeal}, {Austin},
  {Avendano}, {Babak}, {Bacon}, {Badaracco}, {Bader}, {Bae}, {Baer}, {Baird},
  {Baldaccini}, {Ballardin}, {Ballmer}, {Bals}, {Balsamo}, {Baltus},
  {Banagiri}, {Bankar}, {Bankar}, {Barayoga}, {Barbieri}, {Barish}, {Barker},
  {Barkett}, {Barneo}, {Barone}, {Barr}, {Barsotti}, {Barsuglia}, {Barta},
  {Bartlett}, {Bartos}, {Bassiri}, {Basti}, {Bawaj}, {Bayley}, {Bazzan},
  {B{\'e}csy}, {Bejger}, {Belahcene}, {Bell}, {Beniwal}, {Benjamin}, {Bentley},
  {Bergamin}, {Berger}, {Bergmann}, {Bernuzzi}, {Berry}, {Bersanetti},
  {Bertolini}, {Betzwieser}, {Bhandare}, {Bhandari}, {Bidler}, {Biggs},
  {Bilenko}, {Billingsley}, {Birney}, {Birnholtz}, {Biscans}, {Bischi},
  {Biscoveanu}, {Bisht}, {Bissenbayeva}, {Bitossi}, {Bizouard}, {Blackburn},
  {Blackman}, {Blair}, {Blair}, {Blair}, {Bobba}, {Bode}, {Boer}, {Boetzel},
  {Bogaert}, {Bondu}, {Bonilla}, {Bonnand}, {Booker}, {Boom}, {Bork}, {Boschi},
  {Bose}, {Bossilkov}, {Bosveld}, {Bouffanais}, {Bozzi}, {Bradaschia}, {Brady},
  {Bramley}, {Branchesi}, {Brau}, {Breschi}, {Briant}, {Briggs}, {Brighenti},
  {Brillet}, {Brinkmann}, {Brockill}, {Brooks}, {Brooks}, {Brown}, {Brunett},
  {Bruno}, {Bruntz}, {Buikema}, {Bulik}, {Bulten}, {Buonanno}, {Buscicchio},
  {Buskulic}, {Byer}, {Cabero}, {Cadonati}, {Cagnoli}, {Cahillane},
  {Calder{\'o}n Bustillo}, {Callaghan}, {Callister}, {Calloni}, {Camp},
  {Canepa}, {Cannon}, {Cao}, {Cao}, {Carapella}, {Carbognani}, {Caride},
  {Carney}, {Carullo}, {Casanueva Diaz}, {Casentini}, {Casta{\~n}eda},
  {Caudill}, {Cavagli{\`a}}, {Cavalier}, {Cavalieri}, {Cella},
  {Cerd{\'a}-Dur{\'a}n}, {Cesarini}, {Chaibi}, {Chakravarti}, {Chan}, {Chan},
  {Chandra}, {Chao}, {Charlton}, {Chase}, {Chassande-Mottin}, {Chatterjee},
  {Chaturvedi}, {Chatziioannou}, {Chen}, {Chen}, {Chen}, {Cheng}, {Cheong},
  {Chia}, {Chiadini}, {Chierici}, {Chincarini}, {Chiummo}, {Cho}, {Cho}, {Cho},
  {Christensen}, {Chu}, {Chua}, {Chung}, {Chung}, {Ciani}, {Ciecielag},
  {Cie{\'s}lar}, {Ciobanu}, {Ciolfi}, {Cipriano}, {Cirone}, {Clara}, {Clark},
  {Clearwater}, {Clesse}, {Cleva}, {Coccia}, {Cohadon}, {Cohen}, {Colleoni},
  {Collette}, {Collins}, {Colpi}, {Constancio}, {Conti}, {Cooper}, {Corban},
  {Corbitt}, {Cordero-Carri{\'o}n}, {Corezzi}, {Corley}, {Cornish}, {Corre},
  {Corsi}, {Cortese}, {Costa}, {Cotesta}, {Coughlin}, {Coughlin}, {Coulon},
  {Countryman}, {Couvares}, {Covas}, {Coward}, {Cowart}, {Coyne}, {Coyne},
  {Creighton}, {Creighton}, {Cripe}, {Croquette}, {Crowder}, {Cudell},
  {Cullen}, {Cumming}, {Cummings}, {Cunningham}, {Cuoco}, {Curylo}, {Canton},
  {D{\'a}lya}, {Dana}, {Daneshgaran-Bajastani}, {D'Angelo}, {Danilishin},
  {D'Antonio}, {Danzmann}, {Darsow-Fromm}, {Dasgupta}, {Datrier}, {Dattilo},
  {Dave}, {Davier}, {Davies}, {Davis}, {Daw}, {DeBra}, {Deenadayalan},
  {Degallaix}, {De Laurentis}, {Del{\'e}glise}, {Delfavero}, {De Lillo}, {Del
  Pozzo}, {DeMarchi}, {D'Emilio}, {Demos}, {Dent}, {De Pietri}, {De Rosa}, {De
  Rossi}, {DeSalvo}, {de Varona}, {Dhurandhar}, {D{\'\i}az}, {Diaz-Ortiz},
  {Dietrich}, {Di Fiore}, {Di Fronzo}, {Di Giorgio}, {Di Giovanni}, {Di
  Giovanni}, {Di Girolamo}, {Di Lieto}, {Ding}, {Di Pace}, {Di Palma}, {Di
  Renzo}, {Divakarla}, {Dmitriev}, {Doctor}, {Donovan}, {Dooley}, {Doravari},
  {Dorrington}, {Downes}, {Drago}, {Driggers}, {Du}, {Ducoin}, {Dupej},
  {Durante}, {D'Urso}, {Dwyer}, {Easter}, {Eddolls}, {Edelman}, {Edo}, {Edy},
  {Effler}, {Ehrens}, {Eichholz}, {Eikenberry}, {Eisenmann}, {Eisenstein},
  {Ejlli}, {Errico}, {Essick}, {Estelles}, {Estevez}, {Etienne}, {Etzel},
  {Evans}, {Evans}, {Ewing}, {Fafone}, {Fairhurst}, {Fan}, {Farinon}, {Farr},
  {Farr}, {Fauchon-Jones}, {Favata}, {Fays}, {Fazio}, {Feicht}, {Fejer},
  {Feng}, {Fenyvesi}, {Ferguson}, {Fernandez-Galiana}, {Ferrante}, {Ferreira},
  {Ferreira}, {Fidecaro}, {Fiori}, {Fiorucci}, {Fishbach}, {Fisher},
  {Fittipaldi}, {Fitz-Axen}, {Fiumara}, {Flaminio}, {Floden}, {Flynn}, {Fong},
  {Font}, {Forsyth}, {Fournier}, {Frasca}, {Frasconi}, {Frei}, {Freise},
  {Frey}, {Frey}, {Fritschel}, {Frolov}, {Fronz{\`e}}, {Fulda}, {Fyffe},
  {Gabbard}, {Gadre}, {Gaebel}, {Gair}, {Galaudage}, {Ganapathy}, {Ganguly},
  {Gaonkar}, {Garc{\'\i}a-Quir{\'o}s}, {Garufi}, {Gateley}, {Gaudio},
  {Gayathri}, {Gemme}, {Genin}, {Gennai}, {George}, {George}, {Gergely},
  {Ghonge}, {Ghosh}, {Ghosh}, {Ghosh}, {Giacomazzo}, {Giaime}, {Giardina},
  {Gibson}, {Gier}, {Gill}, {Glanzer}, {Gniesmer}, {Godwin}, {Goetz}, {Goetz},
  {Gohlke}, {Goncharov}, {Gonz{\'a}lez}, {Gopakumar}, {Gossan}, {Gosselin},
  {Gouaty}, {Grace}, {Grado}, {Granata}, {Grant}, {Gras}, {Grassia}, {Gray},
  {Gray}, {Greco}, {Green}, {Green}, {Gretarsson}, {Griggs}, {Grignani},
  {Grimaldi}, {Grimm}, {Grote}, {Grunewald}, {Gruning}, {Guidi}, {Guimaraes},
  {Guix{\'e}}, {Gulati}, {Guo}, {Gupta}, {Gupta}, {Gupta}, {Gustafson},
  {Gustafson}, {Haegel}, {Halim}, {Hall}, {Hamilton}, {Hammond}, {Haney},
  {Hanke}, {Hanks}, {Hanna}, {Hannam}, {Hannuksela}, {Hansen}, {Hanson},
  {Harder}, {Hardwick}, {Haris}, {Harms}, {Harry}, {Harry}, {Hasskew},
  {Haster}, {Haughian}, {Hayes}, {Healy}, {Heidmann}, {Heintze}, {Heinze},
  {Heitmann}, {Hellman}, {Hello}, {Hemming}, {Hendry}, {Heng}, {Hennes},
  {Hennig}, {Heurs}, {Hild}, {Hinderer}, {Hoback}, {Hochheim}, {Hofgard},
  {Hofman}, {Holgado}, {Holland}, {Holt}, {Holz}, {Hopkins}, {Horst}, {Hough},
  {Howell}, {Hoy}, {Huang}, {H{\"u}bner}, {Huerta}, {Huet}, {Hughey}, {Hui},
  {Husa}, {Huttner}, {Huxford}, {Huynh-Dinh}, {Idzkowski}, {Iess}, {Inchauspe},
  {Ingram}, {Intini}, {Isac}, {Isi}, {Iyer}, {Jacqmin}, {Jadhav}, {Jadhav},
  {James}, {Jani}, {Janthalur}, {Jaranowski}, {Jariwala}, {Jaume}, {Jenkins},
  {Jiang}, {Johns}, {Johnson-McDaniel}, {Jones}, {Jones}, {Jones}, {Jones},
  {Jones}, {Jonker}, {Ju}, {Junker}, {Kalaghatgi}, {Kalogera}, {Kamai},
  {Kandhasamy}, {Kang}, {Kanner}, {Kapadia}, {Karki}, {Kashyap}, {Kasprzack},
  {Kastaun}, {Katsanevas}, {Katsavounidis}, {Katzman}, {Kaufer}, {Kawabe},
  {K{\'e}f{\'e}lian}, {Keitel}, {Keivani}, {Kennedy}, {Key}, {Khadka},
  {Khalili}, {Khan}, {Khan}, {Khan}, {Khazanov}, {Khetan}, {Khursheed},
  {Kijbunchoo}, {Kim}, {Kim}, {Kim}, {Kim}, {Kim}, {Kim}, {Kim}, {Kimball},
  {King}, {Kinley-Hanlon}, {Kirchhoff}, {Kissel}, {Kleybolte}, {Klimenko},
  {Knowles}, {Knyazev}, {Koch}, {Koehlenbeck}, {Koekoek}, {Koley},
  {Kondrashov}, {Kontos}, {Koper}, {Korobko}, {Korth}, {Kovalam}, {Kozak},
  {Kringel}, {Krishnendu}, {Kr{\'o}lak}, {Krupinski}, {Kuehn}, {Kumar},
  {Kumar}, {Kumar}, {Kumar}, {Kumar}, {Kuo}, {Kutynia}, {Lackey}, {Laghi},
  {Lalande}, {Lam}, {Lamberts}, {Landry}, {Lane}, {Lang}, {Lange}, {Lantz},
  {Lanza}, {La Rosa}, {Lartaux-Vollard}, {Lasky}, {Laxen}, {Lazzarini},
  {Lazzaro}, {Leaci}, {Leavey}, {Lecoeuche}, {Lee}, {Lee}, {Lee}, {Lee}, {Lee},
  {Lehmann}, {Leroy}, {Letendre}, {Levin}, {Li}, {Li}, {li}, {Li}, {Li},
  {Linde}, {Linker}, {Linley}, {Littenberg}, {Liu}, {Liu},
  {Llorens-Monteagudo}, {Lo}, {Lockwood}, {London}, {Longo}, {Lorenzini},
  {Loriette}, {Lormand}, {Losurdo}, {Lough}, {Lousto}, {Lovelace}, {L{\"u}ck},
  {Lumaca}, {Lundgren}, {Ma}, {Macas}, {Macfoy}, {MacInnis}, {Macleod},
  {MacMillan}, {Macquet}, {Maga{\~n}a Hernandez}, {Maga{\~n}a-Sandoval},
  {Magee}, {Majorana}, {Maksimovic}, {Malik}, {Man}, {Mandic}, {Mangano},
  {Mansell}, {Manske}, {Mantovani}, {Mapelli}, {Marchesoni}, {Marion},
  {M{\'a}rka}, {M{\'a}rka}, {Markakis}, {Markosyan}, {Markowitz}, {Maros},
  {Marquina}, {Marsat}, {Martelli}, {Martin}, {Martin}, {Martinez}, {Martynov},
  {Masalehdan}, {Mason}, {Massera}, {Masserot}, {Massinger}, {Masso-Reid},
  {Mastrogiovanni}, {Matas}, {Matichard}, {Mavalvala}, {Maynard}, {McCann},
  {McCarthy}, {McClelland}, {McCormick}, {McCuller}, {McGuire}, {McIsaac},
  {McIver}, {McManus}, {McRae}, {McWilliams}, {Meacher}, {Meadors}, {Mehmet},
  {Mehta}, {Mejuto Villa}, {Melatos}, {Mendell}, {Mercer}, {Mereni}, {Merfeld},
  {Merilh}, {Merritt}, {Merzougui}, {Meshkov}, {Messenger}, {Messick},
  {Metzdorff}, {Meyers}, {Meylahn}, {Mhaske}, {Miani}, {Miao}, {Michaloliakos},
  {Michel}, {Middleton}, {Milano}, {Miller}, {Millhouse}, {Mills}, {Milotti},
  {Milovich-Goff}, {Minazzoli}, {Minenkov}, {Mishkin}, {Mishra}, {Mistry},
  {Mitra}, {Mitrofanov}, {Mitselmakher}, {Mittleman}, {Mo}, {Mogushi},
  {Mohapatra}, {Mohite}, {Molina-Ruiz}, {Mondin}, {Montani}, {Moore}, {Moraru},
  {Morawski}, {Moreno}, {Morisaki}, {Mours}, {Mow-Lowry}, {Mozzon},
  {Muciaccia}, {Mukherjee}, {Mukherjee}, {Mukherjee}, {Mukherjee}, {Mukund},
  {Mullavey}, {Munch}, {Mu{\~n}iz}, {Murray}, {Nagar}, {Nardecchia},
  {Naticchioni}, {Nayak}, {Neil}, {Neilson}, {Nelemans}, {Nelson}, {Nery},
  {Neunzert}, {Ng}, {Ng}, {Nguyen}, {Nguyen}, {Nichols}, {Nichols}, {Nissanke},
  {Nitz}, {Nocera}, {Noh}, {North}, {Nothard}, {Nuttall}, {Oberling},
  {O'Brien}, {Oganesyan}, {Ogin}, {Oh}, {Oh}, {Ohme}, {Ohta}, {Okada},
  {Oliver}, {Olivetto}, {Oppermann}, {Oram}, {O'Reilly}, {Ormiston}, {Ortega},
  {O'Shaughnessy}, {Ossokine}, {Osthelder}, {Ottaway}, {Overmier}, {Owen},
  {Pace}, {Pagano}, {Page}, {Pagliaroli}, {Pai}, {Pai}, {Palamos}, {Palashov},
  {Palomba}, {Pan}, {Panda}, {Pang}, {Pankow}, {Pannarale}, {Pant}, {Paoletti},
  {Paoli}, {Parida}, {Parker}, {Pascucci}, {Pasqualetti}, {Passaquieti},
  {Passuello}, {Patricelli}, {Payne}, {Pearlstone}, {Pechsiri}, {Pedersen},
  {Pedraza}, {Pele}, {Penn}, {Perego}, {Perez}, {P{\'e}rigois}, {Perreca},
  {Perri{\`e}s}, {Petermann}, {Pfeiffer}, {Phelps}, {Phukon}, {Piccinni},
  {Pichot}, {Piendibene}, {Piergiovanni}, {Pierro}, {Pillant}, {Pinard},
  {Pinto}, {Piotrzkowski}, {Pirello}, {Pitkin}, {Plastino}, {Poggiani}, {Pong},
  {Ponrathnam}, {Popolizio}, {Porter}, {Powell}, {Prajapati}, {Prasai},
  {Prasanna}, {Pratten}, {Prestegard}, {Principe}, {Prodi}, {Prokhorov},
  {Punturo}, {Puppo}, {P{\"u}rrer}, {Qi}, {Quetschke}, {Quinonez}, {Raab},
  {Raaijmakers}, {Radkins}, {Radulesco}, {Raffai}, {Rafferty}, {Raja}, {Rajan},
  {Rajbhandari}, {Rakhmanov}, {Ramirez}, {Ramos-Buades}, {Rana}, {Rao},
  {Rapagnani}, {Raymond}, {Razzano}, {Read}, {Regimbau}, {Rei}, {Reid},
  {Reitze}, {Rettegno}, {Ricci}, {Richardson}, {Richardson}, {Ricker},
  {Riemenschneider}, {Riles}, {Rizzo}, {Robertson}, {Robinet}, {Rocchi},
  {Rodriguez-Soto}, {Rolland}, {Rollins}, {Roma}, {Romanelli}, {Romano},
  {Romel}, {Romero-Shaw}, {Romie}, {Rose}, {Rose}, {Rose}, {Rosi{\'n}ska},
  {Rosofsky}, {Ross}, {Rowan}, {Rowlinson}, {Roy}, {Roy}, {Roy}, {Ruggi},
  {Rutins}, {Ryan}, {Sachdev}, {Sadecki}, {Sakellariadou}, {Salafia},
  {Salconi}, {Saleem}, {Salemi}, {Samajdar}, {Sanchez}, {Sanchez},
  {Sanchis-Gual}, {Sanders}, {Santiago}, {Santos}, {Sarin}, {Sassolas},
  {Sathyaprakash}, {Sauter}, {Savage}, {Savant}, {Sawant}, {Sayah}, {Schaetzl},
  {Schale}, {Scheel}, {Scheuer}, {Schmidt}, {Schnabel}, {Schofield},
  {Sch{\"o}nbeck}, {Schreiber}, {Schulte}, {Schutz}, {Schwarm}, {Schwartz},
  {Scott}, {Scott}, {Seidel}, {Sellers}, {Sengupta}, {Sennett}, {Sentenac},
  {Sequino}, {Sergeev}, {Setyawati}, {Shaddock}, {Shaffer}, {Sharifi},
  {Shahriar}, {Sharma}, {Sharma}, {Shawhan}, {Shen}, {Shikauchi}, {Shink},
  {Shoemaker}, {Shoemaker}, {Shukla}, {ShyamSundar}, {Siellez}, {Sieniawska},
  {Sigg}, {Singer}, {Singh}, {Singh}, {Singha}, {Singhal}, {Sintes}, {Sipala},
  {Skliris}, {Slagmolen}, {Slaven-Blair}, {Smetana}, {Smith}, {Smith},
  {Somala}, {Son}, {Soni}, {Sorazu}, {Sordini}, {Sorrentino}, {Souradeep},
  {Sowell}, {Spencer}, {Spera}, {Srivastava}, {Srivastava}, {Staats},
  {Stachie}, {Standke}, {Steer}, {Steinke}, {Steinlechner}, {Steinlechner},
  {Steinmeyer}, {Stevenson}, {Stocks}, {Stops}, {Stover}, {Strain}, {Stratta},
  {Strunk}, {Sturani}, {Stuver}, {Sudhagar}, {Sudhir}, {Summerscales}, {Sun},
  {Sunil}, {Sur}, {Suresh}, {Sutton}, {Swinkels}, {Szczepa{\'n}czyk}, {Tacca},
  {Tait}, {Talbot}, {Tanasijczuk}, {Tanner}, {Tao}, {T{\'a}pai}, {Tapia},
  {Tapia San Martin}, {Tasson}, {Taylor}, {Tenorio}, {Terkowski},
  {Thirugnanasambandam}, {Thomas}, {Thomas}, {Thompson}, {Thondapu}, {Thorne},
  {Thrane}, {Tinsman}, {Saravanan}, {Tiwari}, {Tiwari}, {Tiwari}, {Toland},
  {Tonelli}, {Tornasi}, {Torres-Forn{\'e}}, {Torrie}, {Tosta e Melo},
  {T{\"o}yr{\"a}}, {Travasso}, {Traylor}, {Tringali}, {Tripathee}, {Trovato},
  {Trudeau}, {Tsang}, {Tse}, {Tso}, {Tsukada}, {Tsuna}, {Tsutsui}, {Turconi},
  {Ubhi}, {Udall}, {Ueno}, {Ugolini}, {Unnikrishnan}, {Urban}, {Usman},
  {Utina}, {Vahlbruch}, {Vajente}, {Valdes}, {Valentini}, {van Bakel}, {van
  Beuzekom}, {van den Brand}, {Van Den Broeck}, {Vander-Hyde}, {van der
  Schaaf}, {Van Heijningen}, {van Veggel}, {Vardaro}, {Varma}, {Vass},
  {Vas{\'u}th}, {Vecchio}, {Vedovato}, {Veitch}, {Veitch}, {Venkateswara},
  {Venugopalan}, {Verkindt}, {Veske}, {Vetrano}, {Vicer{\'e}}, {Viets},
  {Vinciguerra}, {Vine}, {Vinet}, {Vitale}, {Vivanco}, {Vo}, {Vocca},
  {Vorvick}, {Vyatchanin}, {Wade}, {Wade}, {Wade}, {Walet}, {Walker},
  {Wallace}, {Wallace}, {Walsh}, {Wang}, {Wang}, {Wang}, {Ward}, {Warden},
  {Warner}, {Was}, {Watchi}, {Weaver}, {Wei}, {Weinert}, {Weinstein}, {Weiss},
  {Wellmann}, {Wen}, {We{\ss}els}, {Westhouse}, {Wette}, {Whelan}, {Whiting},
  {Whittle}, {Wilken}, {Williams}, {Willis}, {Willke}, {Winkler}, {Wipf},
  {Wittel}, {Woan}, {Woehler}, {Wofford}, {Wong}, {Wright}, {Wu}, {Wysocki},
  {Xiao}, {Yamamoto}, {Yang}, {Yang}, {Yang}, {Yap}, {Yazback}, {Yeeles}, {Yu},
  {Yu}, {Yuen}, {Zadro{\.Z}ny}, {Zadro{\.Z}ny}, {Zanolin}, {Zelenova},
  {Zendri}, {Zevin}, {Zhang}, {Zhang}, {Zhang}, {Zhao}, {Zhao}, {Zhou}, {Zhou},
  {Zhu}, {Zimmerman}, {Zucker}, {Zweizig}, {LIGO Scientific Collaboration}, \&
  {Virgo Collaboration}}]{2020PhRvL.125j1102A}
---. 2020{\natexlab{c}}, \prl, 125, 101102,
  \dodoi{10.1103/PhysRevLett.125.101102}

\bibitem[{{Abrams} \& {Takada}(2020)}]{2020ApJ...905..121A}
{Abrams}, N.~S., \& {Takada}, M. 2020, \apj, 905, 121,
  \dodoi{10.3847/1538-4357/abc6aa}

\bibitem[{{Aihara} {et~al.}(2018){Aihara}, {Arimoto}, {Armstrong}, {Arnouts},
  {Bahcall}, {Bickerton}, {Bosch}, {Bundy}, {Capak}, {Chan}, {Chiba}, {Coupon},
  {Egami}, {Enoki}, {Finet}, {Fujimori}, {Fujimoto}, {Furusawa}, {Furusawa},
  {Goto}, {Goulding}, {Greco}, {Greene}, {Gunn}, {Hamana}, {Harikane},
  {Hashimoto}, {Hattori}, {Hayashi}, {Hayashi}, {He{\l}miniak}, {Higuchi},
  {Hikage}, {Ho}, {Hsieh}, {Huang}, {Huang}, {Ikeda}, {Imanishi}, {Inoue},
  {Iwasawa}, {Iwata}, {Jaelani}, {Jian}, {Kamata}, {Karoji}, {Kashikawa},
  {Katayama}, {Kawanomoto}, {Kayo}, {Koda}, {Koike}, {Kojima}, {Komiyama},
  {Konno}, {Koshida}, {Koyama}, {Kusakabe}, {Leauthaud}, {Lee}, {Lin}, {Lin},
  {Lupton}, {Mandelbaum}, {Matsuoka}, {Medezinski}, {Mineo}, {Miyama},
  {Miyatake}, {Miyazaki}, {Momose}, {More}, {More}, {Moritani}, {Moriya},
  {Morokuma}, {Mukae}, {Murata}, {Murayama}, {Nagao}, {Nakata}, {Niida},
  {Niikura}, {Nishizawa}, {Obuchi}, {Oguri}, {Oishi}, {Okabe}, {Okamoto},
  {Okura}, {Ono}, {Onodera}, {Onoue}, {Osato}, {Ouchi}, {Price}, {Pyo}, {Sako},
  {Sawicki}, {Shibuya}, {Shimasaku}, {Shimono}, {Shirasaki}, {Silverman},
  {Simet}, {Speagle}, {Spergel}, {Strauss}, {Sugahara}, {Sugiyama}, {Suto},
  {Suyu}, {Suzuki}, {Tait}, {Takada}, {Takata}, {Tamura}, {Tanaka}, {Tanaka},
  {Tanaka}, {Tanaka}, {Terai}, {Terashima}, {Toba}, {Tominaga}, {Toshikawa},
  {Turner}, {Uchida}, {Uchiyama}, {Umetsu}, {Uraguchi}, {Urata}, {Usuda},
  {Utsumi}, {Wang}, {Wang}, {Wong}, {Yabe}, {Yamada}, {Yamanoi}, {Yasuda},
  {Yeh}, {Yonehara}, \& {Yuma}}]{HSCOverView:17}
{Aihara}, H., {Arimoto}, N., {Armstrong}, R., {et~al.} 2018, \pasj, 70, S4,
  \dodoi{10.1093/pasj/psx066}

\bibitem[{{Alcock} {et~al.}(1995){Alcock}, {Allsman}, {Alves}, {Axelrod},
  {Bennett}, {Cook}, {Freeman}, {Griest}, {Guern}, {Lehner}, {Marshall},
  {Peterson}, {Pratt}, {Quinn}, {Rodgers}, {Stubbs}, \&
  {Sutherland}}]{1995ApJ...454L.125A}
{Alcock}, C., {Allsman}, R.~A., {Alves}, D., {et~al.} 1995, \apjl, 454, L125,
  \dodoi{10.1086/309783}

\bibitem[{{Alcock} {et~al.}(2000){Alcock}, {Allsman}, {Alves}, {Axelrod},
  {Becker}, {Bennett}, {Cook}, {Dalal}, {Drake}, {Freeman}, {Geha}, {Griest},
  {Lehner}, {Marshall}, {Minniti}, {Nelson}, {Peterson}, {Popowski}, {Pratt},
  {Quinn}, {Stubbs}, {Sutherland}, {Tomaney}, {Vandehei}, \&
  {Welch}}]{Alcocketal:00}
{Alcock}, C., {Allsman}, R.~A., {Alves}, D.~R., {et~al.} 2000, \apj, 542, 281,
  \dodoi{10.1086/309512}

\bibitem[{{Antonini} \& {Perets}(2012)}]{2012ApJ...757...27A}
{Antonini}, F., \& {Perets}, H.~B. 2012, \apj, 757, 27,
  \dodoi{10.1088/0004-637X/757/1/27}

\bibitem[{{Astropy Collaboration} {et~al.}(2013){Astropy Collaboration},
  {Robitaille}, {Tollerud}, {Greenfield}, {Droettboom}, {Bray}, {Aldcroft},
  {Davis}, {Ginsburg}, {Price-Whelan}, {Kerzendorf}, {Conley}, {Crighton},
  {Barbary}, {Muna}, {Ferguson}, {Grollier}, {Parikh}, {Nair}, {Unther},
  {Deil}, {Woillez}, {Conseil}, {Kramer}, {Turner}, {Singer}, {Fox}, {Weaver},
  {Zabalza}, {Edwards}, {Azalee Bostroem}, {Burke}, {Casey}, {Crawford},
  {Dencheva}, {Ely}, {Jenness}, {Labrie}, {Lim}, {Pierfederici}, {Pontzen},
  {Ptak}, {Refsdal}, {Servillat}, \& {Streicher}}]{2013A&A...558A..33A}
{Astropy Collaboration}, {Robitaille}, T.~P., {Tollerud}, E.~J., {et~al.} 2013,
  \aap, 558, A33, \dodoi{10.1051/0004-6361/201322068}

\bibitem[{{Astropy Collaboration} {et~al.}(2018){Astropy Collaboration},
  {Price-Whelan}, {Sip{\H{o}}cz}, {G{\"u}nther}, {Lim}, {Crawford}, {Conseil},
  {Shupe}, {Craig}, {Dencheva}, {Ginsburg}, {VanderPlas}, {Bradley},
  {P{\'e}rez-Su{\'a}rez}, {de Val-Borro}, {Aldcroft}, {Cruz}, {Robitaille},
  {Tollerud}, {Ardelean}, {Babej}, {Bach}, {Bachetti}, {Bakanov}, {Bamford},
  {Barentsen}, {Barmby}, {Baumbach}, {Berry}, {Biscani}, {Boquien}, {Bostroem},
  {Bouma}, {Brammer}, {Bray}, {Breytenbach}, {Buddelmeijer}, {Burke},
  {Calderone}, {Cano Rodr{\'\i}guez}, {Cara}, {Cardoso}, {Cheedella}, {Copin},
  {Corrales}, {Crichton}, {D'Avella}, {Deil}, {Depagne}, {Dietrich}, {Donath},
  {Droettboom}, {Earl}, {Erben}, {Fabbro}, {Ferreira}, {Finethy}, {Fox},
  {Garrison}, {Gibbons}, {Goldstein}, {Gommers}, {Greco}, {Greenfield},
  {Groener}, {Grollier}, {Hagen}, {Hirst}, {Homeier}, {Horton}, {Hosseinzadeh},
  {Hu}, {Hunkeler}, {Ivezi{\'c}}, {Jain}, {Jenness}, {Kanarek}, {Kendrew},
  {Kern}, {Kerzendorf}, {Khvalko}, {King}, {Kirkby}, {Kulkarni}, {Kumar},
  {Lee}, {Lenz}, {Littlefair}, {Ma}, {Macleod}, {Mastropietro}, {McCully},
  {Montagnac}, {Morris}, {Mueller}, {Mumford}, {Muna}, {Murphy}, {Nelson},
  {Nguyen}, {Ninan}, {N{\"o}the}, {Ogaz}, {Oh}, {Parejko}, {Parley}, {Pascual},
  {Patil}, {Patil}, {Plunkett}, {Prochaska}, {Rastogi}, {Reddy Janga},
  {Sabater}, {Sakurikar}, {Seifert}, {Sherbert}, {Sherwood-Taylor}, {Shih},
  {Sick}, {Silbiger}, {Singanamalla}, {Singer}, {Sladen}, {Sooley},
  {Sornarajah}, {Streicher}, {Teuben}, {Thomas}, {Tremblay}, {Turner},
  {Terr{\'o}n}, {van Kerkwijk}, {de la Vega}, {Watkins}, {Weaver}, {Whitmore},
  {Woillez}, {Zabalza}, \& {Astropy Contributors}}]{2018AJ....156..123A}
{Astropy Collaboration}, {Price-Whelan}, A.~M., {Sip{\H{o}}cz}, B.~M., {et~al.}
  2018, \aj, 156, 123, \dodoi{10.3847/1538-3881/aabc4f}

\bibitem[{{Bahcall}(1986)}]{1986ARA&A..24..577B}
{Bahcall}, J.~N. 1986, \araa, 24, 577,
  \dodoi{10.1146/annurev.aa.24.090186.003045}

\bibitem[{{Barkat} {et~al.}(1967){Barkat}, {Rakavy}, \&
  {Sack}}]{1967PhRvL..18..379B}
{Barkat}, Z., {Rakavy}, G., \& {Sack}, N. 1967, \prl, 18, 379,
  \dodoi{10.1103/PhysRevLett.18.379}

\bibitem[{{Belczynski} {et~al.}(2016){Belczynski}, {Holz}, {Bulik}, \&
  {O'Shaughnessy}}]{2016Natur.534..512B}
{Belczynski}, K., {Holz}, D.~E., {Bulik}, T., \& {O'Shaughnessy}, R. 2016,
  \nat, 534, 512, \dodoi{10.1038/nature18322}

\bibitem[{{Belczynski} {et~al.}(2002){Belczynski}, {Kalogera}, \&
  {Bulik}}]{2002ApJ...572..407B}
{Belczynski}, K., {Kalogera}, V., \& {Bulik}, T. 2002, \apj, 572, 407,
  \dodoi{10.1086/340304}

\bibitem[{{Bennett} {et~al.}(2002){Bennett}, {Becker}, {Quinn}, {Tomaney},
  {Alcock}, {Allsman}, {Alves}, {Axelrod}, {Calitz}, {Cook}, {Drake},
  {Fragile}, {Freeman}, {Geha}, {Griest}, {Johnson}, {Keller}, {Laws},
  {Lehner}, {Marshall}, {Minniti}, {Nelson}, {Peterson}, {Popowski}, {Pratt},
  {Quinn}, {Rhie}, {Stubbs}, {Sutherland }, {Vandehei}, {Welch}, {MACHO
  Collaboration}, \& {MPS Collaboration}}]{2002ApJ...579..639B}
{Bennett}, D.~P., {Becker}, A.~C., {Quinn}, J.~L., {et~al.} 2002, \apj, 579,
  639, \dodoi{10.1086/342225}

\bibitem[{{Bethe} \& {Brown}(1998)}]{1998ApJ...506..780B}
{Bethe}, H.~A., \& {Brown}, G.~E. 1998, \apj, 506, 780, \dodoi{10.1086/306265}

\bibitem[{{Bird} {et~al.}(2016){Bird}, {Cholis}, {Mu{\~n}oz},
  {Ali-Ha{\"i}moud}, {Kamionkowski}, {Kovetz}, {Raccanelli}, \&
  {Riess}}]{Birdetal:16}
{Bird}, S., {Cholis}, I., {Mu{\~n}oz}, J.~B., {et~al.} 2016, Physical Review
  Letters, 116, 201301, \dodoi{10.1103/PhysRevLett.116.201301}

\bibitem[{{Foreman-Mackey} {et~al.}(2013){Foreman-Mackey}, {Hogg}, {Lang}, \&
  {Goodman}}]{2013PASP..125..306F}
{Foreman-Mackey}, D., {Hogg}, D.~W., {Lang}, D., \& {Goodman}, J. 2013, \pasp,
  125, 306, \dodoi{10.1086/670067}

\bibitem[{{Godines} {et~al.}(2020){Godines}, {Bachelet}, {Narayan}, \&
  {Street}}]{2020arXiv200414347G}
{Godines}, D., {Bachelet}, E., {Narayan}, G., \& {Street}, R.~A. 2020, arXiv
  e-prints, arXiv:2004.14347.
\newblock \doarXiv{2004.14347}

\bibitem[{{Gould}(2000)}]{2000ApJ...535..928G}
{Gould}, A. 2000, \apj, 535, 928, \dodoi{10.1086/308865}

\bibitem[{{Gould}(2004)}]{2004ApJ...606..319G}
---. 2004, \apj, 606, 319, \dodoi{10.1086/382782}

\bibitem[{{Grieger} {et~al.}(1986){Grieger}, {Kayser}, \&
  {Refsdal}}]{1986Natur.324..126G}
{Grieger}, B., {Kayser}, R., \& {Refsdal}, S. 1986, \nat, 324, 126,
  \dodoi{10.1038/324126a0}

\bibitem[{{Griest} {et~al.}(1991){Griest}, {Alcock}, {Axelrod}, {Bennett},
  {Cook}, {Freeman}, {Park}, {Perlmutter}, {Peterson}, {Quinn}, {Rodgers},
  {Stubbs}, \& {MACHO Collaboration}}]{Griestetal:91}
{Griest}, K., {Alcock}, C., {Axelrod}, T.~S., {et~al.} 1991, \apjl, 372, L79,
  \dodoi{10.1086/186028}

\bibitem[{{Han} \& {Gould}(1995)}]{1995ApJ...447...53H}
{Han}, C., \& {Gould}, A. 1995, \apj, 447, 53, \dodoi{10.1086/175856}

\bibitem[{{Han} \& {Gould}(1996)}]{1996ApJ...467..540H}
---. 1996, \apj, 467, 540, \dodoi{10.1086/177631}

\bibitem[{Harris {et~al.}(2020)}]{Harris:2020xlr}
Harris, C.~R., {et~al.} 2020, Nature, 585, 357,
  \dodoi{10.1038/s41586-020-2649-2}

\bibitem[{{Hunter}(2007)}]{4160265}
{Hunter}, J.~D. 2007, Computing in Science Engineering, 9, 90,
  \dodoi{10.1109/MCSE.2007.55}

\bibitem[{{Kent}(1992)}]{1992ApJ...387..181K}
{Kent}, S.~M. 1992, \apj, 387, 181, \dodoi{10.1086/171070}

\bibitem[{{Kroupa}(2001)}]{2001MNRAS.322..231K}
{Kroupa}, P. 2001, \mnras, 322, 231, \dodoi{10.1046/j.1365-8711.2001.04022.x}

\bibitem[{{Kusenko} {et~al.}(2020){Kusenko}, {Sasaki}, {Sugiyama}, {Takada},
  {Takhistov}, \& {Vitagliano}}]{2020arXiv200109160K}
{Kusenko}, A., {Sasaki}, M., {Sugiyama}, S., {et~al.} 2020, arXiv e-prints,
  arXiv:2001.09160.
\newblock \doarXiv{2001.09160}

\bibitem[{{Lam} {et~al.}(2020){Lam}, {Lu}, {Hosek}, {Dawson}, \&
  {Golovich}}]{2020ApJ...889...31L}
{Lam}, C.~Y., {Lu}, J.~R., {Hosek}, Matthew~W., J., {Dawson}, W.~A., \&
  {Golovich}, N.~R. 2020, \apj, 889, 31, \dodoi{10.3847/1538-4357/ab5fd3}

\bibitem[{{Lewis}(2019)}]{2019arXiv191013970L}
{Lewis}, A. 2019, arXiv e-prints, arXiv:1910.13970.
\newblock \doarXiv{1910.13970}

\bibitem[{{Lu} {et~al.}(2016){Lu}, {Sinukoff}, {Ofek}, {Udalski}, \&
  {Kozlowski}}]{2016ApJ...830...41L}
{Lu}, J.~R., {Sinukoff}, E., {Ofek}, E.~O., {Udalski}, A., \& {Kozlowski}, S.
  2016, \apj, 830, 41, \dodoi{10.3847/0004-637X/830/1/41}

\bibitem[{{Ma} {et~al.}(2021){Ma}, {Hopkins}, {Ma}, {Angl{\'e}s-Alc{\'a}zar},
  {Faucher-Gigu{\`e}re}, \& {Kelley}}]{2021arXiv210102727M}
{Ma}, L., {Hopkins}, P.~F., {Ma}, X., {et~al.} 2021, arXiv e-prints,
  arXiv:2101.02727.
\newblock \doarXiv{2101.02727}

\bibitem[{{Mao} \& {Paczynski}(1991)}]{1991ApJ...374L..37M}
{Mao}, S., \& {Paczynski}, B. 1991, \apjl, 374, L37, \dodoi{10.1086/186066}

\bibitem[{{Mao} \& {Paczynski}(1996)}]{1996ApJ...473...57M}
---. 1996, \apj, 473, 57, \dodoi{10.1086/178126}

\bibitem[{{Mao} {et~al.}(2002){Mao}, {Smith}, {Wo{\'z}niak}, {Udalski},
  {Szyma{\'n}ski}, {Kubiak}, {Pietrzy{\'n}ski}, {Soszy{\'n}ski}, \&
  {{\.Z}ebru{\'n}}}]{2002MNRAS.329..349M}
{Mao}, S., {Smith}, M.~C., {Wo{\'z}niak}, P., {et~al.} 2002, \mnras, 329, 349,
  \dodoi{10.1046/j.1365-8711.2002.04986.x}

\bibitem[{{McKernan} {et~al.}(2012){McKernan}, {Ford}, {Lyra}, \&
  {Perets}}]{2012MNRAS.425..460M}
{McKernan}, B., {Ford}, K.~E.~S., {Lyra}, W., \& {Perets}, H.~B. 2012, \mnras,
  425, 460, \dodoi{10.1111/j.1365-2966.2012.21486.x}

\bibitem[{{Minowa} {et~al.}(2020){Minowa}, {Koyama}, {Yanagisawa}, {Motohara},
  {Tanaka}, {Ono}, {Hattori}, {Clergeon}, {Hayano}, {Akiyama}, {Kodama},
  {d'Orgeville}, {Rigaut}, {Wang}, \& {Yoshida}}]{2020SPIE11450E..0OM}
{Minowa}, Y., {Koyama}, Y., {Yanagisawa}, K., {et~al.} 2020, in Society of
  Photo-Optical Instrumentation Engineers (SPIE) Conference Series, Vol. 11450,
  Society of Photo-Optical Instrumentation Engineers (SPIE) Conference Series,
  114500O, \dodoi{10.1117/12.2561950}

\bibitem[{{Motohara} {et~al.}(2020){Motohara}, {Minowa}, {Tanaka}, {Hattori},
  {Koyama}, {Konishi}, {Yanagisawa}, {Iwata}, {Wang}, {Chou}, {Kimura}, \&
  {Pazder}}]{2020SPIE11447E..0NM}
{Motohara}, K., {Minowa}, Y., {Tanaka}, I., {et~al.} 2020, in Society of
  Photo-Optical Instrumentation Engineers (SPIE) Conference Series, Vol. 11447,
  Society of Photo-Optical Instrumentation Engineers (SPIE) Conference Series,
  114470N, \dodoi{10.1117/12.2560324}

\bibitem[{{Mroz} {et~al.}(2018){Mroz}, {Udalski}, {Bennett}, {Ryu}, {Sumi},
  {Shvartzvald}, {Skowron}, {Poleski}, {Pietrukowicz}, {Kozlowski},
  {Szymanski}, {Wyrzykowski}, {Soszynski}, {Ulaczyk}, {Rybicki}, {Iwanek},
  {Albrow}, {Chung}, {Gould}, {Han}, {Hwang}, {Jung}, {Shin}, {Yee}, {Zang},
  {Cha}, {Kim}, {Kim}, {Kim}, {Lee}, {Lee}, {Lee}, {Park}, {Pogge}, {Abe},
  {Barry}, {Bhattacharya}, {Bond}, {Donachie}, {Fukui}, {Hirao}, {Itow},
  {Kawasaki}, {Kondo}, {Koshimoto}, {Li}, {Matsubara}, {Muraki}, {Miyazaki},
  {Nagakane}, {Ranc}, {Rattenbury}, {Suematsu}, {Sullivan}, {Suzuki},
  {Tristram}, {Yonehara}, {Maoz}, {Kaspi}, \&
  {Friedmann}}]{2018arXiv181100441M}
{Mroz}, P., {Udalski}, A., {Bennett}, D.~P., {et~al.} 2018, arXiv e-prints.
\newblock \doarXiv{1811.00441}

\bibitem[{{Niikura} {et~al.}(2019{\natexlab{a}}){Niikura}, {Takada},
  {Yokoyama}, {Sumi}, \& {Masaki}}]{2019PhRvD..99h3503N}
{Niikura}, H., {Takada}, M., {Yokoyama}, S., {Sumi}, T., \& {Masaki}, S.
  2019{\natexlab{a}}, \prd, 99, 083503, \dodoi{10.1103/PhysRevD.99.083503}

\bibitem[{{Niikura} {et~al.}(2019{\natexlab{b}}){Niikura}, {Takada}, {Yasuda},
  {Lupton}, {Sumi}, {More}, {Kurita}, {Sugiyama}, {More}, {Oguri}, \&
  {Chiba}}]{2017arXiv170102151N}
{Niikura}, H., {Takada}, M., {Yasuda}, N., {et~al.} 2019{\natexlab{b}}, Nature
  Astronomy, 3, 524, \dodoi{10.1038/s41550-019-0723-1}

\bibitem[{{Paczynski}(1986)}]{Paczynski:86}
{Paczynski}, B. 1986, \apj, 304, 1, \dodoi{10.1086/164140}

\bibitem[{{Perez} \& {Granger}(2007)}]{4160251}
{Perez}, F., \& {Granger}, B.~E. 2007, Computing in Science Engineering, 9, 21,
  \dodoi{10.1109/MCSE.2007.53}

\bibitem[{{Portegies Zwart} \& {McMillan}(2000)}]{2000ApJ...528L..17P}
{Portegies Zwart}, S.~F., \& {McMillan}, S. L.~W. 2000, \apjl, 528, L17,
  \dodoi{10.1086/312422}

\bibitem[{{Qiu} {et~al.}(2020){Qiu}, {Wang}, {Takada}, {Yasuda}, {Ivezi{\'c}},
  {Lupton}, {Chiba}, {Ishigaki}, \& {Komiyama}}]{2020arXiv200412899Q}
{Qiu}, T., {Wang}, W., {Takada}, M., {et~al.} 2020, arXiv e-prints,
  arXiv:2004.12899.
\newblock \doarXiv{2004.12899}

\bibitem[{{Refsdal}(1966)}]{1966MNRAS.134..315R}
{Refsdal}, S. 1966, \mnras, 134, 315, \dodoi{10.1093/mnras/134.3.315}

\bibitem[{{Samsing} \& {Hotokezaka}(2020)}]{2020arXiv200609744S}
{Samsing}, J., \& {Hotokezaka}, K. 2020, arXiv e-prints, arXiv:2006.09744.
\newblock \doarXiv{2006.09744}

\bibitem[{{Sasaki} {et~al.}(2016){Sasaki}, {Suyama}, {Tanaka}, \&
  {Yokoyama}}]{Sasakietal:16}
{Sasaki}, M., {Suyama}, T., {Tanaka}, T., \& {Yokoyama}, S. 2016, Physical
  Review Letters, 117, 061101, \dodoi{10.1103/PhysRevLett.117.061101}

\bibitem[{{Schneider} {et~al.}(1992){Schneider}, {Ehlers}, \&
  {Falco}}]{1992grle.book.....S}
{Schneider}, P., {Ehlers}, J., \& {Falco}, E.~E. 1992, {Gravitational Lenses},
  \dodoi{10.1007/978-3-662-03758-4}

\bibitem[{{Street} {et~al.}(2018){Street}, {Lund}, {Donachie}, {Khakpash},
  {Golovich}, {Penny}, {Bennett}, {Dawson}, {Pepper}, {Rabus}, {Szkody},
  {Clarkson}, {Di Stefano}, {Rattenbury}, {Hundertmark}, {Tsapras}, {Ridgway},
  {Stassun}, {Bozza}, {Bhattacharya}, {Calchi Novati}, \&
  {Shvartzvald}}]{2018arXiv181204445S}
{Street}, R.~A., {Lund}, M.~B., {Donachie}, M., {et~al.} 2018, arXiv e-prints,
  arXiv:1812.04445.
\newblock \doarXiv{1812.04445}

\bibitem[{{Sumi} {et~al.}(2003){Sumi}, {Abe}, {Bond}, {Dodd}, {Hearnshaw},
  {Honda}, {Honma}, {Kan-ya}, {Kilmartin}, {Masuda}, {Matsubara}, {Muraki},
  {Nakamura}, {Nishi}, {Noda}, {Ohnishi}, {Petterson}, {Rattenbury}, {Reid},
  {Saito}, {Saito}, {Sato}, {Sekiguchi}, {Skuljan}, {Sullivan}, {Takeuti},
  {Tristram}, {Wilkinson}, {Yanagisawa}, \& {Yock}}]{Sumietal:03}
{Sumi}, T., {Abe}, F., {Bond}, I.~A., {et~al.} 2003, \apj, 591, 204,
  \dodoi{10.1086/375212}

\bibitem[{{Sumi} {et~al.}(2011){Sumi}, {Kamiya}, {Bennett}, {Bond}, {Abe},
  {Botzler}, {Fukui}, {Furusawa}, {Hearnshaw}, {Itow}, {Kilmartin}, {Korpela},
  {Lin}, {Ling}, {Masuda}, {Matsubara}, {Miyake}, {Motomura}, {Muraki},
  {Nagaya}, {Nakamura}, {Ohnishi}, {Okumura}, {Perrott}, {Rattenbury}, {Saito},
  {Sako}, {Sullivan}, {Sweatman}, {Tristram}, {Udalski}, {Szyma{\'n}ski},
  {Kubiak}, {Pietrzy{\'n}ski}, {Poleski}, {Soszy{\'n}ski}, {Wyrzykowski},
  {Ulaczyk}, \& {Microlensing Observations in Astrophysics (MOA)
  Collaboration}}]{Sumietal:11}
{Sumi}, T., {Kamiya}, K., {Bennett}, D.~P., {et~al.} 2011, \nat, 473, 349,
  \dodoi{10.1038/nature10092}

\bibitem[{Virtanen {et~al.}(2020)}]{Virtanen:2019joe}
Virtanen, P., {et~al.} 2020, Nature Meth., 17, 261,
  \dodoi{10.1038/s41592-019-0686-2}

\bibitem[{{Wiktorowicz} {et~al.}(2019){Wiktorowicz}, {Wyrzykowski},
  {Chruslinska}, {Klencki}, {Rybicki}, \& {Belczynski}}]{2019ApJ...885....1W}
{Wiktorowicz}, G., {Wyrzykowski}, {\L}., {Chruslinska}, M., {et~al.} 2019,
  \apj, 885, 1, \dodoi{10.3847/1538-4357/ab45e6}

\bibitem[{{Williams} {et~al.}(2009){Williams}, {Bolte}, \&
  {Koester}}]{2009ApJ...693..355W}
{Williams}, K.~A., {Bolte}, M., \& {Koester}, D. 2009, \apj, 693, 355,
  \dodoi{10.1088/0004-637X/693/1/355}

\bibitem[{{Wood} \& {Mao}(2005)}]{2005MNRAS.362..945W}
{Wood}, A., \& {Mao}, S. 2005, \mnras, 362, 945,
  \dodoi{10.1111/j.1365-2966.2005.09357.x}

\bibitem[{{Wyrzykowski} \& {Mandel}(2020)}]{2020A&A...636A..20W}
{Wyrzykowski}, L., \& {Mandel}, I. 2020, \aap, 636, A20,
  \dodoi{10.1051/0004-6361/201935842}

\bibitem[{{Wyrzykowski} {et~al.}(2009){Wyrzykowski}, {Koz{\l}owski}, {Skowron},
  {Belokurov}, {Smith}, {Udalski}, {Szyma{\'n}ski}, {Kubiak},
  {Pietrzy{\'n}ski}, {Soszy{\'n}ski}, {Szewczyk}, \&
  {{\.Z}ebru{\'n}}}]{OGLE:09}
{Wyrzykowski}, {\L}., {Koz{\l}owski}, S., {Skowron}, J., {et~al.} 2009, \mnras,
  397, 1228, \dodoi{10.1111/j.1365-2966.2009.15029.x}

\bibitem[{{Wyrzykowski} {et~al.}(2010){Wyrzykowski}, {Koz{\l}owski}, {Skowron},
  {Belokurov}, {Smith}, {Udalski}, {Szyma{\'n}ski}, {Kubiak},
  {Pietrzy{\'n}ski}, {Soszy{\'n}ski}, \& {Szewczyk}}]{OGLE:10}
---. 2010, \mnras, 407, 189, \dodoi{10.1111/j.1365-2966.2010.16936.x}

\bibitem[{{Wyrzykowski} {et~al.}(2016){Wyrzykowski}, {Kostrzewa-Rutkowska},
  {Skowron}, {Rybicki}, {Mr{\'o}z}, {Koz{\l}owski}, {Udalski}, {Szyma{\'n}ski},
  {Pietrzy{\'n}ski}, {Soszy{\'n}ski}, {Ulaczyk}, {Pietrukowicz}, {Poleski},
  {Pawlak}, {I{\l}kiewicz}, \& {Rattenbury}}]{2016MNRAS.458.3012W}
{Wyrzykowski}, {\L}., {Kostrzewa-Rutkowska}, Z., {Skowron}, J., {et~al.} 2016,
  \mnras, 458, 3012, \dodoi{10.1093/mnras/stw426}

\bibitem[{{Yang} {et~al.}(2019){Yang}, {Bartos}, {Gayathri}, {Ford}, {Haiman},
  {Klimenko}, {Kocsis}, {M{\'a}rka}, {M{\'a}rka}, {McKernan}, \&
  {O'Shaughnessy}}]{2019PhRvL.123r1101Y}
{Yang}, Y., {Bartos}, I., {Gayathri}, V., {et~al.} 2019, \prl, 123, 181101,
  \dodoi{10.1103/PhysRevLett.123.181101}

\end{thebibliography}
